\documentclass[12pt,a4paper]{article}

\usepackage[hidelinks]{hyperref}

\usepackage{afterpage, amsfonts, amsmath, amssymb, amsthm, bbm, caption, color, epstopdf, enumitem, float, graphicx, lipsum, longtable, lscape, mathrsfs, multirow, nccmath, nonfloat, pdflscape, rotating, tocloft, threeparttable}

\usepackage[margin=1in]{geometry}

\usepackage[round]{natbib}
\usepackage[doublespacing]{setspace}
\usepackage[table]{xcolor}

\def\be{\begin{equation}}
\def\ee{\end{equation}}
\def\bea{\begin{eqnarray}}
\def\eea{\end{eqnarray}}

\author{}
\title{}

\DeclareMathOperator*{\argmin}{\arg\!\min}

\DeclareMathOperator*{\plim}{p\!\lim}

\def\*#1{\mathbf{#1}}
\def\+#1{\boldsymbol{#1}}

\begin{document}
\newcommand\blfootnote[1]{
\begingroup
\renewcommand\thefootnote{}\footnote{#1}
\addtocounter{footnote}{-1}
\endgroup
}

\newtheorem{theorem}{Theorem}[section]
\newtheorem{corollary}{Corollary}[section]
\newtheorem{lemma}{Lemma}[section]
\theoremstyle{definition}
\newtheorem{assumption}{Assumption}
\newtheorem{remark}{Remark}
\newtheorem{step}{Step}
\newtheorem{dgp}{DGP}

\numberwithin{corollary}{section}
\numberwithin{equation}{section}
\numberwithin{lemma}{section}
\numberwithin{remark}{section}
\numberwithin{theorem}{section}

\allowdisplaybreaks[4]

\begin{titlepage}

{\small

\begin{center}
{\Large \bf Interactive Effects Panel Data Models with General Factors and Regressors\blfootnote{Previous versions of the paper were presented at seminars at Aarhus University, Lund University and Michigan State University. Westerlund would like to thank seminar participants, and in particular Richard Baillie, Nicholas Brown, David Edgerton, Yousef Kaddoura, Yana Petrova, Simon Reese, Tim Vogelsang, Jeffrey Wooldridge, and Morten {\O}rregaard Nielsen. Westerlund thanks the Knut and Alice Wallenberg Foundation for financial support through a Wallenberg Academy Fellowship. Peng and Westerlund also acknowledge the Australian Research Council Discovery Grants Program for its financial support under grant number DP210100476.

\textit{Correspondence:} Department of Econometrics and Business Statistics, Monash University, Australia. Email: Bin.Peng@monash.edu.
}

} 
\medskip

{\sc Bin Peng$^{\star}$, Liangju Su$^\ddagger$, Joakim Westerlund$^{\sharp}$ and Yanrong Yang$^{\dag}$}

\medskip

$^\star$Monash University, $^\ddagger$Tsinghua University, $^\sharp$Lund University,  $^{\dag}$Australian National University

\bigskip\bigskip

\today

\begin{abstract}
This paper considers a model with general regressors and unobservable factors. An estimator based on iterated principal components is proposed, which is shown to be not only asymptotically normal and oracle efficient, but under certain conditions also free of the otherwise so common asymptotic incidental parameters bias. Interestingly, the conditions required to achieve unbiasedness become weaker the stronger the trends in the factors, and if the trending is strong enough unbiasedness comes at no cost at all. In particular, the approach does not require any knowledge of how many factors there are, or whether they are deterministic or stochastic. The order of integration of the factors is also treated as unknown, as is the order of integration of the regressors, which means that there is no need to pre-test for unit roots, or to decide on which deterministic terms to include in the model.
\end{abstract}

\end{center}


\medskip


}

\end{titlepage}

\section{Introduction}\label{sect:intro}

The use of panel data in regression analysis has attracted considerable attention in the empirical literature in economics and elsewhere. A major reason for this is the ability to deal with the presence of unobserved heterogeneity, and the problem that this causes when said heterogeneity is correlated with the regressors. The simplest approach is to assume that the unobserved heterogeneity is made up of additive individual- and time-specific constants, or ``fixed effects''. Such effects can add hugely to the fit of the model, and are typically found to be much more important than the regressors predicted by economic theory. However, in many applications fixed effects are unlikely to be enough to capture the unobserved heterogeneity (see, for example, \citealp{Lemmonetal2008}, and \citealp{DeAngeloRoll2015}), and this has in turn spurred much interest in so-called ``interactive effects'' models, in which the individual and time effects enter in an multiplicative way. One of the most common approaches to such interactive effects models by far is the principal components (PC) approach of \cite{Bai}. In fact, it is so common that it has given rise to a separate strand of literature (see \citealp{Ando}, \citealp{bai2014}, \citealp{LiQianSu}, \citealp{MoonWeidner2015}, to mention a few).\footnote{While popular, PC is not the only approach that can be used to estimate interactive effects models. One alternative is the so-called ``common correlated effects'' (CCE) estimator of \cite{Pesaran2006}. The problem with this approach is that it requires that the number of time effects, or factors, is bounded by the number of observables, and that the regressors load on the same factors as the dependent variable, which is not necessary in PC. There is also the generalized method of moments (GMM) approach of  \cite{ALS13}. However, this approach supposes that the number of time periods is ``small'', and is not suitable for the type of large panel data sets that we have in mind.} The present paper aims to contribute to this strand, and it does so in at least three ways.

The first contribution of the paper is to consider a very general data generating process (DGP) that includes most of the specifications considered previously in the literature as special cases. The only requirement is that suitably normalized sample second moment matrices of the factors and regressors have positive definite limits. This is noteworthy because the existing literature is almost exclusively based on the assumption that both the factors and regressors are stationary. The only exceptions known to us are \cite{BaiKaoNg}, \cite{DGP2020} and \cite{HJPS}, but they assume instead that either the regressors or the factors and the regressors are pure unit root processes, which is also not realistic. Indeed, regressors and factors of different order of magnitude are likely to be the rule rather than the exception, especially in economic and financial data, due to differences in
persistence over time.

The unrestricted DGP is important in itself but also because it can be accommodated without requiring any knowledge thereof. Hence, not only do we treat the factors and their number as unknown, but we also do not require any knowledge of the order of magnitude of both factors and regressors. An important implication of this is that there is no need to distinguish between deterministic and stochastic factors, or stationary and non-stationary factors. In the existing literature, deterministic factors are often treated as known, and are projected out prior to the application of PC (see, for example, \citealp{MoonWeidner2015}). The problem here is that there is typically great uncertainty over which deterministic terms to include, which raises the issue of model misspecification. The fact that in the present paper deterministic terms are treated as additional factors means that the problem of deciding on which terms to include does not arise. Similarly, while the regressors can be tested for unit roots, and the estimation can be made conditional on the test outcome, this raises the issue of pre-testing bias. In the present paper we do not require any knowledge about the order of integration of the regressors, which means that there is no need for any pre-testing.

Equally as important as the general model formulation and its empirical appeal is the extension of the existing econometric theory, which has not yet ventured much outside the stationary or pure unit root environments. This is our second major contribution. The main difficulty here is not the unrestricted specification of the factors and regressors per se, but rather that the order of magnitude of the factors may differ. In particular, the problem is that the nonlinearity of the PC estimator distorts the signal coming from the factors, just as it does in estimation of nonlinear regression models with mixtures of integrated regressors (see, for example, \citealp{ParkPhillips2000}). This is true if both the number and order of the factors are known, and the problem does not become any simpler when these quantities are treated as unknown, as they are here. An additional problem that then arises is that existing studies on the selection of the number of factors all require that the data are stationary (see, for example, \citealp{BaiNg2002}, and \citealp{AH13}), and it is not obvious how one should go about this when the order of magnitude of the factor is unknown.

Intuitively, the factors whose order is largest should dominate the PC estimator. This motivates the use of an iterative estimation procedure in which the factors and their number are estimated in order according to their magnitude with relatively larger factors being estimated first. We begin by prescribing a large number of factors, and estimate the resulting model by PC. The estimated factors only capture the most dominating factors whose order of magnitude is largest. In spite of this, we can show that the estimator is consistent, albeit at a relatively low rate of convergence. The rate is, however, high enough to ensure that the number of dominating factors can be consistently estimated using a version of the eigenvalue ratio approach of \cite{LY12}, and \cite{AH13}. We then apply PC conditional on the first-step factor estimates, and estimate the second most dominating factors. This procedure continues until we cannot identify any more factors, and we can show that both the estimated factors and their number are consistent. Because of the iterative fashion in which the factors are estimated, we refer to the new estimation procedure as ``iterative PC'' (IPC), which is shown to be asymptotically normal and ``oracle efficient''.

Our third contribution is to point to a ``blessing'' of trending factors. The blessing occurs if the magnitude of the factors is sufficiently large, in which case the otherwise so common asymptotic bias of the PC approach can be completely eliminated without imposing any additional restrictions on the cross-sectional and time series dependencies of the regression errors. This is noteworthy, because the conclusion made in the previous literature suggests that in order to eliminate the asymptotic bias, the errors have to be independent.

The reminder of the paper is organized as follows. We begin by describing the model that we will be considering and the proposed IPC approach that we will use to estimate it. This is done in Section \ref{sect:est}. Section \ref{sect:asy} presents the formal assumptions and our main asymptotic results, whose small-sample accuracy is evaluated using Monte Carlo simulations in
Section \ref{sect:mc}. Section \ref{sect:ill} reports the results obtained from an empirical application to the long-run relationship between US house prices and income. For the sake of space, we present an extra empirical study on the returns to scale in the US banking industry in the supplementary appendix. Section \ref{sect:concl} concludes. All proofs are relegated to the online appendix.

\section{Model and estimation procedure}\label{sect:est}

Consider the panel data variable $y_{i,t}$, observable for $i=1,\ldots,N$ cross-sectional units and $t=1,\ldots,T$ time periods. The model of this variable that we will be considering is given by
\begin{eqnarray}  \label{eq:yit}
y_{i,t} = \*x_{i,t}^{\prime} \+\beta^0+ \+\gamma_{i}^{0\prime} \*f_{t}^0 + \varepsilon_{i,t},
\end{eqnarray}
where $\*x_{i,t} = (x_{1,i,t},\ldots,x_{d_x,i,t})^{\prime }$ is a $d_x\times 1$ vector of regressors, $\*f_{t}^0=(f_{1,t}^{0},\ldots,f_{d_f,t}^{0})^{\prime }$ is a $d_f\times 1$ vector of unobservable common factors with $\+\gamma_{i}^0 = (\gamma_{1,i}^0,\ldots,\gamma_{d_f,i}^0)^{\prime }$ being a conformable vector of factor loadings, and $\varepsilon_{i,t}$ is an idiosyncratic error term. Moreover, the factors are divided into groups according to their order of magnitude. There are $G$ groups of size $d_1,\ldots,d_G$, which means that $d_1+\cdots+d_G = d_f$. Because the grouping is unknown, we may without loss of generality assume that the factors are ordered, such that the first $d_1$ factors have the highest order of magnitude, the next $d_2$ factors have the second highest order, and so on. Hence, if we denote by $\*f_{g,t}^0$ and $\+\gamma_{g,i}^0$ the $d_g\times 1$ vectors of factors and loadings associated with group $g$, respectively, then $\+\gamma_{i}^{0\prime} \*f_{t}^0 = \sum_{g=1}^G \+\gamma_{g,i}^{0\prime}\*f_{g,t}^0$, where $\*f_{t}^0 =(\* f_{1,t}^{0\prime},\ldots,\*f_{G,t}^{0\prime})^{\prime}=(f_{1,t}^{0},\ldots,f_{d_f,t}^{0})^{\prime }$ and $\+\gamma_{i}^0 = (\+\gamma_{1,i}^{0\prime},\ldots,\+\gamma_{G,i}^{0\prime})^{\prime }= (\gamma_{1,i}^0,\ldots,\gamma_{d_f,i}^0)^{\prime }$. If $d_f = 0$, then $G=0$ and $\+\gamma_{i}^{0\prime} \*f_{t}^0 = 0$.

\begin{remark}
The (implicit) assumption of a common slope coefficient $\+\beta^0$ is standard in the literature (see, for example, \citealp{Bai}, and \citealp{MoonWeidner2015}). We speculate that our results can be extended to accommodate heterogeneous slope coefficients as in, for example, \cite{Castagnetti2015}, \cite{Castagnetti2019},
\cite{LCL2020}, and  \cite{HJPS}.
\end{remark}

It is useful to write \eqref{eq:yit} on stacked vector form. Let us therefore introduce the $T\times 1$ vectors $\*y_i =(y_{i,1},\ldots,
y_{i,T})^{\prime }$ and $\+\varepsilon_i = (\varepsilon_{i,1},\ldots, \varepsilon_{i,T})^{\prime }$, the $T\times d_f$ matrix $\*F^0 = (\*f_{1}^0,\ldots,\*f_{T}^0)^{\prime }$, and the $T\times d_x$ matrix $\*X_i =(\*x_{i,1},\ldots, \*x_{i,T})^{\prime }$. Analogous to $\*f_{t}^0$, $\*F^0$ is further partitioned as $\*F^0 =(\*F_{1}^{0},\ldots,\*F_{G}^{0})^{\prime }$, where the $T\times d_g$ matrix $\*F^0_g = (\*f_{g,1}^0,\ldots,\*f_{g,T}^0)^{\prime }$ contains the stacked factor observations for group $g$. In this notation, \eqref{eq:yit} can be written as
\begin{eqnarray}
\*y_{i} = \*X_{i}\+\beta^0 + \*F^{0}\+\gamma_{i}^0 + \+\varepsilon_{i}.
\end{eqnarray}

The goal of this paper is to infer $\+\beta^0$. As mentioned in Section \ref{sect:intro}, however, because of the generality of the model being considered, the main difficulty in the estimation process is how to control for $\*F^0$. Our proposed estimation procedure consists of three steps. We first initialize the estimation procedure by applying the PC estimator of  \cite{Bai}. However, because the first group of factors dominates all the other groups in terms of order of magnitude, the first-step PC factor estimator will only be estimating (the space spanned by) $\*F_{1}^0$. The second step of the procedure therefore involves iteratively applying PC conditional on previous factor estimates to estimate all subsequent groups of factors; hence, the ``I'' in IPC. In the third and final step, we estimate $\+\beta^0$ conditional on the second-step IPC estimator of $\*F^0$ and the first-step PC estimator of $\+\beta^0$.

\begin{step}[Initial estimation]\ \label{step1}
The (concentrated) ordinary least squares (OLS) objective function that we consider is the same as in \cite{Bai}. It is given by
\begin{eqnarray}
\mathrm{SSR}(\+\beta,\*F)= \sum_{i=1}^N (\*y_i-\*X_i\+\beta)^{\prime }\*M_F(\*y_i-\*X_i\+\beta),
\end{eqnarray}
where $\*F$ is a $T\times d_{max}$ matrix satisfying $T^{-\delta}\*F^{\prime}\*F = \*I_{d_{max}}$, $d_{max} \in \mathbb{N} \ge d_f$ and $\delta\in[0,\infty)$ are user-specified numbers, and $\*M_A=\*I_T- \*A(\*A^{\prime }\*A)^{-1}\*A^{\prime }=\*I_T- \*P_A$ for any $T$-rowed full column rank matrix $\*A$. As we explain in Remark \ref{rem:delta}, the IPC estimator of $\+\beta^0$ is invariant to the choice of $\delta$ and the need to select $d_{max}$ is standard. The initial estimator that we consider is the
minimizer of $\mathrm{SSR}(\+\beta,\*F)$:
\begin{eqnarray}  \label{eq:step1est}
(\widehat{\+\beta}_0, \widehat{\*F}_0) = \argmin_{(\+\beta,\*F)\in \mathbb{D}}\, \mathrm{SSR}(\+\beta,\*F),
\end{eqnarray}
where $\mathbb{D} = \mathbb{R}^{d_x} \times \mathbb{D}_F $ and $\mathbb{D}_F= \{ \*F : T^{-\delta}\*F^{\prime }\*F = \*I_{d_{max}} \}$. It is useful to note that $\widehat{\+\beta}_0$ satisfies $\widehat{\+\beta}_0 = \widehat{\+\beta}(\widehat{\*F}_0)$, where
\begin{align}  \label{defbetaf}
\widehat{\+\beta}(\*F) = \left(\sum_{i=1}^N \*X_i^{\prime }\*M_{F}\*X_i\right)^{-1} \sum_{i=1}^N \*X_i^{\prime }\*M_{F}\*y_i.
\end{align}
\end{step}

\begin{step}[Iterative estimation of factors]\ \label{step2}
As already pointed out, the factors are estimated in order according to magnitude. Therefore, $\widehat{\*F}_0$ is estimating $\*F_1^0$. Since $d_1 \leq d_f \leq d_{max}$, in general the dimension of $\widehat{\*F}_0$ will be larger than that of $\*F_1^0$. We therefore begin this step of the estimation procedure by estimating $d_1$, and for this purpose we employ a version of the ratio of eigenvalue-based estimator considered by, for example, \cite{LY12}, and \cite{AH13}, which is given by
\begin{eqnarray}
\widehat{d}_1 = \argmin_{0\leq d\le d_{max}}\left\{ \frac{\widehat{\lambda}_{1,d+1}}{\widehat{\lambda}_{1,d}}\cdot \mathbb{I}\left( \frac{\widehat{\lambda}_{1,d}}{\widehat{\lambda}_{1,0}}\ge \tau_{N}\right) + \mathbb{I}\left( \frac{\widehat{\lambda}_{1,d}}{\widehat{\lambda}_{1,0}} <\tau_{N}\right)\right\},
\end{eqnarray}
where $\mathbb{I}(A)$ is the indicator function for the event $A$ taking the value one if $A$ is true and zero otherwise, $\tau_{N} = 1/\ln\, (\max\{\widehat{\lambda}_{1,0},N\})$, $\widehat{\lambda}_{1,0} = N^{-1}\sum_{i=1}^N (\*y_i -\*X_i\widehat{\+\beta}_0)^{\prime }\*M_{\widehat{F}_0}(\*y_i -\*X_i \widehat{\+\beta}_0)$ and $\widehat{\lambda}_{1,1}\geq \cdots \geq \widehat{\lambda}_{1,d_{max}}$ are the $d_{max}$ largest eigenvalues of the following  $T\times T$ matrix:
\begin{eqnarray}  \label{defSig1}
\widehat{\+\Sigma}_1 = \frac{1}{N}\sum_{i=1}^N (\*y_i -\*X_i\widehat{\+\beta}_0)(\*y_i -\*X_i\widehat{\+\beta}_0)^{\prime }.
\end{eqnarray}
The threshold $\tau_{N}$, the ``mock'' eigenvalue $\widehat{\lambda}_{1,0}$, and the indicator function are there to ensure that the estimator is consistent. The need for these will be explained later. Given $\widehat d_1$, we update the estimate of $\*F_{1}^0$ by setting $\widehat{\*F}_1$ equal to $T^{\delta/2}$ times the eigenvectors associated with $\widehat{\lambda}_{1,1},\ldots, \widehat{\lambda}_{1,\widehat{d}_1}$, and estimate $\+\gamma_{1,i}^0$ by $\widehat{\+\gamma}_{1,i} = T^{-\delta}\widehat{\*F}_1^{\prime }(\*y_i -\*X_i\widehat{\+\beta}_0)$.

The estimation of $\*F_2^0,\ldots,\*F_G^0$ is analogous to that of $\*F_1^0$. The main difference is that we have to condition on all previous estimates. Let us therefore use $\widehat{\*F}_{-g} = (\widehat{\*F}_1,\ldots,\widehat{\*F}_{g-1})$ and $\widehat{\+\gamma}_{-g,i} = (\widehat{\+\gamma}_{1,i}^{\prime },\ldots,\widehat{\+\gamma}_{g-1,i}^{\prime})^{\prime }$ to denote the matrices containing the previously estimated factors and loadings, respectively, when estimating group $g > 1$. The estimator of $d_g$ is given quite naturally by
\begin{eqnarray}
\widehat{d}_g = \argmin_{0\leq d\le d_{max}}\left\{ \frac{\widehat{\lambda}_{g,d+1}}{\widehat{\lambda}_{g,d}}\cdot \mathbb{I}\left( \frac{\widehat{\lambda}_{g,d}}{\widehat{\lambda}_{g,0}}\ge \tau_{N}\right) + \mathbb{I}\left( \frac{\widehat{\lambda}_{g,d}}{\widehat{\lambda}_{g,0}} < \tau_{N}\right)\right\},
\end{eqnarray}
where $\widehat{\lambda}_{g,0} = N^{-1}\sum_{i=1}^N (\*y_i -\*X_i\widehat{\+\beta}_0)^{\prime }\*M_{\widehat{F}_{-g}}(\*y_i -\*X_i\widehat{\+\beta}_0)$ and $\widehat{\lambda}_{g,1}\geq \cdots \geq \widehat{\lambda}_{g,d_{max}}$ are the $d_{max}$ largest eigenvalues of
\begin{eqnarray}
\widehat{\+\Sigma}_g = \frac{1}{N}\sum_{i=1}^N (\*y_i -\*X_i\widehat{\+\beta}_0 - \widehat{\*F}_{-g}\widehat{\+\gamma}_{-g,i})(\*y_i -\*X_i\widehat{\+\beta}_0- \widehat{\*F}_{-g}\widehat{\+\gamma}_{-g,i})^{\prime }.
\end{eqnarray}
The resulting estimator $\widehat{\*F}_g$ of $\*F_{g}^0$ is given by the eigenvectors associated with $\widehat{\lambda}_{g,1},\ldots, \widehat{\lambda}_{g,\widehat{d}_g}$ and $\widehat{\+\gamma}_{g,i} = T^{-\delta}\widehat{\*F}_g^{\prime }(\*y_i -\*X_i\widehat{\+\beta}_0 - \widehat{\*F}_{-g}\widehat{\+\gamma}_{-g,i})$. New groups of factors are estimated until $\widehat{d}_{g} =0$. At this point, we set $\widehat G = g-1$ and define $\widehat{\*F} = (\widehat{\*F}_1,\ldots,\widehat{\*F}_{\widehat{G}})$. This is the IPC estimator of $\*F^0$.
\end{step}

\begin{step}[Estimation of $\+\protect\beta^0$]\ \label{step3} Given $\widehat{\*F}$, we compute $\widehat{\+\beta}_1 = \widehat{\+\beta}(\widehat{\*F})$ using \eqref{defbetaf}. The IPC-based estimator of $\+\beta^0$ is given by
\begin{eqnarray}\label{estbhat}
\widehat{\+\beta} = \widehat{\+\beta}_0+\left(\sum_{i=1}^N \widehat{\*Z}_i^{\prime }\widehat{\*Z}_i \right)^{-1}\sum_{i=1}^N \*X_i^{\prime }\*M_{\widehat{F}}\*X_i ( \widehat{\+\beta}_1-\widehat{\+\beta}_0) ,
\end{eqnarray}
where $\widehat{\*Z}_i = \*M_{\widehat{F}}\*X_i - \sum_{j=1}^{N}\*M_{\widehat{F}}\*X_j\widehat a_{ij}$ with $\widehat a_{ij} = \widehat{\+\gamma}_{i}^{\prime }(\widehat{\+\Gamma}^{\prime }\widehat{\+\Gamma})^{-1}\widehat{\+\gamma}_{j}$, $\widehat{\+\gamma}_i =(\widehat{\+\gamma}_{1,i}^{\prime},\ldots,\widehat{\+\gamma}_{\widehat G,i}^{\prime })^{\prime }$ and $\widehat{\+\Gamma} =(\widehat{\+\gamma}_{1},\ldots,\widehat{\+\gamma}_{N})^{\prime }$.
\end{step}

\begin{remark}\label{rem:delta}
The reason for why in Steps \ref{step1} and \ref{step2} $\delta$ can be set arbitrarily is that the only use of this parameter is to normalize $\widehat{\*F}$ and $T^{-\delta/2}\|\widehat{\*F}_0\| = \sqrt{d_{max}}$ for any $\delta\in [0,\infty)$. This result is nice from both applied and theoretical points of view. In the bulk of the previous literature, the appropriate value of $\delta$ to use depends on whether $\* F^0$ is stationary or unit root non-stationary (see, for example, \citealp{Bai2004}). The assumed knowledge of $\delta$ is therefore tantamount to assuming that the order of integration of $\*F^0$ is known, which is again not a requirement here. The need to specify a maximum $d_{max}$ for the number of factors is standard in the literature (see, for example, \citealp{BaiNg2002}). In Sections \ref{sect:mc} and \ref{sect:ill}, we explain how $d_{max}$ is set in the Monte Carlo and empirical studies, respectively.
\end{remark}

\begin{remark}
Many papers use information criteria to select the number of factors (see, for example, \citealp{BaiNg2002}). We do not. The reason is that the eigenvalue ratio is self-normalizing, which makes it possible to handle factors that are of different order of magnitude. The general idea behind the eigenvalue ratio approach in Step \ref{step2} is similar to the one used in ocular inspection of ``scree plots"; one looks for the point at which the ordered eigenvalues drop substantially, and set the number of estimated factors accordingly. The most natural way to mimic this decision rule is to minimize $\widehat{\lambda }_{g,d+1}/\widehat{\lambda }_{g,d}$ over $1\leq d\leq d_{max}$. However, this raises two issues. One is that since $\widehat{\lambda }_{g,d+1}/\widehat{\lambda }_{g,d}$ is not defined for $d=0$, we cannot have $d_{f}=0$, and we want to be able to entertain the possibility that there are no factors. The use of the mock eigenvalue $\widehat{\lambda }_{g,0}$ allows us to do just that. The other problem is that under the conditions of this paper (laid out in Section \ref{sect:asy}) the limiting behavior of $\widehat{\lambda }_{g,d+1}/\widehat{\lambda }_{g,d}$ when $d>d_{g}$ is unknown. \cite{LY12} discuss this issue at length. They conjecture that $\widehat{\lambda }_{g,d+1}/\widehat{\lambda }_{g,d}\asymp 1$, where $a\asymp b$ means that $a=O_{P}(b)$ and $b=O_{P}(a)$, but there is no proof. The challenge when studying $\widehat{\lambda }_{g,d+1}/\widehat{\lambda }_{g,d}$ for $d>d_{g}$ is to bound this ratio from below, as both of the numerator and denominator converge to zero for a certain choice of $\delta $ at the same rate. The use of the indicator function allows us to circumvent this problem. The idea is to look at $\widehat{\lambda }_{g,d}$ only. If this eigenvalue is ``small", we take it as a sign of $d>d_{g}$ and set $\widehat{\lambda }_{g,d+1}/\widehat{\lambda }_{g,d}$ to one. However, because the order of $\*f_{t}^{0}$ is assumed to be unknown, we cannot look at $\widehat{\lambda }_{g,d}$ directly but rather we look at $\widehat{\lambda }_{g,d}/\widehat{\lambda }_{g,0}$, which in contrast to $\widehat{\lambda }_{g,d}$ is self-normalizing. \cite{BT2021} also look at ratios of eigenvalues as a way to determine the number of factors when their order is unknown. However, they only consider factors that are either trending linearly, unit root non-stationary or stationary, and the implementation of their procedure is quite involved. We therefore do not consider it here.
\end{remark}

\begin{remark}
Intuition suggests to take $\widehat{\+\beta}_1$, the OLS estimator conditional on $\widehat{\*F}$, as the final estimator of $\+\beta^0$ in Step \ref{step3}. Interestingly, while consistent, because of the stepwise estimation of the factors, the asymptotic distribution of $\widehat{\+\beta}_1$ is generally not (mixed) normal and nuisance parameter-free. In fact, as we show later in Lemma \ref{lem:beta0hat}, even $\widehat{\+\beta}_0$ is consistent but again with non-normal asymptotic distribution. In Section \ref{sect:mc}, we use Monte Carlo simulations to shed some light on the consistency and non-normality of $\widehat{\+\beta}_0$ and $\widehat{\+\beta}_1$.
\end{remark}

\section{Assumptions and asymptotic results}\label{sect:asy}

Before we state our assumptions and asymptotic results, we introduce some notation. Specifically, if $\*A$ is a matrix, $\lambda_{min}(\*A)$ and $\lambda_{max}(\*A)$ signify its smallest and largest eigenvalues, respectively, $\text{tr}\,\*A$ signifies its trace, and $\|\*A\| = \sqrt{\text{tr}\,\*A'\*A}$ and $\|\*A\|_2 = \sqrt{\lambda_{max}(\*A)}$ signify its Frobenius and spectral norms, respectively. We write $\*A > 0$ to signify that $\*A$ is positive definite. If $\*B$ is also a matrix, then $\mathrm{diag}(\*A, \*B)$ denotes the block-diagonal matrix that takes $\*A$ ($\*B$) as the upper left (lower right) block. The symbols $\to_D$, $\to_P$ and $MN(\cdot, \cdot)$ signify convergence in distribution, convergence in probability and a mixed normal distribution, respectively. We use $N,\,T\to\infty$ to indicate that the limit has been taken while passing both $N$ and $T$ to infinity. We use w.p.a.1 to denote with probability approaching one.

Assumption \ref{ass:mom} is concerned with the order of magnitude of $\*f_t^0$ and $\*x_{i,t}$. It is therefore key. The assumption is stated in terms of the required moment conditions rather than primitive assumptions on $\*f_t^0$ and $\*x_{i,t}$. The reason is that we would like to be agnostic about these variables.

\begin{assumption}[Moments]\ \label{ass:mom}

\begin{itemize}
\item[(a)] There exists a matrix $\+\Sigma_X$ such that $E\|(NT)^{-1}\sum_{i=1}^N \*D_T \*X_i^{\prime }\*M_{F^0}\*X_i \*D_T -\+\Sigma_X \|^2 =o(1)$, where $\*D_T = \mathrm{diag}( T^{ -\kappa_1/2} ,\ldots, T^{-\kappa_{d_x}/2 } )$ with $0\le \kappa_j< \infty$ for $j=1,\ldots,d_x$, $E\|\+\Sigma_X\|^2<\infty$, and $0<\lambda_{min}(\+\Sigma_X) \le \lambda_{max}(\+\Sigma_X) <\infty$ w.p.a.1.

\item[(b)] $\|(NT)^{-1}\sum_{i=1}^N \*D_T\*X_i^{\prime }\+\varepsilon_i\| =O_P ( 1/\min\{\sqrt{N}, \sqrt{T}\})$ and $\|\+\varepsilon\|_{2} =O_P (\max\{\sqrt{N}, \sqrt{T}\} )$ with $\+\varepsilon = (\+\varepsilon_1,\ldots,\+\varepsilon_N)$.

\item[(c)] There exists a matrix $\+\Sigma_{F^0}$ such that $E\|\*C_T \*F^{0\prime}\*F^0 \*C_T- \+\Sigma_{F^0} \|^2 =o(1)$, where $\*C_T = \mathrm{diag}( T^{ -\nu_1/2} \*I_{d_1} ,\ldots, T^{-\nu_G/2 } \*I_{d_G})$ with $\nu_1>\cdots >\nu_G> 1/2$, $E\| \+\Sigma_{F^0}\|^2<\infty$ and $0<\lambda_{min}(\+\Sigma_{F^0} )\le \lambda_{max}(\+\Sigma_{F^0} )<\infty$ w.p.a.1.

\item[(d)] There exists a matrix $\+\Sigma_{\Gamma^0}$ such that $\|N^{-1}\+\Gamma^{0\prime}\+\Gamma^0- \+\Sigma_{\Gamma^0}\| =o_P(1)$ and $\max_{i\ge 1}E\| \+\gamma_{i}^0\|^4<\infty$, where $\+\Gamma^0 =(\+\gamma_{1}^0,\ldots,\+\gamma_{N}^0)^{\prime }$ is $N\times d_f$ and $ 0<\lambda_{min}(\+\Sigma_{\Gamma^0} )\le \lambda_{max}(\+\Sigma_{\Gamma^0})<\infty$.
\end{itemize}
\end{assumption}

Assumption \ref{ass:mom} (a) is very general in that it imposes almost no restrictions on the type of trending behaviour that $\*x_{i,t}$ may have. The trending can be deterministic but it can also be stochastic, as in the presence of unit roots. Either way, the degree of the trending is not restricted in any way, provided that it is finite. The only requirement is that $0<\lambda _{min}(\+\Sigma _{X})\leq \lambda _{max}(\+\Sigma _{X})<\infty$ w.p.a.1, which means that the elements of $\*x_{i,t}$ cannot be asymptotically collinear. Note that $\+\Sigma _{X}$ is not required to be a constant matrix, as this would rule out regressors that are stochastically integrated. This is very different when compared to the bulk of the previous PC-based interactive effects literature in which $\*x_{i,t}$ is assumed to be either stationary (see, for example, \citealp{Bai}, and \citealp{MoonWeidner2015}), such that $\kappa _{1}=\cdots =\kappa _{d_{x}}=0$, or unit-root non-stationary (see \citealp{BaiKaoNg}, and \citealp{DGP2020}), such that $\kappa _{1}=\cdots =\kappa _{d_{x}}=1$. Assumption \ref{ass:mom} (a) is analogous to the work of, for example, \cite{DongLinton}, wherein time series of different order of magnitude are considered.

The first requirement of Assumption \ref{ass:mom} (b) is quite mild and holds if a central limit theorem in only one of the two panel dimensions applies to the normalized sum of $\mathbf{D}_{T}\mathbf{X}_{i}^{\prime }\+\varepsilon_{i}$. The second requirement is quite common in the literature, and is expected to hold as long as $\varepsilon_{i,t}$ has zero mean, and weak serial and cross-sectional correlation (see \citealp{MoonWeidner2015}, for a discussion).

Assumption \ref{ass:mom} (c) is similar to (a) in that it leaves the trending behavior of the factors essentially unrestricted. A majority of previous PC-based works assume that $T^{-1}\*F^{0\prime }\*F^{0}$ converges to positive definite matrix (see, for example, \citealp{Bai}, and \citealp{MoonWeidner2015}). Notable exceptions include \cite{Bai2004}, \cite{BaiKaoNg}, and \cite{Choi2017}, in which $\*f_{t}^{0}$ is assumed to follow pure unit root process, and \cite{BaiNg2004}, who allow for a mix of
stationary and unit root factors. The only study that comes close to ours in terms of the generality of the factors is that of \cite{Westerlund2018}. However, he assumes that $\*x_{i,t}$ has a factor structure that loads on the same set of factors as $y_{i,t}$, which is not required here. Also, unlike \cite{Westerlund2018}, we do not require $\nu _{G}\geq 1$, but also allow $1/2<\nu _{G}<1$, such that $T^{-1}\*F_{G}^{0\prime }\*F_{G}^{0}$ is asymptotically singular. The signal coming from $\*F_{G}^{0}$ is therefore even weaker than under stationarity, and we will therefore refer to these factors as ``signal-weak", as opposed to ``weak" in usual sense (see, for example, \citealp{Chudiketal2011} and \cite{UY2020}, for discussions). One example of such signal-weak factors is when $\*f_{G,t}^{0}$ is stationary and sparse.

Assumption \ref{ass:mom} (d) is standard and ensures that each factor has a nontrivial contribution to the variance of $y_{i,t}$.

\begin{assumption}[Identification]\ \label{ass:id}
$\inf_{\*F\in \mathbb{D}_F} \lambda_{min}(\*B(\*F))\geq c_0 >0$ for all $N$ and $T$, where
\begin{align}
\*B(\*F) = \frac{1}{NT}\sum_{i=1}^{N}\*D_T \*Z_i(\*F)^{\prime }\*Z_i(\*F)\*D_T
\end{align}
with $\*Z_i(\*F) = \*M_{F}\*X_i - \sum_{j=1}^{N}\*M_F\*X_ja_{ij}$ and $a_{ij} =\+\gamma_{i}^{0\prime}(\+\Gamma^{0\prime}\+\Gamma^0)^{-1}\+\gamma_{j}^0$.
\end{assumption}

The requirement that the smallest eigenvalue of $\*B(\*F)$ should be positive is equivalent to requiring that $\*B(\*F)$ be positive definite uniformly in $\*F$, which is a non-collinearity condition that rules out low-rank regressors (see \citealp{MoonWeidner2015}, for a discussion). It demands that the regressors in $\*X_{i}$ have enough variation after projecting out all variation that can be explained by $\*F^{0}$ and $\+\gamma _{i}^{0}$. Time-invariant regressors are therefore not allowed, which is similar to the conventional cross-section fixed effects-only OLS condition. Assumption \ref{ass:id} is a high-level condition, just like Assumption \ref{ass:mom} (see, for example, \citealp{Bai}, and \citealp{MoonWeidner2015}, for similar high-level conditions). The reason for formulating it in this way is again that we want to avoid making highly specific and potentially invalid assumptions about the DGP of $\*x_{i,t}$, $\*f_{t}^{0}$ and $\+\gamma _{i}^{0}$. If all the regressors are stationary, Assumption \ref{ass:id} reduces to Assumption A in \cite{Bai}.

Assumptions \ref{ass:mom} and \ref{ass:id} are enough to ensure that the initial estimator $\widehat{\+\beta}_0$ of $\+\beta^0$ is consistent.

\begin{lemma}[Consistency of $\protect\widehat{\+\protect\beta}_0$]\ \label{lem:beta0hat}
Under Assumptions \ref{ass:mom} and \ref{ass:id}, as $N,T\to \infty$,
\begin{eqnarray}
\min\{\sqrt{N}, \sqrt{T}\}\*D_T^{-1}(\widehat{\+\beta}_0-\+\beta^0) =O_P(1).
\end{eqnarray}
\end{lemma}

Lemma \ref{lem:beta0hat} establishes that $\widehat{\+\beta}_0$ is consistent for $\+\beta^0$ and that the rate of convergence is $\|\*D_T\|/\min\{\sqrt{N}, \sqrt{T}\} = \max\{T^{-\kappa_1/2}, \ldots, T^{-\kappa_{d_x}/2}\}/\min\{\sqrt{N}, \sqrt{T}\}$. To put this into
perspective, suppose that $\*x_{i,t}$ is stationary, such that $\kappa_1 = \cdots = \kappa_{d_x} = 0$. In this case, $\*D_T = \*I_{d_x}$ and the rate of convergence is given by $1/\min\{\sqrt{N}, \sqrt{T}\}$, which is the slowest of the regular rates in pure time series and cross-section regressions. Still, the rate is fast enough for the estimation of the number of factors. This brings us to Step \ref{step2} of the estimation procedure. In order to be able to show that $\widehat d_1,\ldots,\widehat d_{G+1}$ and $\widehat{\*F}$ are consistent, however, we need more structure.

\begin{assumption}[Errors]\ \label{ass:eps}

\begin{itemize}
\item[(a)] $E(\varepsilon_{i,t})=0$ and $E(\+\varepsilon_{i}\+\varepsilon_i^{\prime })=\+\Sigma_{\varepsilon,i} $.

\item[(b)] Let $\+\varepsilon_{t} = (\varepsilon_{1,t},\ldots,\varepsilon_{N,t})^{\prime }$ in this assumption only. $\{\+\varepsilon_{t}
: \ t\ge 1\}$ is strictly stationary and $\alpha$-mixing such that $\max_{i\geq 1}E|\varepsilon_{i,1}|^{4+\mu} <\infty$ for some $\mu >0$ and the mixing coefficient $\alpha(t) = \sup_{A\in \mathcal{F}_{-\infty}^0, B\in  \mathcal{F}_t^\infty} |P(A)P(B) -P(AB)|$ satisfies $\sum_{t=1}^\infty \alpha(t)^{\mu/(2+\mu)}<\infty$, where $\mathcal{F}_{-\infty}^t$ and $\mathcal{F}_{t}^\infty$ are the sigma-algebras generated by $\{\+\varepsilon_{s}: s \leq 0\}$ and $\{\+\varepsilon_{s}: s \geq t\}$,
respectively.

\item[(c)] Suppose that $\sum_{i=1}^N\sum_{j=1}^N\sum_{t=1}^T\sum_{s=1}^T|E\varepsilon_{i,t}\varepsilon_{j,s}|=O(NT)$ and $\sum_{i=1}^N \sum_{j\ne i}|\sigma_{\varepsilon,ij}|=O(N)$, where $\sigma_{\varepsilon,ij}=E(\varepsilon_{i,t}\varepsilon_{j,t})$.

\item[(d)] $\varepsilon_{i,t}$ is independent of $\+\gamma_j^0$, $\*f_s^0$ and $\*x_{j,s}$ for all $i$, $j$, $t$ and $s$.
\end{itemize}
\end{assumption}

Assumption \ref{ass:eps} is similar to Assumptions C and D in \cite{Bai}. Assumption \ref{ass:eps} (b) and (c) ensure that the serial and cross-sectional dependencies of $\varepsilon_{i,t}$ are at most weak. Assumption \ref{ass:eps} (d) requires that the regressors are exogenous, which rules out the presence of lagged dependent variables in $\*x_{i,t}$. Assumption \ref{ass:eps} (d) can be relaxed by instead requiring that $\varepsilon_{i,t}$ satisfies a martingale type condition similar to Assumption 3 in \cite{DGP2020}. However, since this will add to the already lengthy derivations and heavy notation, in the present paper we maintain Assumption \ref{ass:eps}.

If $\nu_G \geq 1$, Assumptions \ref{ass:mom}--\ref{ass:eps} are enough to ensure that $\widehat d_1$ is consistent. If, however, $\nu_G<1$, so that some of the factors are signal-weak, then we also need Assumption \ref{ass:weak}.

\begin{assumption}[Signal-weak factors]
\ \label{ass:weak} If $\nu_G<1$, then $T/N^2\to c_1\in [0,\infty)$.
\end{assumption}

As already pointed out, $\*F_{1}^0$ dominates all other factors and is thus easiest to estimate. It is therefore not surprising to find that $\widehat{d}_1$ is consistent for $d_1$. Lemma \ref{lem:d1hat} states this result formally.

\begin{lemma}[Consistency of $\protect\widehat{d}_1$]\ \label{lem:d1hat}
Under Assumptions \ref{ass:mom}--\ref{ass:weak}, as $N,\,T\to \infty$,
\begin{eqnarray}
P (\widehat{d}_1= d_1 )\to 1.
\end{eqnarray}
\end{lemma}

Hence, $\widehat{d}_1$ is consistent. In order to ensure that also $\widehat d_2,\ldots,\widehat d_{G+1}$ are consistent, the groups have to be ``distinct''. The next assumption formalizes this requirement. The assumption is stated in terms of $\*F_{g}^0$ and $\+\Gamma_{g}^0$, where $\*F_{g}^0$ is as before and $\+\Gamma_{g}^0 = (\+\gamma_{g,1}^{0},\ldots,\+\gamma_{g,N}^{0})^{\prime }$ is the $N\times d_g$ matrix of stacked factor loadings for group $g$.

 \begin{assumption}[Orthogonal groups]\label{ass:ortho}
$\max_{ g\ne h}\|\+\Gamma_{g}^{0\prime}\+\Gamma_{h}^0\|  =O_P(N^p)$ and $\max_{ g\ne h}\|\*F_{g}^{0\prime}\*F_{h}^0\| =O_P(T^q)$, where $g,h = 1,\ldots, G$, $G>1$, $p < 1$, $q<(\nu_G+\nu_{G-1})/2$ and $\nu_{G-1}\ge 1$.
\end{assumption}

The condition that $\nu _{G-1}\geq 1$ means that we only allow for one group of weak factors. This can be seen as a form of normalization and is not particularly restrictive. Let us therefore instead consider $\max_{g\neq h}\Vert \*F_{g}^{0\prime }\*F_{h}^{0}\Vert =O_{P}(T^{q})$, which is less restrictive that the exact orthogonality condition typically required in papers on grouped factor structures (see, for example, \citealp{Ando}). As is well known, $\+\gamma _{i}^{0\prime }\*f_{t}^{0}=\+\gamma _{i}^{0\prime }\*W^{-1}\*W\*f_{t}^{0}$ for any positive definite matrix $\*W$. Now set $\*W=(\*F^{0\prime }\*F^{0}\*C_{T}^{2})^{-1/2}$. This implies
\begin{equation}
\mathbf{C}_{T}\*W\mathbf{F}^{0 \prime }\mathbf{F}^{0 }\*W'\mathbf{C}_{T}=\mathbf{I}_{d_{f}}
\end{equation}
which means that the renormalized factors are exactly orthogonal, and hence that Assumption \ref{ass:ortho} is satisfied for $\mathbf{F}^{0 }\*W'$ with $q=-\infty $. This is one way in which $\max_{g\neq h}\Vert \*F_{g}^{0\prime }\*F_{h}^{0}\Vert =O_{P}(T^{q})$ can be justified. Another way is through orthogonal basis functions on $[0,1]$. For example, if $\*f_{t}^{0}=(1,t)^{\prime }$ and $\*C_{T}=\mathrm{diag}(T^{-1/2},T^{-3/2})$, then $\sqrt{T}\*C_{T}\*f_{t}^{0}\rightarrow \*f^{0}(w)=(1,w)^{\prime }$ and $\*C_{T}\*F^{0\prime }\*F^{0}\*C_{T}\rightarrow \int_{0}^{1}\*f^{0}(w)\*f^{0}(w)^{\prime }dw$ as $T\rightarrow \infty $ with $w\in \lbrack 0,1]$. Because they are square integrable, the elements of $\*f^{0}(w)$ can be represented in terms of orthogonal basis functions (see, for example, \citealp{DongLinton}). This is true not only for the simple example given here but for most commonly used deterministic trend terms. The loading condition, $\max_{g\neq h}\Vert \+\Gamma _{g}^{0\prime }\+\Gamma _{h}^{0}\Vert
=O_{P}(N^{p})$, can be rationalized in the same way as the factor condition. Another possibility is that $\+\Gamma _{1}^{0},\ldots ,\+\Gamma _{G}^{0}$ are independent and at most one of them has non-zero mean. Independence is often assumed and we therefore do not justify it here (see, for example, \citealp{Chudiketal2011}, and \citealp{Pesaran2006}). In order to justify the zero mean assumption, suppose for simplicity that $G=2$, that $d_{1}=\+\gamma _{1,i}^{0}=1$ and that $\+\gamma_{2,i}^{0}=\+\gamma _{2}^{0}+\+\eta _{i}$ with $E(\+\eta _{i})=\mathbf{0}_{d_{2}\times 1}$. Hence, $\+\gamma _{i}^{0}=(\+\gamma _{1,i}^{0}, \+\gamma_{2,i}^{0\prime })^{\prime }=(1,\+\gamma _{2}^{0\prime }+\+\eta _{i}^{\prime})^{\prime }$ and $\*f_{t}^{0}=(\*f_{1,t}^{0},\*f_{2,t}^{0\prime })^{\prime }$, which in turn implies that
\begin{equation}
\+\gamma _{i}^{0\prime }\*f_{t}^{0}=\*f_{1,t}^{0}+(\+\gamma _{2}^{0\prime}+\+\eta _{i}^{\prime })\*f_{2,t}^{0}=(\*f_{1,t}^{0}+\+\gamma _{2}^{0\prime }\*f_{2,t}^{0})+\+\eta _{i}^{\prime }\*f_{2,t}^{0}.
\end{equation}
The common component with $\+\gamma _{i}^{0}$ as loading and $\*f_{t}^{0}$ as factor can therefore be written equivalently as one in which $(\*f_{1,t}^{0}+\+\gamma _{2}^{0\prime }\*f_{2,t}^{0},\*f_{2,t}^{0\prime})^{\prime }$ is the factor and $(1,\+\eta _{i}^{\prime })^{\prime }$ is the
loading.

\begin{lemma}[Consistency of $\widehat d_2,\ldots,\widehat d_{G+1}$ and $\widehat{\*F}$]\ \label{lem:dghat}
Suppose that Assumptions \ref{ass:mom}--\ref{ass:ortho} are satisfied. Then, the following results hold as $N,\,T\to \infty$:
\begin{itemize}
\item[$\mathrm{(a)}$] $P (\widehat{d}_g = d_g)\to 1$ for $g=2,\ldots, G+1$, where $G> 1$ and  $d_{G+1} =0$;

\item[$\mathrm{(b)}$] $\| \*P_{\widehat{F}} -\*P_{F^0}\| =o_P(1)$.
\end{itemize}
\end{lemma}

Together with Lemma \ref{lem:d1hat}, Lemma \ref{lem:dghat} (a) implies that $\widehat d_1,\ldots,\widehat d_{G+1}$ are all consistent. The consistency of  $\widehat d_1,\ldots,\widehat d_G$ is important for obvious reasons. The consistency of $\widehat d_{G+1}$ ensures that the stopping rule in Step \ref{step2} of the estimation procedure is asymptotically valid, which in turn implies that $\widehat G$ is consistent. Hence, under the conditions of Lemma \ref{lem:dghat}, we have
\begin{eqnarray}
P (\widehat G =G )\to 1
\end{eqnarray}
as $N,\,T\to \infty$. The consistency of $\widehat G$ further implies that $d_f$ can be consistently estimated using $\widehat d_f = \widehat d_1 +\cdots+ \widehat d_{\widehat G}$.

As we alluded to in the discussion of Assumption \ref{ass:ortho} above, $\*F^{0}$ and $\+\gamma _{i}^{0}$ are only identified up to a rotation matrix (see, for example, \citealp{BaiNg2002}). However, we cannot claim that $\widehat{\*F}$ is rotationally consistent for $\*F^{0}$, as the number of rows of both objects is growing with $T$. We therefore have to resort to alternative consistency concepts. This is where Lemma \ref{lem:dghat} (b) comes in. It shows that the spaces spanned by $\widehat{\*F}$ and $\*F^{0}$ are asymptotically the same. This establishes that all the Step \ref{step2} estimates are consistent. We therefore move on to investigate the asymptotic properties of the IPC estimator $\widehat{\+\beta }$ of $\+\beta ^{0}$ obtained in Step \ref{step3}.

\begin{assumption}[Rates]\ \label{ass:rates}
\begin{itemize}
\item[(a)] $N/T^{\nu_G}\to \rho_1 \in [0,\infty)$;

\item[(b)] $T^{2-\nu_G}/N \to \rho_2 \in [0,\infty)$.
\end{itemize}
\end{assumption}

It is important to note that if all the factors in $\*f_t^0$ are stationary such that $G=1$ and $\nu_G  = \nu_1 = 1$, then Assumption \ref{ass:rates} (a) and (b) require that $N/T \to \rho_1 = 1/\rho_2  \in (0,\infty)$, which is the condition considered in \cite{Bai}, and \cite{MoonWeidner2015}. Assumption \ref{ass:rates} (a) and (b) are therefore more general than the condition considered in these other papers.

\begin{assumption}[Asymptotic normality]\ \label{ass:norm}
\begin{align}
 \frac{1}{\sqrt{NT}}\sum_{i=1}^N\*D_T\*Z_i(\*F^0)'\+\varepsilon_i \to_D MN(\*0_{d_x\times 1},\+\Omega)
\end{align}
as $N,\,T\to\infty$, where
\begin{align}
\+\Omega = \plim_{N,T\to\infty} \frac{1}{NT}\sum_{i=1}^N\sum_{j=1}^N\*D_T E[\*Z_i(\*F^0)'\+\varepsilon_i\+\varepsilon_j\*Z_j(\*F^0) | \mathcal{C}] \*D_T,
\end{align}
with $\mathcal{C}$ being the sigma-algebra generated by $\*F^0$.
\end{assumption}

Assumption \ref{ass:norm} is a central limit theorem that is analogous to Assumption E of \cite{Bai}. The reason for requiring that the asymptotic distribution is mixed normal as opposed to normal is that by doing so we can accommodate stochastically integrated factors (see, for example, \citealp{BaiKaoNg}, and \citealp{HJPS}). In the absence of such integrated factors, the mixed normal becomes normal. Either way, Assumption \ref{ass:norm} ensures that standard normal and chi-squared inference based on $\widehat{\+\beta }$ is possible.

\begin{theorem}[Asymptotic distribution of $\widehat{\+\beta}$]\ \label{thm:betahat}
Under Assumptions \ref{ass:mom}--\ref{ass:norm}, as $N,\, T\to \infty$,
\begin{align}
\sqrt{NT}\*D_T^{-1}(\widehat{\+\beta}-\+\beta^0) \to_D MN( \*B_0^{-1}(\sqrt{\rho_1}\*A_{1} + \sqrt{\rho_2}\*A_{2})  ,\*B_0^{-1}\+\Omega\*B_0^{-1})
\end{align}
where
\begin{align}
\*B_0 & = \plim_{N,T\to\infty} E[\*B(\*F^0) | \mathcal{C}], \\
\*A_{1} & = -\plim_{N,T\to\infty} \frac{1}{T^{(1-\nu_G)/2}}\sum_{i=1}^N \*D_T E[\*X_i'\*M_{F^0} \+\Sigma_{\varepsilon} \*F_{G}^{0} (\*F_{G}^{0\prime}\*F_{G}^{0})^{-1} (\+\Gamma_G^{0\prime}\+\Gamma_G^0)^{-1} \+\gamma_{G,i}^0 | \mathcal{C}],\\
\*A_{2} & = -\plim_{N,T\to\infty} \frac{1}{T^{(3-\nu_G)/2}} \sum_{i=1}^N \sum_{j=1}^N \*D_T E[\*Z_i(0)'\*F_G^0  (\*F_G^{0\prime} \*F_G^0 )^{-1} (\+\Gamma_G^{0\prime}\+\Gamma_G^0)^{-1} \+\gamma_{G,j}^{0} \+\varepsilon_j ' \+\varepsilon_i | \mathcal{C}],
\end{align}
with $\+\Sigma_{\varepsilon} = N^{-1}\sum_{i=1}^N \+\Sigma_{\varepsilon,i} $ and $\*Z_i(0) = \*X_i - \sum_{j=1}^{N}\*X_ja_{ij}$.
\end{theorem}

According to Theorem \ref{thm:betahat}, the asymptotic bias is driven by the factors and loadings of group $G$, which is intuitive as the factors of this group are smallest in order of magnitude. They therefore dominate the asymptotic bias. By bounding $\nu_G$ from below Assumption \ref{ass:rates} ensures that $\*B_0^{-1}(\sqrt{\rho_1}\*A_{1} + \sqrt{\rho_2}\*A_{2})$ is not diverging. It is important to note that $\rho_1$ and $\rho_2$ may both be zero, which will be the case if $\nu_G>1$ and $T/N$ converges to a constant. In general, the larger $\nu_{G}$ is, the less restrictive the condition on $T/N$ has to be for $\rho_1$ and $\rho_2$ to be zero. For example, if $\nu _{G}=2$, then we only require $N/T^{2}\to 0$. In this sense, trending factors are a blessing. In order to put this into perspective, suppose again that all the factors are stationary, such that $G=1$, $\nu_G  = \nu_1 = 1$ and $\rho_1 = 1/\rho_2  \in (0,\infty)$, and that the regressors are stationary, too, such that $\*D_T = \*I_{d_x}$. In this case, the bias in Theorem \ref{thm:betahat} reduces to $\*B_0^{-1}(\sqrt{\rho_1}\*A_{1} + \rho_1^{-1/2}\*A_{2})$, which is identical to the bias reported in Theorem 3 of \citet{Bai}. In fact, it is not difficult to show that under these conditions, $\sqrt{NT} (\widehat{\+\beta}-\widehat{\+\beta}_0) = o_P(1)$, so that the IPC estimator is asymptotically equivalent to the first step PC estimator. The point about the bias is that while $\sqrt{\rho_1}\*A_{1}$ can be made arbitrarily small (large) by just taking $\rho_1$ to zero (infinity), this will make $\rho_1^{-1/2}\*A_{2}$ divergent (negligible). It follows that unless $\rho_1 \in (0,\infty)$ the bias will diverge, which in turn means that there is no way to make the bias disappear by just manipulating $\rho_1$, which in practical terms means restricting $T/N$. By allowing $\nu_G > 1$, we break the inverse relationship between $\rho_1$ and $\rho_2$, which means that one can be zero without for that matter forcing the other to infinity.

\begin{remark}\label{remarkBai}
\citet[Corollary 1]{Bai} shows that PC can be unbiased even if $\*f_t^0$ and $\*x_{i,t}$ are stationary. However, this requires either that $\varepsilon_{i,t}$ is independent and identically distributed across both $i$ and $t$, or that it is uncorrelated and homoskedastic in $t$ ($i$) with $T/N\to 0$ ($N/T \to 0$).
\end{remark}

\begin{remark}
Note that while biased, $\widehat{\+\beta }$ is still consistent at the best achievable rate. Again, if $\*x_{i,t}$ is stationary, $\*D_{T}=\*I_{d_{x}}$ and the rate of convergence is given by $1/\sqrt{NT}$, which is the same as in \cite{Bai}. This is in contrast to Lemma \ref{lem:beta0hat} and the relatively slow rate of convergence reported there. The reason for this difference is that unlike Theorem \ref{thm:betahat}, which requires that Assumptions \ref{ass:mom}--\ref{ass:norm} all hold, Lemma \ref{lem:beta0hat} only requires Assumptions \ref{ass:mom} and \ref{ass:id}, and under these very relaxed conditions the Theorem \ref{thm:betahat} rate is not attainable. If, however, the conditions of Theorem \ref{thm:betahat} are met, then we can show that both $\widehat{\+\beta }_{0}$ and $\widehat{\+\beta}_{1}$ are consistent at the same rate as $\widehat{\+\beta }$ (see the appendix for a formal proof). However, unlike $\widehat{\+\beta }$, they are generally not mixed normal, which we verify using Monte Carlo simulations in Section \ref{sect:mc}.
\end{remark}

\begin{corollary}[Unbiased asymptotic distribution]\ \label{cor:betahat}
Suppose that the conditions of Theorem \ref{thm:betahat} are met and that $\rho_1 = \rho_2 =0$. Then, as $N,\, T\to \infty$,
\begin{align}
\sqrt{NT}\*D_T^{-1}(\widehat{\+\beta}-\+\beta^0) \to_D MN( \*0_{d_x\times 1} ,\*B_0^{-1}\+\Omega\*B_0^{-1}).
\end{align}
\end{corollary}

In the appendix, we provide some alternative conditions that ensure $\*A_{1} = \*A_{2} = \*0_{d_x\times 1}$. If $\rho_1$, $\rho_2$, $\*A_{1}$ and $\*A_{2}$ are all different from zero, one possibility is to use bias correction. In the appendix, we explain how the Jackknife approach can be used for this purpose.

Theorem \ref{thm:betahat} imposes only minimal conditions on the correlation and heteroskedasticity of $\varepsilon_{i,t}$, and is in this sense very general. Such generality is, however, not possible if we also want to ensure consistent estimation of $\+\Omega$. Let us therefore assume for a moment that $E(\varepsilon_{i,t}\varepsilon_{j,s}) = 0$ for all $(i,t)\ne (j,s)$, so that $\varepsilon_{i,t}$ is serially and cross-sectionally uncorrelated. In this case,
\begin{eqnarray}
\sum_{i=1}^N\sum_{j=1}^N E[\*Z_i(\*F^0)^{\prime}\+\varepsilon_i\+\varepsilon_j\*Z_j(\*F^0) | \mathcal{C}] = \sum_{i=1}^N
\sigma_{\varepsilon,i}^2 E[\*Z_i(\*F^0)^{\prime }\*Z_i(\*F^0) | \mathcal{C}].
\end{eqnarray}
A natural estimator of this matrix is given by
\begin{eqnarray}
\sum_{i=1}^N \widehat \sigma_{\varepsilon,i}^2 \widehat{\*Z}_i^{\prime }\widehat{\*Z}_i,
\end{eqnarray}
where $\widehat{\sigma}_{\varepsilon,i}^2 = T^{-1}\sum_{t=1}^T (\*y_i -\*X_i\widehat{\+\beta})^{\prime }\*M_{\widehat F}(\*y_i -\*X_i\widehat{\+\beta})$
and $\widehat{\*Z}_i$ is as in the definition of $\widehat{\+\beta}$. It is
not difficult to show that under the conditions of Theorem \ref{thm:betahat},
\begin{eqnarray}
\left\|\frac{1}{NT}\sum_{i=1}^N \widehat \sigma_{\varepsilon,i}^2 \*D_T\widehat{\*Z}_i^{\prime }\widehat{\*Z}_i\*D_T - \+\Omega\right\| = o_P(1).
\end{eqnarray}
Of course, in this paper we do not assume knowledge of the order of the regressors, which in practice means that the appropriate normalization matrix $\*D_T$ to use is unknown. This is not a problem, however, as the usual Wald and $t$-test statistics are self-normalizing. As an illustration, consider testing the null hypothesis of $H_0: \*R \+\beta^0 = \*r$, where $\*R$ is a $r_0 \times d_x$ matrix of rank $r_0 \leq d_x$ and $\*r$ is a $r_0 \times 1$ vector. The Wald test statistic for testing this hypothesis is
given by
\begin{eqnarray}  \label{waldtest}
W_{\widehat{\+\beta}} = (\*R\widehat{\+\beta}-\*r)^{\prime }\left[\*R\left(\sum_{i=1}^N \widehat{ \*Z}_i^{\prime }\widehat{\*Z}_i \right)^{-1}\sum_{i=1}^N \widehat\sigma_{\varepsilon,i}^2 \widehat{\*Z}_i^{\prime }\widehat{\*Z}_i\left(\sum_{i=1}^N \widehat{\*Z}_i^{\prime }\widehat{\*Z}_i \right)^{-1} \*R\right]^{-1} (\*R\widehat{\+\beta}-\*r),
\end{eqnarray}
which has a limiting chi-squared distribution with $r_0$ degrees of freedom under $H_0$, as is clear from
\begin{align}
W_{\widehat{\+\beta}} &= [\*R\*D_T\sqrt{NT} \*D_T^{-1} (\widehat{\+\beta}-\+\beta^0)]^{\prime }\Bigg[\*R\*D_T\left(\frac{1}{NT}\sum_{i=1}^N \*D_T\widehat{\*Z}_i^{\prime }\widehat{\*Z}_i \*D_T \right)^{-1}\frac{1}{NT}\sum_{i=1}^N \widehat \sigma_{\varepsilon,i}^2 \*D_T\widehat{\*Z}_i^{\prime }\widehat{\*Z}_i\*D_T \notag \\
& \times \left(\frac{1}{NT}\sum_{i=1}^N \*D_T\widehat{\*Z}_i^{\prime }\widehat{\*Z}_i\*D_T \right)^{-1}\*D_T\*R \Bigg]^{-1} \*R\*D_T \sqrt{NT}\*D_T^{-1} (\widehat{\+\beta}-\+\beta^0) \to_D \chi^2(r_0).
\end{align}
In the next section, we use Monte Carlo simulations as a means to evaluate
the accuracy of this last results in small samples.

The above results are for the case when $\varepsilon _{i,t}$ is serially and cross-sectionally uncorrelated. If $\varepsilon _{i,t}$ is serially and/or cross-sectionally correlated, we recommend following \cite{Bai}, who discusses the issue of consistent covariance matrix estimation at length. The same arguments can be applied without change in current context.

\section{Monte Carlo results}\label{sect:mc}

This section reports the results obtained from a small-scale Monte Carlo simulation exercise. One important goal of the simulations is to examine the finite sample performance of $\widehat{\+\beta}_0$, $\widehat{\+\beta}_1$, and $\widehat{\+\beta}$, and justify the fact that $\widehat{\+\beta}$ is superior to $\widehat{\+\beta}_0$ and $\widehat{\+\beta}_1$.

We consider two DGPs, henceforth denoted ``DGP \ref{dgp:1}'' and ``DGP \ref{dgp:2}'', which we now describe.

\begin{dgp}[Artificial data]\ \label{dgp:1}
In this DGP, $y_{i,t}$ is generated according to a restricted version of \eqref{eq:yit} that sets $d_x = 2$, $\+\beta^0 = \*1_{2\times 1}$, $\varepsilon_{i,t}\sim N(0,1)$ and $N,\,T \in \{40, 80, 160, 320\}$. We further set $d_f = 3$ and generate the factor loadings as $\gamma_{1,i}^0 \sim N(1,1)$, $\gamma_{2,i}^0 \sim N(0,1)$ and $\gamma_{3,i}^0 \sim N(0,1)$. The factors themselves are generated as $f_{1,t}^0=t$, $f_{2,t}^0= \mu_t$ and $f_{3,t}^0= c_t$, where $\mu_t= \mu_{t-1} + \xi_{t}$, $\mu_0 =0$, $\xi_{t} \sim N(0,1/4)$ and $c_{t} = \sin (8\pi t/T )$. Hence, in this DGP, the common component is a random walk with drift and cycle. Also, $d_1= d_2 = d_3=1$ and $(\nu_1,\nu_2,\nu_3)=(3,2,1)$. Let us denote by $x_{j,i,t}$ the $j$-th element of $\*x_{i,t}$. The following specification makes $x_{j,i,t}$ correlated with the common component of $y_{i,t}$:
\begin{eqnarray}
x_{j,i,t} =\frac{1}{d_x}\left(\sum_{j=1}^{d_f}| \gamma_{j,i}^0| +|\xi_t | +|c_t|\right)+ \left(\frac{t}{4}\right)^{(j-1)/4}+v_{j,i,t},
\end{eqnarray}
where $v_{j,i,t}$ is the $i$-th element of the $N\times 1$ vector $\*v_{j,t} = ( v_{j,1,t}, \ldots ,v_{j,N,t} )'$, which we generate as
\begin{eqnarray}
\*v_{j,t} = 0.5 \*v_{j,t-1} + \+\omega_{j,t},
\end{eqnarray}
where $\+\omega_{j,t} \sim N(\*0,\+\Sigma_\omega)$ and $\+\Sigma_\omega$ has $0.5^{|m-n|}$ in row $m$ and column $n$. Thus, $v_{j,i,t}$ is weakly correlated across both $i$ and $t$.
\end{dgp}

\begin{dgp}[Calibrated data]\ \label{dgp:2}
This DGP is calibrated by using the US commercial banks data set studied in Section \ref{ExEE} in which $N=466$ and $T=80$. The $d_x=5$ regressors are the same as those included in Section \ref{sect:ill} and $\+\beta^0 = (0.4063, 0.0199, 0.3805, 0.0912, 0.04726)'$ is calibrated using the empirical estimates taken from Table \ref{tab:emp1}. The factors are the estimated ones reported in Figure \ref{fig:fact1}. Hence, in this DGP, $d_f=2$. The loadings are generated as in DGP 1 with $\gamma_{1,i}^0 \sim N(1,1)$ and $\gamma_{2,i}^0 \sim N(0,1)$.
\end{dgp}

For each combination of $N$ and $T$, we report the correct selection frequency for $(\widehat d_1,\ldots,\widehat d_{\widehat G})$ when seen as an estimator of $(d_1,\ldots,d_{G})$ and for $\widehat d_g$ individually for each group $g = 1,\ldots,G$. Hence, while the former frequency captures the accuracy of the estimation of both $(d_1,\ldots,d_{G})$ and $G$, the latter frequency only captures the accuracy of the estimation of $d_g$ for each $g = 1,\ldots,G$. We also report the root mean squared error (RMSE) of $\widehat{\+\beta}$ and $\*P_{\widehat{F}}$, as measured by the square root of the average of $\|\widehat{\+\beta}-\+\beta^0\|^2$ and $\|\*P_{\widehat{F}} -\*P_{F^0}\|^2$, respectively, over the replications. In interest of comparison, RMSEs of $\widehat{\+\beta}$ is compared to that of $\widehat{\+\beta}_0$, $\widehat{\+\beta}_1$ and the infeasible OLS estimator of $\+\beta^0$ based on taking $\*F^0$ as known, $\widehat{\+\beta}(\*F^0)$. Some results on the 5\% size of the Wald test based on $\*R =\*I_{d_x}$ and $\*r =\+\beta^0$ are also reported. In particular, we report the size of $W_{\widehat{\+\beta}}$, $W_{\widehat{\+\beta}_0}$ and $W_{\widehat{\+\beta}_1}$, which are computed in an obvious fashion by replacing $(\widehat{\+\beta},\widehat{\*F})$ with $(\widehat{\+\beta}_0,\widehat{\*F}_0)$ and $(\widehat{\+\beta}_1,\widehat{\*F})$, respectively, and $W_{\widehat{\+\beta} (\+F^0)}$, which is calculated in the same way as $W_{\widehat{\+\beta}}$ but with $\*M_F\*X_i$ in place of $\widehat{\*Z}_i$. The critical values are taken from $\chi^2(d_x)$. As pointed out in Section \ref{sect:est}, the IPC estimator is invariant with respect to $\delta$. The results reported here are based on $\delta=1$. In time series analysis, rules such as (the integer part of) $8 (T/100)^{1/4}$ are sometimes used for the maximum model dimension to be considered. In our theory, however, $d_{max}$ is fixed and some of the arguments used in the proofs fail if $d_{max}$ is allowed to increase with $N$ and $T$. In this section, we therefore follow the bulk of the previous literature and set $d_{max}$ to a fixed large number (see, for example, \citealp{AH13}, \citealp{BaiNg2002}, and \citealp{MoonWeidner2015}). We choose $d_{max}=10$, which led to the same results as some of the other values we tried. The number of replications is set to 1,000.

We begin by considering the results for DGP \ref{dgp:1}, which are reported in Tables \ref{tab:freDGP1} and \ref{tab:rmseDGP1}. In particular, while Table \ref{tab:freDGP1} contains the results for the estimated common component, Table \ref{tab:rmseDGP1} contains the results for the estimated slopes. We first look at the correct selection frequencies for all and each individual group of factors reported in Table \ref{tab:freDGP1}. The results for the individual groups suggest that the accuracy is generally degreasing in $g$, which is partly expected because the signal strength of the factors, as measured by $\nu_g$, decreases with increasing values of $g$. The accuracy of $(\widehat d_1,\ldots,\widehat d_{\widehat G})$ is almost identical to that of $\widehat d_{G}$, suggesting that the accuracy of $\widehat G$ is driven by the accuracy of the group whose factors has the weakest signal. Looking next at the RMSE results reported in Table \ref{tab:rmseDGP1} for estimating $\+\beta^0$, we see that there is a clear improvement as the sample size increases. The best overall performance is generally obtained when taking $\*F^0$ as known, which is in accordance with our priori expectations. However, the improvement is not very large and it decreases with increases in $N$ and $T$. The reason for this is the accuracy of the estimated factors, which according to the RMSE of $\*P_{\widehat{F}}$ reported Table \ref{tab:freDGP1} is high and increasing in $N$ and $T$. The second best performance is obtained by using $\widehat{\+\beta}$, followed by $\widehat{\+\beta}_1$, and then $\widehat{\+\beta}_0$. If our asymptotic theory is correct, while the size of $W_{\widehat{\+\beta}}$ and $W_{\widehat{\+\beta} (\+F^0)}$ should converge to 5\% as the sample size increases, that of $W_{\widehat{\+\beta}_0}$ and $W_{\widehat{\+\beta}_1}$ should not, and this is exactly what we see in Table \ref{tab:rmseDGP1}. Note in particular how the size of $W_{\widehat{\+\beta}_0}$ and $W_{\widehat{\+\beta}_1}$ is not only nonconvergent but that it is in fact increasing in $N$ and $T$.

The RMSEs of $\widehat{\+\beta}_0$, $\widehat{\+\beta}_1$, $\widehat{\+\beta}$ and $\widehat{\+\beta}(\*F^0)$ in DGP \ref{dgp:2} are given by 0.0444, 0.0405, 0.0405 and 0.0391, respectively, and the sizes of the Wald tests associated with these estimators are given by 0.285, 0.103, 0.078 and 0.061, respectively. Hence, just as in DGP \ref{dgp:1}, the best performance is obtained by using $\widehat{\+\beta}(\*F^0)$ with $\widehat{\+\beta}$, $\widehat{\+\beta}_1$ and $\widehat{\+\beta}_0$ ending up in second, third and fourth place, respectively.

\section{Empirical illustration --- house prices and income}\label{sect:ill}

Economists have become concerned that recently house prices have grown too quickly, and that prices are now too high relative to per capita incomes. If this is correct and there is any truth to the theory on the matter, prices should stagnate or fall until they are better aligned with income, which in statistical terms mean that house prices should be cointegrated with income. The validity of this assumption has important implications for policy, because a failure could be due to a housing bubble.

In this section, we revisit the real house price data set of \cite{Hollyetal2010}, which comprises data on log real house prices ($p_{i,t}$) and log real per capita income ($w_{i,t}$) for 49 US states across the 1975--2003 period. According to theory, $p_{i,t}$ and $w_{i,t}$ should be cointegrated with cointegrating vector $(1,\,-1)'$. The previous empirical evidence of this prediction has, however, been mixed and far from convincing (see, for example, \citealp{Gallin2006}). \cite{Hollyetal2010} argue that this lack of empirical support can be attributed in part to a failure to account for cross-sectional dependence, leading to deceptive conclusions. The authors therefore apply the ``CIPS'' panel unit root test of \cite{Pesaran2007}, which allow for cross-section dependence in the form of a common factor. The test is applied both to $p_{i,t}$ and $w_{i,t}$ separately, and to $p_{i,t}-w_{i,t}$. According to the results, while the variables are unit root non-stationary, their difference is not. \cite{Hollyetal2010} also report CCE results suggesting that the estimated income elasticity is indeed close to one. They therefore conclude that $p_{i,t}$ and $w_{i,t}$ are cointegrated with cointegrating vector $(1,\,-1)'$, just as predicted by theory.

\begin{figure}[h]\caption{The variables in the house prices and income illustration.}\label{fig:vars2}
\centering
\hspace*{-1cm}\includegraphics[trim={1cm 8cm 1cm 6cm},clip,scale=0.9]{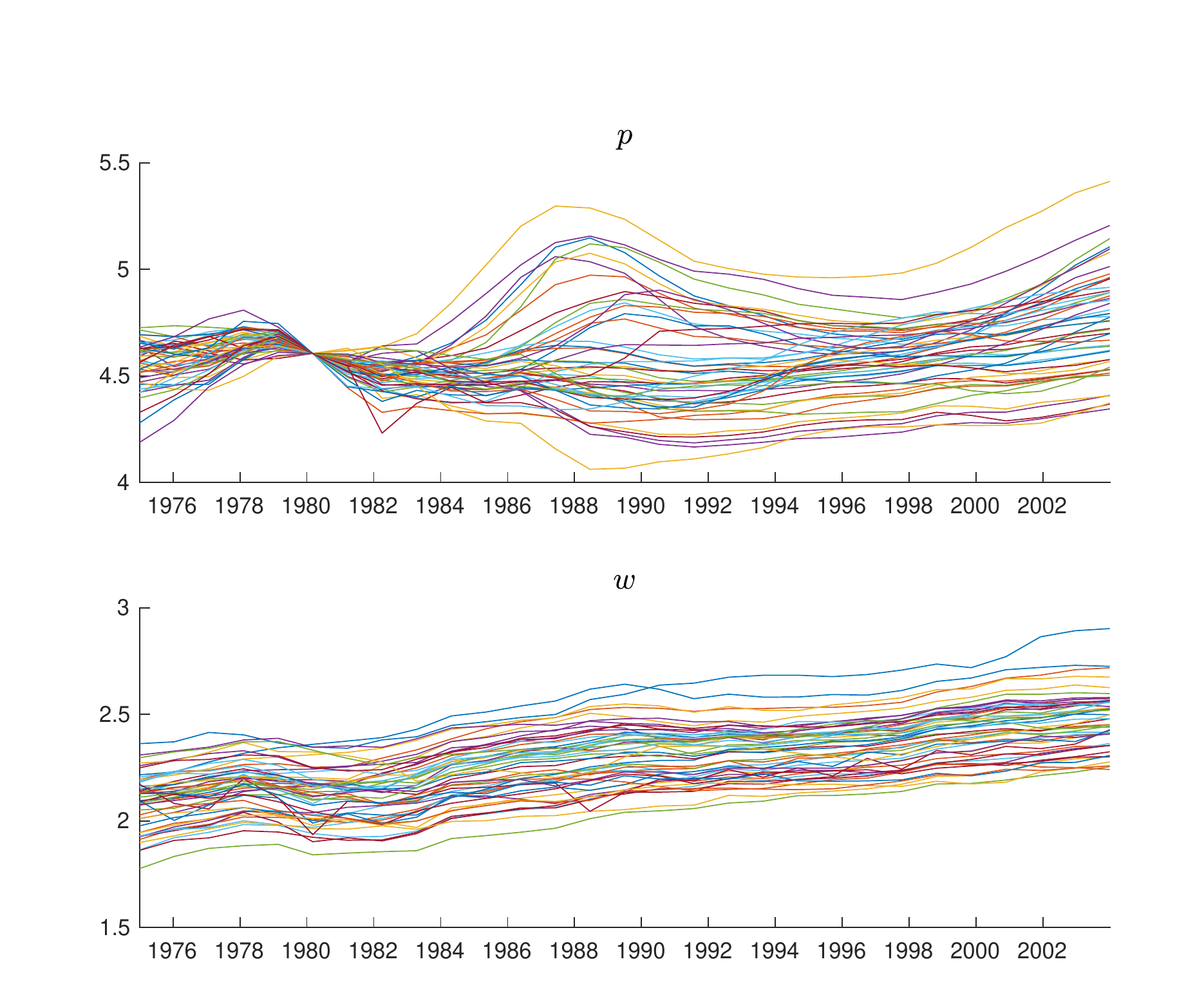}
\end{figure}

Our interest in the work of \cite{Hollyetal2010} stems from their preference to apply the CIPS test, which tests for a unit root in the defactored data. This means that if $p_{i,t}$ and $w_{i,t}$ are not cointegrated by themselves, but only when conditioning on unit root common factors, because of the way that the data are defactored prior to the testing, the unit root null hypothesis is likely to be rejected by the CIPS test. That is, the test is likely to lead to the conclusion of cointegration when in fact there is none. In this section, we use IPC as a means to investigate this possibility.

\begin{figure}[h]\caption{The estimated factors in the house prices and income illustration.}\label{fig:fact2}
\centering
\hspace*{-1cm}\includegraphics[trim={1cm 8cm 1cm 7cm},clip,scale=0.9]{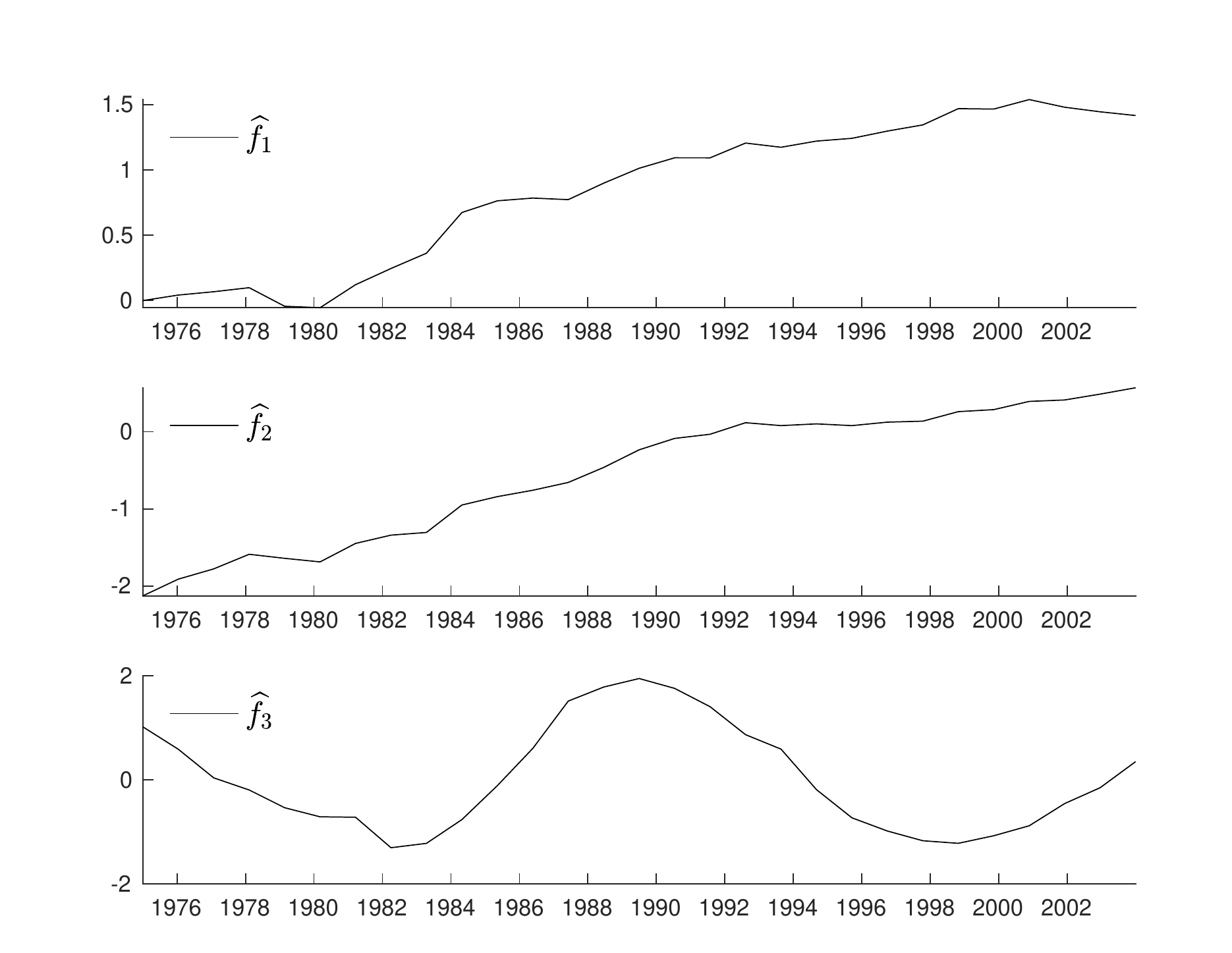}
\end{figure}

As in the RTS illustration, we begin by plotting the variables. This is done in Figure \ref{fig:vars2}. As expected, both variables are highly persistent and the ADF test provides no evidence against the unit root null. This corroborates the unit root test results reported by \cite{Hollyetal2010}. However, we also see that the trending behaviours of $p_{i,t}$ and $w_{i,t}$ are very different, suggesting that their stochastic trends are not the same, which they should be under cointegration. We also see that the trending behaviour is very similar across states, which is suggestive of non-stationary common factors. Of course, the IPC procedure does not require cointegration and it does allow for very general types of factors. We therefore proceed with the estimation of the model. Hence, in this illustration, $y_{i,t}=p_{i,t}$ and $\*x_{i,t} = w_{i,t}$. For comparison purposes, the IPC results are presented together with the results obtained by applying the PC estimator of \cite{Bai}, as well as the usual OLS estimator with time and state fixed effects. The results reported in Table \ref{tab:emp2}. The first thing to note is that the estimated slopes vary a lot depending on the estimator used. Interestingly, the point estimates are increasing in the generality of the estimator with fixed effects OLS (IPC) leading to the lowest (highest) estimate. We therefore begin by considering the IPC results. The point estimate of 2.1024 is far from the theoretically predicted value of one, which is also not included in the reported 95\% confidence interval. As for the factors, we estimate $\widehat d_1 = \widehat d_2 = \widehat d_3 = 1$, implying that $\widehat d_f =3$. The estimated factors, denoted $\widehat f_{1,t}$, $\widehat f_{2,t}$ and $\widehat f_{3,t}$, are plotted in Figure \ref{fig:fact2}. As expected given Figure \ref{fig:vars2}, all three factors are highly persistent, although with clearly distinct trends. We take this as evidence against cointegration between $p_{i,t}$ and $w_{i,t}$, since under cointegration the factors should be stationary.

Because we estimate three time-varying factors, fixed effects OLS is invalid, as it only allows for a common time effect. Moreover, since the factors come from three distinct groups, and are not all stationary, PC is invalid, too. This leaves us with the IPC estimator, which again provides strong evidence against the theoretically predicted one-to-one cointegrated relationship between $p_{i,t}$ and $w_{i,t}$. \citet[page 172]{Hollyetal2010} conclude that ``[o]ur results support the hypothesis that real house prices have been rising in line with fundamentals (real incomes), and there seems little evidence of house price bubbles at the national level.'' The results reported here reveal a completely different picture with housing prices being long run disconnected with real income.

\section{Conclusion}\label{sect:concl}

The PC approach of \cite{Bai} has attracted considerable interest in recent years, so much so that it has given rise to a separate PC literature. A key assumption in this literature is that both the unknown factors and regressors are stationary, which is rarely the case in practice. In the present paper, we relax this assumption by considering a very general specification in which the factors and regressors are essentially unrestricted. In spite of this generality, the proposed IPC estimator can be applied without any input from the practitioner, except for the maximum number of factors to be considered. The fact that in IPC there is no need to distinguish between deterministic and stochastic factors means that the usual problem in applied work of deciding on which deterministic terms to include in the model does not arise, as these are estimated along with the other factors of the model. There is also no need to pre-test the regressors for unit roots, which is otherwise standard practice when using procedures that do not require data to be stationary. In other words, the proposed IPC is not only very general but also extremely user-friendly. It should therefore be a valuable addition to the already existing menu of techniques for panel regression models with interactive effects.

{\small
\bibliographystyle{chicago}
\bibliography{Refs}
}

\begin{table}[H]
\small
\centering
\caption{Monte Carlo results for the estimated common component in DGP \ref{dgp:1}.}\label{tab:freDGP1}
\hskip 12 pt
\par
\begin{tabular}{rrcccccc}
\hline \hline
&  & \multicolumn{4}{c}{Correct selection frequency} & & RMSE \\ \cline{3-6}\cline{8-8}
$N$ & $T$ & $\widehat d_1,...,\widehat d_{\widehat G}$ & $\widehat d_1$ & $\widehat d_2$ & $\widehat d_3$ & & $\*P_{\widehat{F}}$ \\ \hline
40 & 40 & 0.341 & 1.000 & 0.406 & 0.341 & & 0.9453 \\
 & 80 & 0.628 & 1.000 & 0.646 & 0.628 & & 0.5599\\
 & 160 & 0.835 & 1.000 & 0.837 & 0.835 & & 0.3715 \\
 & 320 & 0.954 & 1.000 & 0.954 & 0.954 & & 0.3520\\
80 & 40 & 0.348 & 1.000 & 0.378 & 0.348 & & 0.9452 \\
 & 80 & 0.661 & 1.000 & 0.668 & 0.661 & & 0.4523 \\
 & 160 & 0.867 & 1.000 & 0.867 & 0.867 & & 0.2634 \\
 & 320 & 0.958 & 1.000 & 0.958 & 0.958 & & 0.2401\\
160 & 40 & 0.329 & 1.000 & 0.337 & 0.329 & & 0.6401 \\
 & 80 & 0.684 & 1.000 & 0.686 & 0.684 & & 0.4359 \\
 & 160 & 0.864 & 1.000 & 0.864 & 0.864 & & 0.1761 \\
 & 320 & 0.960 & 1.000 & 0.960 & 0.960 & & 0.1696\\
320 & 40 & 0.220 & 1.000 & 0.225 & 0.220 & & 0.4043 \\
 & 80 & 0.603 & 1.000 & 0.603 & 0.603 & & 0.2083 \\
 & 160 & 0.882 & 1.000 & 0.882 & 0.882 & & 0.1326 \\
 & 320 & 0.988 & 1.000 & 0.988 & 0.988 & & 0.1195\\
 \hline \hline
\end{tabular}
\begin{tablenotes}	
\item \emph{Notes}: The correct selection frequencies are for the estimated factor groups, which is Step \ref{step2} of the IPC procedure. The results for $\widehat d_1,\ldots,\widehat d_{\widehat G}$ treat both the groups and their number, $G$, as unknown, while the results for $\widehat d_1$, $\widehat d_2$ and $\widehat d_3$ take $G$ as given. The reported RMSE results for $\*P_{\widehat{F}}$ refer to the square root of the average of $\|\*P_{\widehat{F}} -\*P_{F^0}\|^2$ across the Monte Carlo replications.
\end{tablenotes}		
\end{table}

\begin{table}[H]
\small \centering
\caption{Monte Carlo results for the estimated slopes in DGP \ref{dgp:1}.}\label{tab:rmseDGP1}
\hskip 12 pt
\par
\begin{tabular}{crccccrcccc}
\hline\hline
 &  & \multicolumn{4}{c}{RMSE} & & \multicolumn{4}{c}{5\% size} \\ \cline{3-6}\cline{8-11}
Estimator & $N$\textbackslash $T$ & 40 & 80 & 160 & 320 &  & 40 & 80 & 160 & 320 \\ \hline
$\widehat{\+\beta}_0$  & 40 & 0.0573 & 0.0423 & 0.0380 & 0.0410 &  & 0.6800 & 0.7280 & 0.8410 & 0.9160 \\
 & 80 & 0.0335 & 0.0234 & 0.0212 & 0.0220 &  & 0.5990 & 0.6550 & 0.7940 & 0.9050 \\
 & 160 & 0.0219 & 0.0169 & 0.0158 & 0.0169 &  & 0.5820 & 0.3500 & 0.8680 & 0.9560 \\
 & 320 & 0.0148 & 0.0109 & 0.0105 & 0.0142 &  & 0.5680 & 0.6230 & 0.7750 & 0.9040 \\
$\widehat{\+\beta}_1$  & 40 & 0.0435 & 0.0296 & 0.0252 & 0.0262 &  & 0.3330 & 0.4320 & 0.6080 & 0.7520 \\
 & 80 & 0.0286 & 0.0177 & 0.0144 & 0.0138 &  & 0.2520 & 0.3130 & 0.4720 & 0.6830 \\
 & 160 & 0.0172 & 0.0127 & 0.0104 & 0.0102 &  & 0.2300 & 0.7260 & 0.5610 & 0.7560 \\
 & 320 & 0.0113 & 0.0079 & 0.0069 & 0.0084 &  & 0.2330 & 0.2770 & 0.4580 & 0.7060 \\
$\widehat{\+\beta}$ & 40 & 0.0383 & 0.0212 & 0.0129 & 0.0091 &  & 0.1320 & 0.0950 & 0.0580 & 0.0680 \\
 & 80 & 0.0280 & 0.0146 & 0.0092 & 0.0062 &  & 0.1500 & 0.0670 & 0.0650 & 0.0550 \\
 & 160 & 0.0164 & 0.0105 & 0.0064 & 0.0043 &  & 0.0960 & 0.0750 & 0.0690 & 0.0430 \\
 & 320 & 0.0099 & 0.0064 & 0.0044 & 0.0032 &  & 0.0720 & 0.0480 & 0.0420 & 0.0660 \\
$\widehat{\+\beta}(\*F^0)$ & 40 & 0.0254 & 0.0171 & 0.0114 & 0.0081 &  & 0.0550 & 0.0610 & 0.0440 & 0.0410 \\
 & 80 & 0.0186 & 0.0117 & 0.0084 & 0.0057 &  & 0.0750 & 0.0470 & 0.0560 & 0.0410 \\
 & 160 & 0.0127 & 0.0086 & 0.0059 & 0.0041 &  & 0.0560 & 0.0600 & 0.0530 & 0.0490 \\
 & 320 & 0.0087 & 0.0059 & 0.0041 & 0.0030 &  & 0.0640 & 0.0420 & 0.0520 & 0.0660\\
 \hline\hline
\end{tabular}
\begin{tablenotes}	
\item \emph{Notes}: The RMSE of $\widehat{\+\beta}$ refers to the square root of the average of $\|\widehat{\+\beta}-\+\beta^0\|^2$ across the Monte Carlo replications. The RMSEs of $\widehat{\+\beta}_0$, $\widehat{\+\beta}_1$ and $\widehat{\+\beta}(\*F^0)$ are constructed in an analogous fashion, where $\widehat{\+\beta}_0$ is the initial Step \ref{step1} IPC estimator, $\widehat{\+\beta}_1$ is the regular PC estimator based on the IPC estimator of the factors, $\widehat{\*F}$, and $\widehat{\+\beta}(\*F^0)$ is the infeasible OLS estimator based on the true factors. The 5\% size results are for the Wald test associated with each estimator.
\end{tablenotes}		
\end{table}

\begin{table}[H]
\centering \caption{Empirical results for the house price and income illustration.}\label{tab:emp2}
\hskip 12 pt
\par
\begin{tabular}{ccc}
\hline \hline
Estimator & Point estimate & Wald $p$-value \\ \hline
FE & 0.3453 & 0.0000 \\
PC & 1.1602 & 0.0000\\
IPC & 2.1024 & 0.0000\\
\hline \hline
\end{tabular}
\begin{tablenotes}	
\item \emph{Notes}: ``FE'', ``PC'' and ``IPC'' refer to the two-way fixed effects OLS estimator, the PC estimator of \cite{Bai} and the proposed IPC estimator, respectively, in a regression of $p_{i,t}$ onto $w_{i,t}$. The reported Wald $p$-values test the null hypothesis that the relevant coefficient is zero.
\end{tablenotes}								
\end{table}

\newpage

\begin{center}
{\Large \bf Online Appendix to: ``Interactive Effects Panel Data Models with General Factors and Regressors"}

\medskip

{\sc Bin Peng$^{\star}$, Liangju Su$^\ddagger$, Joakim Westerlund$^{\sharp}$ and Yanrong Yang$^{\dag}$}

\medskip

$^\star$Monash University, $^\ddagger$Tsinghua University, $^\sharp$Lund University,  $^{\dag}$Australian National University

\end{center}

\setcounter{page}{1}
\renewcommand{\theequation}{A.\arabic{equation}}
\renewcommand{\thesection}{A.\arabic{section}}
\renewcommand{\thefigure}{A.\arabic{figure}}
\renewcommand{\thetable}{A.\arabic{table}}
\renewcommand{\thelemma}{A.\arabic{lemma}}
\renewcommand{\theremark}{A.\arabic{remark}}
\renewcommand{\thecorollary}{A.\arabic{corollary}}
\renewcommand{\theassumption}{A.\arabic{assumption}}

\setcounter{equation}{0}
\setcounter{section}{0}
\setcounter{table}{0}
\setcounter{figure}{0}
\setcounter{assumption}{0}

\bigskip

We begin this appendix by laying out the notation that will be used throughout. This is done in Section \ref{appsect:not}, which in Section \ref{appsect:ass} is followed by a discussion of some conditions that ensure that the asymptotic distribution of $\widehat{\+\beta}$ is correctly centered at zero. Section \ref{ExEE} provides an extra empirical study using the US bank data. Section \ref{sec:outline} describes the outline of the proofs. Sections \ref{appsect:lem} and \ref{appsect:lemproof} provide some auxiliary lemmas and their proofs, respectively. Proofs of the main results are provided in Section \ref{appsect:proof}.

\section{Notation}\label{appsect:not}

The matrices $\+\Sigma_{F^0}$ and $\+\Sigma_{\Gamma^0}$ have been defined in Assumption \ref{ass:mom}. In this appendix, we use $\+\Sigma_{F_g^0}$ and $\+\Sigma_{\Gamma_g^0}$ to denote the sub-matrices of $\+\Sigma_{F^0}$ and $\+\Sigma_{\Gamma^0}$ corresponding to $T^{-\nu_g}\*F_{g}^{0\prime}\*F_{g}^0$ and $N^{-1}\+\Gamma_g^{0\prime}\+\Gamma_g^0$, respectively, for $g = 1,\ldots,G$. We also define $\*V_{g} = \mathrm{diag}(\widehat{\lambda}_{g,1},\ldots, \widehat{\lambda}_{g,d_{max}})$, where $\widehat{\lambda}_{g,d}$ has been defined in Step \ref{step2} of the IPC estimation procedure. We partition $\*F^0 = (\*F_{1}^0,\ldots,\*F_g^0,\*F_{+g}^0)$ and $\*C_T = \mathrm{diag} (T^{-\nu_1/2}\*I_{d_1},\ldots, T^{-\nu_g/2}\*I_{d_g}, \*C_{+g,T} )$, where $\*F_{+g}^0 = (\*F_{g+1}^0,\ldots,\*F_G^0)$ is $T\times (d_{g+1} +\cdots+d_G)$, and $\*C_{+g,T} = \mathrm{diag}( T^{ -\nu_{g+1}/2} \*I_{d_{g+1}} ,\ldots, T^{-\nu_G/2 } \*I_{d_G})$ is $(d_{g+1} +\cdots+d_G)\times (d_{g+1} +\cdots+d_G)$. We partition $\widehat{\*F}$, $\+\gamma_{i}^0$ and $\+\Gamma^0$ conformably as $\widehat{\*F} =(\widehat{\*F}_1,\ldots,\widehat{\*F}_g,\widehat{\*F}_{+g})$, $\+\gamma_{i}^0 = (\+\gamma_{1,i}^{0\prime},\ldots,\+\gamma_{g,i}^{0\prime}, \+\gamma_{+g,i}^{0\prime})'$ and $\+\Gamma^0 = (\+\Gamma_{1}^0,\ldots,\+\Gamma_{g}^0, \+\Gamma_{+g}^0)$, respectively.

We introduce $\lambda_{g,d} =  T^{-\nu_g} \*h_{g,d}^{0\prime}\*F_{g}^{0\prime}\+\Sigma_{g}^0\*F_{g}^{0}\*h_{g,d}^0$, where $\+\Sigma_{g}^0= N^{-1} \*F_{g}^{0} \+\Gamma_{g}^{0\prime} \+\Gamma_{g}^0\*F_{g}^{0\prime}$ and $\*h_{g,d}^0$ is the $d$-th column of $\*H_{g}^0 = N^{-1}T^{(\nu_g-\delta)/2}\+\Gamma_{g}^{0\prime}\+\Gamma_{g}^0 \*F_{g}^{0\prime}\widehat{\*F}_{g}^0 (\*V_{g}^0)^{-1}$ with $\widehat{\*F}_{g}^0$ being the $T\times d_g$ matrix consisting of the first $d_g$ columns of $\widehat{\*F}_{g}$ and $\*V_{g}^0$ being the leading $d_g \times d_g$ principal submatrix of $\*V_{g}$. In other words, $\widehat{\*F}_{g}^0$ and $\*V_{g}^0$ are $\widehat{\*F}_{g}$ and $\*V_{g}$ based on treating the number of factors for each group $g$, $d_g$, as known. We also define $\*H_{g} = T^{-(\nu_g-\delta)/2}\*H_{g}^0$. In order to appreciate the implication of the difference in normalization with respect to $T$, let us consider $\*H_{g}^0$. By Assumption \ref{ass:mom}, $N^{-1} \+\Gamma_{g}^{0\prime}\+\Gamma_{g}^0$ is asymptotically of full rank, and hence $\|N^{-1} \+\Gamma_{g}^{0\prime}\+\Gamma_{g}^0\| = O_P(1)$. Hence, since
\begin{align}
\|T^{-(\nu_g+\delta)/2}\*F_{g}^{0\prime} \widehat{\*F}_g\|^2\leq T^{-\delta }\| \widehat{\*F}_{g}\| _{2}^{2}T^{-\nu _{g}}\|\*{F}_{g}^{0}\|^{2}=T^{-\nu _{g}}\|\*{F}_{g}^{0}\|^{2}=O_{P}\left( 1\right) .
\end{align}
and $\|(T^{-\nu_1}\*V_{1}^0)^{-1}\| = O_P(1)$ as explained under \eqref{eqbp2}, we can show that
\begin{align}
\|\*H_{g}^0\| \leq \|N^{-1}\+\Gamma_{g}^{0\prime}\+\Gamma_{g}^0 \|\|T^{-(\nu_g+\delta)/2}\*F_{g}^{0\prime}\widehat{\*F}_{g}^0 \|\| (T^{-\nu_g}\*V_{g}^0)^{-1}\| = O_P(1),
\end{align}
which in turn implies
\begin{align}
\|\*H_{g}\|  = T^{-(\nu_g-\delta)/2}\|\*H_{g}^0\| = O_P(T^{-(\nu_g-\delta)/2}).
\end{align}
We further use $\widehat{\*F}_{g,d}^0$ to refer to the $d$-th column of $\widehat{\*F}_g^0$. In this notation, $\widehat{\lambda}_{g,d} = T^{-\delta}\widehat{\*F}_{g,d}^{0\prime} \widehat{\+\Sigma}_g \widehat{\*F}_{g,d}^0$ for $d = 1,\ldots,d_g$.

We also partition $\*X_i$ as $\*X_i= (\*X_{1,i},\ldots, \*X_{d_x,i})$ with $\*X_{j,i}$ being the $j$-th column of $\*X_i$. The $j$-th column of $\*X_i\*D_T$ is therefore given by $T^{-\kappa_j/2}\*X_{j,i}$. Moreover, $\mathrm{vec}\,\*A$, $\mathrm{rank}\,\*A$, $\mathrm{span}\,\*A$ and $\lambda(\*A)$ denote the vectorized version, rank, span and eigenvalues of $\*A$, respectively, $a\wedge b = \min\{a,b\}$ and $a\vee b = \max\{a,b\}$.

\section{Conditions that ensure asymptotic unbiasedness}\label{appsect:ass}

In this section, we provide a set of assumptions that ensure that the asymptotic distribution of $\sqrt{NT}\*D_T(\widehat{\+\beta} - \+\beta^0)$ given in Theorem \ref{thm:betahat} is free of bias without for that matter requiring that $\varepsilon_{i,t}$ is serially and cross-sectionally independent. One way to accomplish this is to assume that $\rho_1 = \rho_2 = 0$, as in Corollary \ref{cor:betahat}. The assumptions considered here, which are stated in Assumption \ref{ass:bias}, can be seen as alternatives to this last condition. In terms of the notation of Theorem \ref{thm:betahat}, they ensure that $\*A_{1}=\*A_{2} = \*0_{d_x\times 1}$.

\begin{assumption}[No asymptotic bias]\ \label{ass:bias}
One of the following set of conditions is met:
\begin{itemize}
\item[(a)] $T^{1-\nu_g}E(\*f_{g,t}^{0\prime}\*f_{g,s}^0|\*x_t,\*x_s) = \phi_{ts}$ w.p.a.1 and $\sum_{t=1}^T\sum_{s=1}^T |\phi_{ts}| = O(T)$, where $\*x_{t} = (\*x_{1,t},\ldots, \*x_{N,t})'$. If $G\ge 2$, then $q<(\nu_G +\nu_{G-1})/2-1/4$ and $\nu_{g-1} - \nu_{g}>1/2$ for $g=2,\ldots,G$.

\item[(b)] $T/N \to c_4 \in (0,\infty)$ and $\nu_G > 1$. If $G\ge 2$, then $q<(\nu_G +\nu_{G-1}-1)/2$ and $\nu_{g-1} - \nu_{g}>1/2$.

\item[(c)] $T^{1-\nu_g-\kappa_j}E(\sum_{t=1}^T\sum_{s=1}^T x_{j,i,t}x_{j,k,s}\*f_{g,t}^{0\prime}\*f_{g,s}^0) = O(T^{2-r})$, where $r < 2$, $r+\nu_G -1 > 0$ and $x_{j,i,t}$ is the $j$-th row of $\*x_{i,t}$. If $G\ge 2$, then $\nu_{g-1} - \nu_{g}>1/2$.
\end{itemize}
\end{assumption}

Assumption \ref{ass:bias} ensures that the asymptotic distribution of $\sqrt{NT}\*D_T(\widehat{\+\beta} - \+\beta^0)$ is bias-free. It is, however, not necessary, and (a)--(b) should therefore be viewed as examples of conditions under which there is no asymptotic distribution bias. These conditions all have their strengths and weaknesses, and so their suitability will in general depend on the context. Take as an example condition (b), which has the advantage of not requiring any more moment conditions than those that are already in Assumption \ref{ass:mom}. It does, however, require that $\nu_G > 1$, which rules out both signal-weak and stationary factors ($\nu_G \leq 1$). Conditions (a) and (c) are more general in this regard, but then at the expense of requiring additional moment conditions. Condition (c) is a functional central limit theorem style moment condition.

The next corollary to Theorem \ref{thm:betahat} verifies that the asymptotic distribution of $\sqrt{NT}\*D_T^{-1}(\widehat{\+\beta}-\+\beta^0)$ is indeed bias-free under Assumption \ref{ass:bias}.

\begin{corollary}[Unbiased asymptotic distribution]\ \label{appcor:betahat}
Suppose that Assumptions \ref{ass:mom}--\ref{ass:ortho}, \ref{ass:norm}, and \ref{ass:bias} are met and that $N/T^{\nu_G}\to 0$. Then, as $N,\,T\to \infty$,
\begin{eqnarray}
\sqrt{NT}\*D_T^{-1}(\widehat{\+\beta}-\+\beta^0) \to_D MN(\*0_{d_x\times 1} ,\*B_0^{-1}\+\Omega\*B_0^{-1}).
\end{eqnarray}
\end{corollary}

The asymptotic distribution in Corollary \ref{appcor:betahat} is the same as the one given in Corollary \ref{cor:betahat}.

If $\rho_1$, $\rho_2$, $\*A_{1}$ and $\*A_{2}$ are all different from zero, one possibility is to use bias correction. \cite{DJ2015} were first to bring attention to the relevance of the half-panel Jackknife approach for bias correction in panel data. These authors focus on the fixed effects case, but \cite{Chen2020} have shown that the Jackknife can be used to correct for bias also in models with interactive effects. In our setting, the value of $\nu_G$ is unknown, and therefore the standard Jackknife is not directly applicable. We thus turn to the hybrid half-panel Jackknife correction proposed by \cite{FW2018}. The bias-corrected version of the IPC estimator is given by
\begin{eqnarray}
\widehat{\+\beta}_{\text{bc}} = 3\widehat{\+\beta} - \frac{1}{2}(\widehat{\+\beta}_{1,N}+ \widehat{\+\beta}_{2,N} + \widehat{\+\beta}_{1,T}+ \widehat{\+\beta}_{2,T}),
\end{eqnarray}
where $\widehat{\+\beta}$ is the standard IPC estimator defined in \eqref{estbhat}, $\widehat{\+\beta}_{1,N}$ and $\widehat{\+\beta}_{2,N}$ are defined in the same way but applied to cross-sectional units $\{1,\ldots, \lfloor N/2\rfloor \}$ and $\{\lfloor N/2\rfloor +1,\ldots,  N\}$, respectively, and $\widehat{\+\beta}_{1,T}$ and $\widehat{\+\beta}_{2,T}$ are the IPC estimators applied to odd and even numbered time periods, respectively. The splitting in even and odd numbered time periods is needed in order to account for the behaviour of the factors. See \cite{FW2018} for a more detailed discussion of the hybrid Jackknife approach.

\section{Extra empirical study --- Returns to scale}\label{ExEE}

In this section, we estimate the returns to scale (RTS) of US banks, an area that has attracted considerable attention in the empirical production literature (see, for example, \citealp{FengSerletis2008}). Let us therefore denote by $CT_{i,t}$ the observed total cost of bank $i$ during period $t$. The total cost is a function $C(\cdot)$ of a $J\times 1$ vector of input prices $\*p_{i,t} = (p_{1,i,t},\ldots,p_{J,i,t})'$ and a $L\times 1$ vector of output quantities $\*q_{i,t} = (q_{1,i,t},\ldots,q_{L,i,t})'$. Hence, $CT_{i,t} = C(\*p_{i,t},\*q_{i,t})$, or when expressed in logs, $\ln CT_{i,t} = \ln C(\*p_{i,t},\*q_{i,t})$. The RTS is simply the reciprocal of the sum of the output elasticities, which can be estimated for a given $C(\cdot)$. In order to ensure that $C(\cdot)$ is linearly homogenous in prices, however, it is very common to first normalize costs and prices by one of the prices, $p_{J,i,t}$ say. Hence, $\ln(CT_{i,t}/p_{J,i,t}) = \ln C(\*p_{i,t}/p_{J,i,t},\*q_{i,t})$, which in turn implies that the RTS is given by
\begin{eqnarray}
\mathrm{RTS} = \left( \sum_{j=1}^L \frac{\partial\ln C(\*p_{i,t}/p_{J,i,t},\*q_{i,t})}{\partial \ln q_{j,i,t}}\right)^{-1}.
\end{eqnarray}
In order to estimate this quantity, estimates of the partial derivatives are needed. The standard way to obtain such estimates in the literature is to assume that $\ln C(\*p_{i,t}/p_{J,i,t},\*q_{i,t})$ is linear, to estimate the resulting linear relationship between $\ln (CT_{i,t}/p_{J,i,t})$, $\ln (\*p_{i,t}/p_{J,i,t})$ and $\ln \*q_{i,t}$ by OLS, and to estimate the partial derivatives by simply replacing the parameters by their OLS estimates. A number of modelling issues then arise. First, as pointed out by \citet[page 1884]{FengZhang}, ``[t]here are many ... types of ... unobserved heterogeneity in the US banking industry, which include, but are not limited to, location-related corporate income tax rate, property tax rate, and personal income tax rate.'' Hence, there is a need to control for unobserved heterogeneity. Second, the literature typically assumes that the regressors in $\ln(\*p_{i,t}/p_{J,i,t})$ and $\ln\*q_{i,t}$ are stationary, which, as pointed out by \cite{DGP2020}, is highly unlikely to be the case in practice. There is therefore a need to consider approaches that do not require the regressors to be stationary. Third, to account for the fact that costs are typically trending, many researchers fit their models with linear and quadratic trend terms. Of course, deterministic trends can account for some trending behaviour, but not all. Moreover, the results tend to be highly sensitive to how the trending is modelled, suggesting that this is an important issue (see \citealp{FengSerletis2008}, for a discussion).

The discussion of the last paragraph suggests that there is a need for an approach that is general enough to accommodate not only unobserved heterogeneity but also trending behaviour. The proposed IPC approach fits this bill and we will therefore use it in this empirical illustration to the RTS of US commercial banks. The data that we will use for this purpose, which are the same as in \cite{FGPZ2017}, are quarterly and cover the period 1986--2005, which means that $T=80$. To avoid the impact of entry and exit, we only include continuously operating large banks with assets of at least one billion USD. This gives us a total of $N=466$ banks. The data set contains observations on three input prices and output quantities. They are the wage rate of labour ($p_{1,i,t}$), the price of deposits and purchased funds ($p_{2,i,t}$), the price of physical capital ($p_{3,i,t}$), consumer loans ($q_{1,i,t}$), non-consumer loans, which is composed of industrial, commercial and real estate loans ($q_{2,i,t}$), and securities, including non-loan financial assets ($q_{3,i,t}$). All outputs are deflated by the GDP deflator. Following the previous literature (see, for example, \citealp{Stiroh}), total cost ($CT_{i,t}$) is computed as the sum of total salaries and benefits divided by the number of full-time employees, the price of deposits and purchased funds equals total interest expense divided by total deposits and purchased funds, and the price of capital equals expenses on premises and equipment divided by premises and fixed assets. Hence, after normalization by $p_{3,i,t}$, there are six variables included in the model; $\ln (CT_{i,t}/p_{3,i,t})$, $\ln (p_{1,i,t}/p_{3,i,t})$, $\ln (p_{2,i,t}/p_{3,i,t})$, $\ln q_{1,i,t}$, $\ln q_{2,i,t}$ and $\ln q_{3,i,t}$.

\begin{figure}[h]\caption{The variables in the RTS illustration.}\label{fig:vars1}
\centering
\hspace*{-1cm}\includegraphics[trim={1cm 1cm 1cm 0cm},clip,scale=0.9]{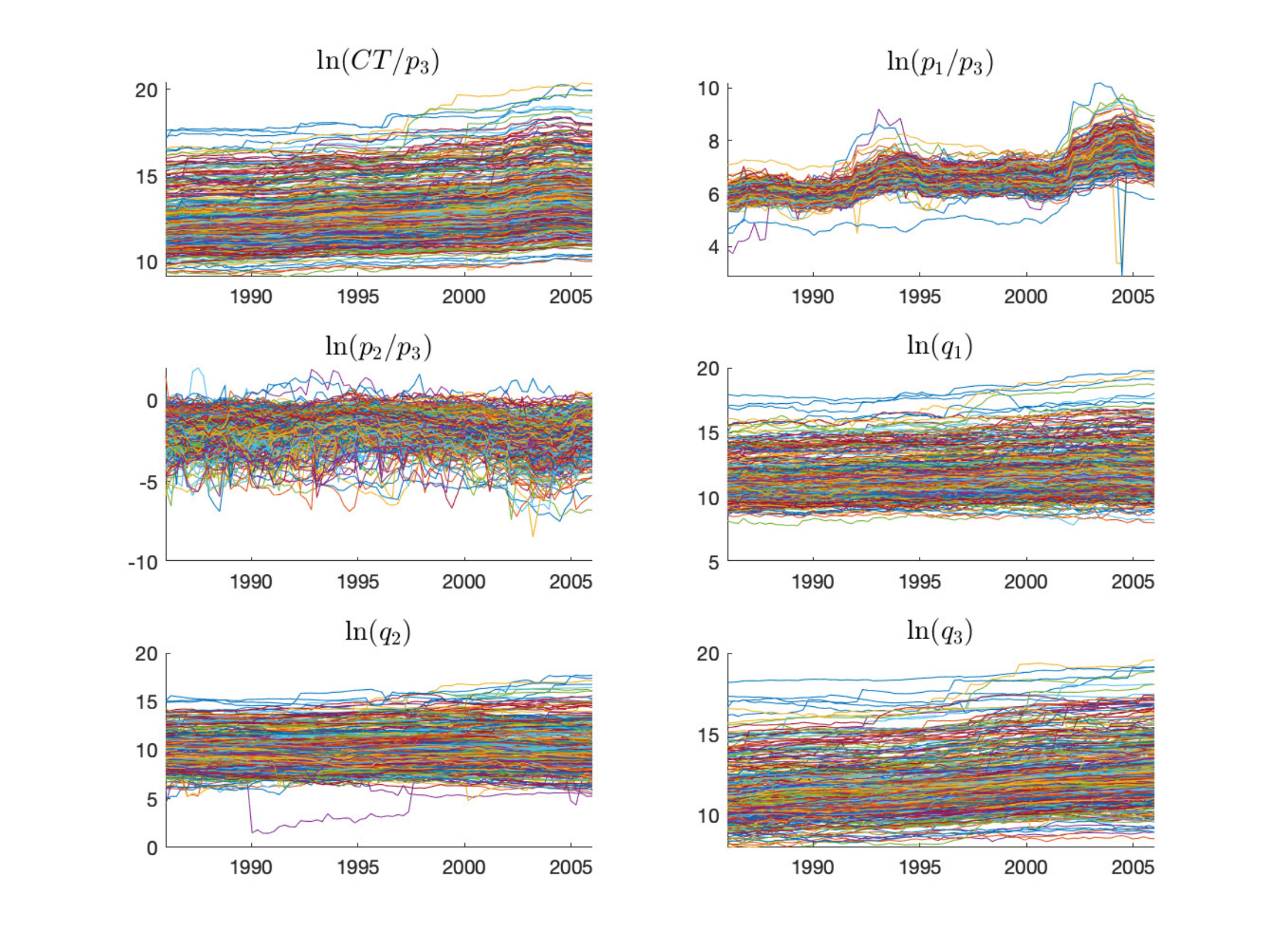}
\end{figure}

In order to get a feeling for the trending behaviour of the data, in Figure \ref{fig:vars1} we plot all $N$ series for each variable. A few observations are noteworthy. First, there is strong co-movement among the series. This is true for all six variables, including the dependent variable, $\ln (CT_{i,t}/p_{3,i,t})$, which we take as evidence in support of our interactive effects specification. Second, the variables are highly persistent and most are trending over time, suggesting that they are non-stationary. This last observation is supported by a formal augmented Dickey--Fuller (ADF) test (with intercept and linear trend) applied to each series. The highest rejection frequency across all $N$ series is obtained for $\ln q_{2,i,t}$ and is given by 6.7\%, which in turn suggests that the evidence against unit root null hypothesis is weak. In fact, since the ADF test does not account for the multiplicity of the testing problem, the true proportion of (trend-)stationary series is likely to be even lower. Third, it is unclear if the trend in $\ln (CT_{i,t}/p_{3,i,t})$ is the same as those in the regressors, which means that it is important to allow the factors to be trending, as otherwise any unaccounted for trend in $\ln (CT_{i,t}/p_{3,i,t})$ will be pushed into the regression errors.

\begin{figure}[h]\caption{The estimated factors in the RTS illustration.}\label{fig:fact1}
\centering
\hspace*{-1cm}\includegraphics[trim={1cm 9cm 1cm 8cm},clip,scale=0.9]{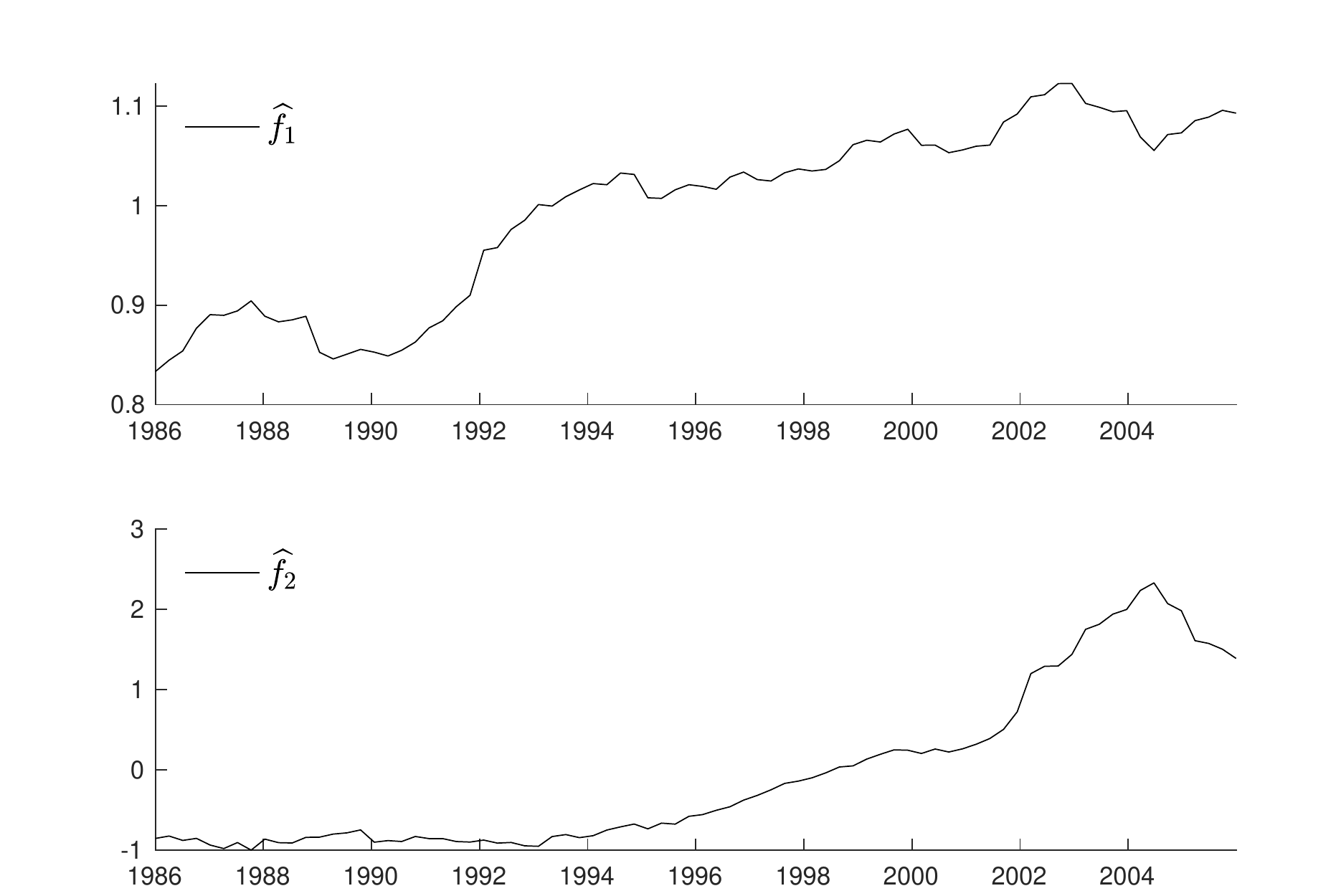}
\end{figure}

We now go on to discuss the actual estimation results, which are reported in Table \ref{tab:emp1}. The estimated model is given by \eqref{eq:yit} with $y_{i,t}=\ln (CT_{i,t}/p_{3,i,t})$, $\+\beta^0 = (\beta_1^0,\ldots,\beta_5^0)'$ and $\*x_{i,t} = (\ln(p_{1,i,t}/p_{3,i,t}), \ln( p_{2,i,t}/p_{3,i,t}), \ln q_{1,i,t } , \ln q_{2,i,t}, \ln q_{3,i,t})'$. Hence, in this notation,
\begin{eqnarray}
\mathrm{RTS} = \frac{1}{\beta_3^0 + \beta_4^0 + \beta_5^0}
\end{eqnarray}
is the object of interest, which we estimate using
\begin{eqnarray}
\widehat{\mathrm{RTS}} = \frac{1}{\widehat\beta_3 + \widehat\beta_4 + \widehat\beta_5},
\end{eqnarray}
where $\widehat\beta_3$, $\widehat\beta_4$ and $\widehat\beta_5$ are from $\widehat{\+\beta} = (\widehat\beta_1,\ldots, \widehat\beta_5)'$. The values of $\delta$ and $d_{max}$ are set as in Section \ref{sect:mc}. The value of $\delta$ is again arbitrary and does not require justification. As for $d_{max}$, 10 factors should be large enough to capture the true number, yet not ``excessively'' large (see \citealp{AH13}, for a discussion). In order to enable inference not only for $\+\beta^0$ but also for the RTS, we follow \cite{FGPZ2017} and report 95\% bootstrap confidence intervals. The estimated factor groups are given by $\widehat{d}_1=\widehat{d}_2=1$, implying that the estimated number of factors equals $\widehat d_f = \widehat{d}_1+\widehat{d}_2=2$. In Figure \ref{fig:fact1}, we plot the estimated factors, denoted $\widehat f_{1,t}$ and $\widehat f_{2,t}$. We see that while both factor estimates are trending, $\widehat f_{2,t}$ is growing much faster. As a measure of this difference in the trends, we look at $\sum_{i=1}^N \widehat{\gamma}_{1,i}^2 /\sum_{i=1}^N \widehat{\gamma}_{2,i}^2$. By using the results provided in the appendix, we can show that this ratio should be $O_P(T^{\nu_1-\nu_2})$, implying $\ln(\sum_{i=1}^N \widehat{\gamma}_{1,i}^2 /\sum_{i=1}^N \widehat{\gamma}_{2,i}^2) /\ln T = \nu_1 - \nu_2 + o_P(1)$. By plugging in the known value of $T$ and $\sum_{i=1}^N \widehat{\gamma}_{1,i}^2 /\sum_{i=1}^N \widehat{\gamma}_{2,i}^2 = 123.4$, we get $\ln(\sum_{i=1}^N \widehat{\gamma}_{1,i}^2 /\sum_{i=1}^N \widehat{\gamma}_{2,i}^2) /\ln T = 1.1$, which is thus an estimate of $\nu_1 - \nu_2$. Hence, there is indeed a substantial difference in the degree of trending of the two factors. But it is not as large as it would have been if $\widehat f_{1,t}$ and $\widehat f_{2,t}$ were made up of a quadratic and a linear trend, say, in which case $\nu_1 = 5$ and $\nu_2 = 3$, such that $\nu_1 - \nu_2 = 2$. Hence, just as expected given Figure \ref{fig:fact1}, while trending, the behaviour of the factors is not just a simple deterministic function of time, which of course casts doubt on much of the existing research.

While our main interest is in the RTS, for completeness in Table \ref{tab:emp1}, we also report the estimated elasticities. The first thing to note is that the RTS is estimated to be significantly larger than one, suggesting that the large US banks considered here have had increasing returns to scale during the period of investigation. This is consistent with the results reported by, for example, \cite{FGPZ2017} and \cite{WheelockWilson2012}. One explanation for this finding is that the banks have been adopting productivity-enhancing internet technologies, leading to increased RTS. The confidence intervals are very narrow not only for the RTS, but also for the elasticities, which are all significant.

\begin{table}[H]
\centering \caption{Empirical results for the RTS illustration.}\label{tab:emp1}
\hskip 12 pt
\par
\begin{tabular}{ccc}
\hline \hline
Estimator & Point estimate & $p$-value \\ \hline
$\widehat\beta_1$ & 0.4063 & 0.0000\\
$\widehat\beta_2$ & 0.0199 & 0.0000\\
$\widehat\beta_3$ & 0.3805 & 0.0000\\
$\widehat\beta_4$ & 0.0912 & 0.0000\\
$\widehat\beta_5$ & 0.4726 & 0.0000\\
$\widehat{\mathrm{RTS}}$ & 1.0590 & 0.0000\\
\hline \hline
\end{tabular}
\begin{tablenotes}	
\item \emph{Notes}: $\widehat\beta_1$, $\widehat\beta_2$, $\widehat\beta_3$, $\widehat\beta_4$ and $\widehat\beta_5$ refer to the estimated IPC slopes of $\ln (p_{i,t}/p_{3,i,t})$, $\ln (p_{2,i,t}/p_{3,i,t})$, $\ln q_{1,i,t}$, $\ln q_{2,i,t}$ and $\ln q_{3,i,t}$, respectively, when the dependent variable is $\ln (CT_{i,t}/p_{3,i,t})$. The reported Wald $p$-values for $\widehat\beta_1$--$\widehat\beta_5$ test the null hypothesis that the relevant coefficient is zero. The $p$-value for RTS test the null hypothesis that $\text{RTS}=1$.
\end{tablenotes}												
\end{table}

\section{Outline of the proofs}\label{sec:outline}

In this section, we describe the outline of the proofs. We assume throughout that $G\ge 2$. The proofs for the cases when $G\in \{0,1\}$ are much simpler, and can be obtained by manipulating the proofs for $G\ge 2$.

Lemma \ref{applem:ortho} is a widely used result for studying the eigenvalues of large dimensional matrices (\citealp{LYB2011}), and is presented here for convenience.

Our own theoretical development is carried out stepwise, starting with Step \ref{step1} of the IPC estimation procedure. Lemma \ref{applem:mom} is very important in this regard. It presents two order results that are used repeatedly throughout the proofs. Given Lemma \ref{applem:mom}, we are able to establish Lemma \ref{lem:beta0hat} of the main text, which provides a lower bound on the rate of convergence of the initial Step \ref{step1} estimator $\widehat{\+\beta}_0$.

As a first step towards establishing the consistency of the Step \ref{step2} estimators of $(d_2, \ldots, d_G)$, in Lemma \ref{applem:lam1} we study limiting behaviour of the eigenvalues associated with $\widehat{\+\Sigma}_1$ in \eqref{defSig1}. The consistency of $\widehat d_1$ reported in Lemma \ref{lem:d1hat} of the main text is a direct consequence of Lemma \ref{applem:lam1}. Lemmas \ref{applem:f} and \ref{applem:lam2} enable consistent estimation of $(d_2, \ldots, d_G)$, and are thus key in proving Lemma \ref{lem:dghat} of the main text. For simplicity, in the proofs of Lemmas \ref{applem:f} and \ref{applem:lam2} we treat $d_1$ as known, which according to Lemma \ref{lem:d1hat} is justified in large samples.

The final step in our theoretical analysis is to prove the asymptotic mixed normality of the Step \ref{step3} IPC estimator $\widehat{\+\beta}$, as stated in Theorem \ref{thm:betahat}. In so doing, we treat $(d_1,\ldots, d_G)$ as known, which is again justified in large samples. We begin by proving Lemma \ref{applem:beta}, which is needed in the proof of Theorem \ref{thm:betahat}. The proofs of Corollaries \ref{cor:betahat} and \ref{appcor:betahat} are immediate consequences of that of Theorem \ref{thm:betahat}. We also derive the rate of convergence of $\widehat{\+\beta}_0$ under the conditions of Theorem \ref{thm:betahat}. This rate is provided as a part of Lemma \ref{applem:beta}.

\section{Auxiliary lemmas}\label{appsect:lem}

\begin{lemma}\label{applem:ortho}
Suppose that $\*A$ and $\*A+\*E$ are $n\times n$ symmetric matrices and that $\*Q = (\*Q_1, \*Q_2)$, where $\*Q_1$ is $n\times r$ and $\*Q_2$ is $n\times (n-r)$, is an orthogonal matrix such that $\text{span}\,\*Q_1$ is an invariant subspace for $\*A$. Decompose $\*Q'\*A\*Q$ and $\*Q'\*E\*Q$ as $\*Q'\*A\*Q = \mathrm{diag}(\*D_1,\*D_2)$ and
\begin{align}
\*Q'\*E\*Q = \left( \begin{array}{cc}
\*E_{11} & \*E_{12} \\
\*E_{21} & \*E_{22}
\end{array} \right).
\end{align}
Let $\mathrm{sep}(\*D_1,\*D_2) = \min_{\lambda_1\in \lambda(\*D_1),\ \lambda_2\in \lambda(\*D_2)} |\lambda_1 -\lambda_2|$. If $\mathrm{sep}(\*D_1,\*D_2) > 0$ and $\|\*E\|_{2} \leq \mathrm{sep}(\*D_1,\*D_2)/5$, then there exists a $(n-r)\times r$ matrix $\*P$ with $\|\*P \|_{2} \leq 4 \|\*E_{21}\|_2/\mathrm{sep}(\*D_1,\*D_2)$, such that the columns of $\*Q_1^0 = (\*Q_1 + \*Q_2\*P)(\*I_r+\*P'\*P)^{-1/2}$ define an orthonormal basis for a subspace that is invariant for $\*A+\*E$.
\end{lemma}

\begin{lemma}\label{applem:mom}
Under Assumption \ref{ass:mom}, as $N,\,T\to\infty$,
\begin{itemize}
\item[$\mathrm{(a)}$] $\sup_{\*F\in \mathbb{D}_F}(NT)^{-1}\sum_{i=1}^{N} \+\varepsilon_i^{\prime }\*P_{F} \+\varepsilon_i =O_P(N^{-1}\vee T^{-1} )$;

\item[$\mathrm{(b)}$] $\sup_{\*F\in \mathbb{D}_F} \|(NT)^{-1}\sum_{i=1}^{N}\*D_T \*X_{i}'\*P_F\+\varepsilon_i\| =O_P(N^{-1/2}\vee T^{-1/2})$.
\end{itemize}
If Assumption \ref{ass:eps} also holds, then

\begin{itemize}
\item[$\mathrm{(c)}$] $(NT)^{-1} \|\+\varepsilon\+\varepsilon'\| =O_P(N^{-1/2}\vee T^{-1/2})$ and $(NT)^{-1} \|\+\varepsilon'\+\varepsilon\| = O_P(N^{-1/2}\vee T^{-1/2})$;

\item[$\mathrm{(d)}$] $\|\+\Gamma^{0\prime}\+\varepsilon\| =O_P(\sqrt{NT})$.
\end{itemize}
\end{lemma}

\begin{lemma}\label{applem:lam1}
Let Assumptions \ref{ass:mom}--\ref{ass:weak} hold. Then, as $N,\,T\to\infty$,
\begin{itemize}
\item[$\mathrm{(a)}$] $T^{-\nu_1}|\widehat{\lambda}_{1,d} - \lambda_{1,d} | = O_P( T^{-(\nu_1-\nu_2)/2})$ for $d=1,\ldots, d_1$;

\item[$\mathrm{(b)}$] $T^{-\nu_1}|\widehat{\lambda}_{1,d} | =O_P( T^{-(\nu_1-\nu_2)})$ for $d=d_1+1,\ldots, d_{max}$.
\end{itemize}
\end{lemma}

\begin{lemma}\label{applem:f}
As $N,\,T\to\infty$, the following hold under Assumptions \ref{ass:mom}--\ref{ass:ortho}:
\begin{itemize}
\item[$\mathrm{(a)}$] $T^{-\delta} \|\*F_{2}^{0\prime} \widehat{\*F}_1\|  =  O_P( T^{-(\delta+\nu_1-\nu_2)/2}+N^{-1/2}T^{-(\delta+\nu_1 -\nu_2 -1)/2} + N^{-(1-p)}T^{-(\delta +\nu_1-2\nu_2)/2}+T^{-(\nu_1+\delta)/2-q})$;

\item[$\mathrm{(b)}$] $\sum_{i=1}^N \| \*F_{1}^{0\prime} \+\gamma_{1,i}^0 - \widehat{\*F}_1 \widehat{\+\gamma}_{1,i} \|^2 = O_P( N\vee T+NT^{2q -\nu_1}+N^{-(1-2p)}T^{\nu_2})$.
\end{itemize}
\end{lemma}

\begin{lemma}\label{applem:lam2}
Let $\tau_{NT} =N^{-1/2}T^{-(\nu_2-1)/2}  +T^{-\nu_2/2}  + T^{-(\nu_2-\nu_3)} + T^{q-(\nu_1+\nu_2)/2}  + N^{-(1-p)}$. As $N,\,T\to\infty$, the following hold under the conditions of Lemma \ref{applem:f}:
\begin{itemize}
\item[$\mathrm{(a)}$] $T^{-\nu_2}|\widehat{\lambda}_{2,d} - \lambda_{2,d} | = O_P( \tau_{NT} )$ for $d=1,\ldots, d_2$;

\item[$\mathrm{(b)}$] $T^{-\nu_2}|\widehat{\lambda}_{2,d} | =O_P( \tau_{NT}^2 )$ for $d=d_2+1,\ldots, d_{max}$.
\end{itemize}
\end{lemma}

\begin{lemma}\label{applem:beta}
Let Assumptions \ref{ass:mom}--\ref{ass:ortho} hold. In addition, let $ NT^{-\nu_G}<\infty $, as $N,\,T\to\infty$,

\begin{itemize}
\item[$\mathrm{(a)}$]  $\|\*D_T^{-1}( \widehat{\+\beta}_1-\widehat{\+\beta}_0 )\| = O_P((NT)^{-1/2}\vee \|\*D_T^{-1} ( \widehat{\+\beta}_0-\+\beta^0)\| )$;
\item[$\mathrm{(b)}$]  $\|\*D_T^{-1} ( \widehat{\+\beta}_0-\+\beta^0)\| = O_P((NT)^{-1/2})$.
\end{itemize}
\end{lemma}

\section{Proofs of auxiliary lemmas}\label{appsect:lemproof}

\noindent \textbf{Proof of Lemma \ref{applem:ortho}.}

\bigskip

\noindent This is Lemma 3 of \cite{LYB2011}. The proof is therefore omitted.\hspace*{\fill}{$\blacksquare$}

\bigskip

\noindent \textbf{Proof of Lemma \ref{applem:mom}.}

\bigskip

\noindent Consider (a). We have
\begin{align}
\sup_{\*F\in \mathbb{D}_F}\frac{1}{NT}\sum_{i=1}^{N} \+\varepsilon_i^{\prime }\*P_{F} \+\varepsilon_i & = \sup_{\*F\in \mathbb{D}_F}(NT)^{-1}\text{tr}( \*P_{F} \+\varepsilon' \+\varepsilon ) \le O(1)  \sup_{\*F\in \mathbb{D}_F}(NT)^{-1}\|\*P_F \|_{2}\|\+\varepsilon'\+\varepsilon \|_{2} \notag\\
&\le O(1) \sup_{\*F\in \mathbb{D}_F}(NT)^{-1}\| \+\varepsilon\|_{2}^2 = O_P(N^{-1}\vee T^{-1}),
\end{align}
where the first inequality follows from the fact that $|\text{tr}\,\*A| \leq \text{rank}\,\*A \, \| \*A\|_{2}$, the second inequality follows from the fact that $\|\*{P}_{{F}}\| _{2}=1$, and the second equality hold by Assumption \ref{ass:mom} (b).

\medskip

The result in (b) is due to
\begin{align}
\sup_{\*F\in \mathbb{D}_F} \left\|\frac{1}{NT}\sum_{i=1}^{N}\*D_T \*X_i ' \*P_F \+\varepsilon_i\right\|  & \le \frac{1}{NT}\sum_{j=1}^{d_x} \sup_{\*F\in \mathbb{D}_F} \left| \text{tr}\,(T^{-\kappa_j/2}\*X_j \*P_F \+\varepsilon')\right| \notag\\
&\le O(1) \frac{1}{NT}\sum_{j=1}^{d_x}\sup_{\*F\in \mathbb{D}_F} \|T^{-\kappa_j/2}\*X_j\|_{2}\| \*P_F\| _{2}\|  \+\varepsilon\| _{2} \notag\\
&= O(1)(NT)^{-1} O_P(\sqrt{NT}) O_P(\sqrt{N}\vee \sqrt{T}) \notag\\
& =O_P(N^{-1/2}\vee T^{-1/2}),
\end{align}
where, with a slight abuse of notation and in this proof only, $\*X_j= (\*X_{j,1},\ldots, \*X_{j,N})'$, the second inequality follows from $|\text{tr}\,\*A| \leq \text{rank}\,\*A \,\| \*A\|_{2}$, while the first equality is due to Assumption \ref{ass:mom}.

\medskip

For (c), we use
\begin{align}
(NT)^{-2}E\| \+\varepsilon'\+\varepsilon\|^2 &=\frac{1}{(NT)^2} \sum_{t=1}^T\sum_{s=1}^T  \left(\sum_{i=1}^NE[\varepsilon_{i,t}^2 \varepsilon_{i,s}^2] +\sum_{i=1}^N\sum_{j\ne i}
E[\varepsilon_{i,t}\varepsilon_{i,s}\varepsilon_{j,t}\varepsilon_{j,s}]\right) \notag\\
&= \frac{1}{(NT)^2}\sum_{t=1}^T \left(\sum_{i=1}^N E(\varepsilon_{i,t}^4) +\sum_{i=1}^N\sum_{j\ne i}
E[(\varepsilon_{i,t}\varepsilon_{j,t}-\sigma_{\varepsilon,ij})^2]  \right) \notag\\
&+ \frac{1}{(NT)^2}\sum_{t=1}^T\sum_{s\ne t}\left(\sum_{i=1}^N E[\varepsilon_{i,t}^2\varepsilon_{i,s}^2] +\sum_{i=1}^N\sum_{j\ne i}
E[(\varepsilon_{i,t}\varepsilon_{j,t}-\sigma_{\varepsilon,ij})(\varepsilon_{i,s}\varepsilon_{j,s}-\sigma_{\varepsilon,ij})] \right)\notag\\
& +\frac{1}{N^2} \sum_{i=1}^N\sum_{j\ne i} \sigma_{\varepsilon,ij}^2  = O(N^{-1})+O(T^{-1}),
\end{align}
where the third equality follows from using the mixing condition on $\varepsilon_{i,t}\varepsilon_{j,t}$ across $t$. The above result implies that $(NT)^{-1}\| \+\varepsilon'\+\varepsilon\| =O_P(N^{-1/2})+O_P(T^{-1/2})$ as required for the first result in (c). The second follows from
\begin{align}
(NT)^{-2} E \|\+\varepsilon\+\varepsilon'\|^2 &= \frac{1}{N^2T^2} \sum_{i=1}^N \sum_{j=1}^N \sum_{t=1}^T\sum_{s=1}^T  E [ \varepsilon_{i,t} \varepsilon_{j,t} \varepsilon_{i,s} \varepsilon_{j,s}] \notag \\
&= \frac{1}{(NT)^2} \sum_{t=1}^T\sum_{s=1}^T\left(\sum_{i=1}^N E[\varepsilon_{i,t}^2\varepsilon_{i,s}^2] + \sum_{i=1}^N\sum_{j\ne i} E[\varepsilon_{i,t}\varepsilon_{i,s}\varepsilon_{j,t}\varepsilon_{j,s})\right)\notag\\
& = O(N^{-1})+O(T^{-1}),
\end{align}
where the last step follows by the same arguments used to establish the first result of (c).

\medskip

It remains to prove (d), which is a direct consequence of Assumptions \ref{ass:mom} and \ref{ass:eps}, as seen from
\begin{eqnarray}
E\| \+\Gamma^{0\prime}\+\varepsilon \|^2 = \sum_{t=1}^T\sum_{i=1}^N \sum_{j=1}^N E[\+\gamma_{i}^{0\prime}\+\gamma_{j}^0\varepsilon_{i,t}\varepsilon_{j,t}]\le O(T)\sum_{i=1}^N \sum_{j=1}^N |\sigma_{\varepsilon,ij}| =O(NT).
\end{eqnarray}
This establishes (d) and hence the proof of the lemma is complete.\hspace*{\fill}{$\blacksquare$}

\bigskip

\noindent \textbf{Proof of Lemma \ref{applem:lam1}.}

\bigskip

\noindent Consider (a). As in Appendix \ref{appsect:not}, decompose $\*F^0 = (\*F_{1}^0,\*F_{+1}^0)$ and $\*C_T = \mathrm{diag} (T^{-\nu_1/2}\*I_{d_1}, \*C_{+1,T} )$, where $\*F_{+1}^0 = (\*F_2^0,\ldots,\*F_G^0)$ is $T\times (d_f -d_1)$ and $\*C_{+1,T} = \mathrm{diag}( T^{ -\nu_2/2} \*I_{d_2} ,\ldots, T^{-\nu_G/2 } \*I_{d_G})$ is $(d_f -d_1)\times (d_f -d_1)$. We partition $\widehat{\*F}$, $\+\gamma_{i}^0$ and $\+\Gamma^0$ conformably as $\widehat{\*F} =(\widehat{\*F}_1,\widehat{\*F}_{+1})$, $\+\gamma_{i}^0 = (\+\gamma_{1,i}^{0\prime}, \+\gamma_{+1,i}^{0\prime})'$ and $\+\Gamma^0 = (\+\Gamma_{1}^0, \+\Gamma_{+1}^0)$, respectively.

By the definition of the eigenvectors and eigenvalues, $\widehat{\+\Sigma}_1 \widehat{\*F}_1 = \widehat{\*F}_1\*V_{1}$. By using this and the definition of $\widehat{\+\Sigma}_1$,
\begin{align}\label{B16}
\lefteqn{ T^{-(\nu_1+\delta)/2}\widehat{\*F}_1 \*V_{1} }\notag\\
&=\frac{1}{NT^{(\nu_1+\delta)/2}} \sum_{i=1}^N \*X_i (\+\beta^0 - \widehat{\+\beta}_0) (\+\beta^0 - \widehat{\+\beta}_0)' \*X_i' \widehat{\*F}_1 \notag\\
&  + \frac{1}{NT^{(\nu_1+\delta)/2}} \sum_{i=1}^N \*X_i (\+\beta^0 - \widehat{\+\beta}_0)\+\gamma_{1,i}^{0\prime} \*F_{1}^{0\prime}\widehat{\*F}_1 +  \frac{1}{NT^{(\nu_1+\delta)/2}} \sum_{i=1}^N \*F_{1}^{0}\+\gamma_{1,i}^0 (\+\beta^0 - \widehat{\+\beta}_0)' \*X_i' \widehat{\*F}_1 \notag\\
&  + \frac{1}{NT^{(\nu_1+\delta)/2}} \sum_{i=1}^N \*X_i (\+\beta^0 - \widehat{\+\beta}_0) ( \*F_{+1}^0 \+\gamma_{+1,i}^0 + \+\varepsilon_i)' \widehat{\*F}_1  \notag\\
&  +  \frac{1}{NT^{(\nu_1+\delta)/2}} \sum_{i=1}^N ( \*F_{+1}^0 \+\gamma_{+1,i}^0+ \+\varepsilon_i) (\+\beta^0 - \widehat{\+\beta}_0)' \*X_i'  \widehat{\*F}_1  \notag\\
&  + \frac{1}{NT^{(\nu_1+\delta)/2}} \sum_{i=1}^N ( \*F_{+1}^0 \+\gamma_{+1,i}^0  + \+\varepsilon_i)( \*F_{+1}^0 \+\gamma_{+1,i}^0 + \+\varepsilon_i)' \widehat{\*F}_1 \notag\\
& + \frac{1}{NT^{(\nu_1+\delta)/2}} \sum_{i=1}^N \*F_{1}^0 \+\gamma_{1,i}^0 ( \*F_{+1}^0 \+\gamma_{+1,i}^0 + \+\varepsilon_i)' \widehat{\*F}_1 + \frac{1}{NT^{(\nu_1+\delta)/2}} \sum_{i=1}^N (\*F_{+1}^0 \+\gamma_{+1,i}^0 + \+\varepsilon_i)  \+\gamma_{1,i}^{0\prime} \*F_{1}^{0\prime}  \widehat{\*F}_1 \notag\\
&+ \frac{1}{NT^{(\nu_1+\delta)/2}} \sum_{i=1}^N \*F_{1}^0 \+\gamma_{1,i}^0 \+\gamma_{1,i}^{0\prime} \*F_{1}^{0\prime}  \widehat{\*F}_1 \notag\\
&=\sum_{j=1}^9 \*J_j ,
\end{align}
with implicit definitions of $\*J_{1},\ldots,\*J_{9}$. Note that
\begin{align}
\*J_{9} =  \*F_{1}^0(N^{-1}\+\Gamma_{1}^{0\prime}\+\Gamma_{1}^0) (T^{-(\nu_1+\delta)/2}\*F_{1}^{0\prime} \widehat{\*F}_1).
\end{align}
Hence, moving this term over to the left-hand side, the above expression for $T^{-(\nu_1+\delta)/2}\widehat{\*F}_1 \*V_{1}$ becomes
\begin{align}\label{eq:fexp}
T^{-(\nu_1+\delta)/2}\widehat{\*F}_1 \*V_{1} - \*J_{9} = \sum_{j=1}^8 \*J_{j}.
\end{align}
We now evaluate each of the terms on the right-hand side.

Because $T^{-\delta}\|\widehat{\*F}_1\|^2 = d_{max}$ and $(NT)^{-1}\sum_{i=1}^N \| \*X_i \*D_T\|^2 = O_p(1)$ by Assumption \ref{ass:mom}, the order of $\*J_{1}$ is given by
\begin{align}
T^{-\delta/2}\| \*J_{1}\| & \le \left\|\frac{1}{NT^{(\nu_1+\delta)/2}} \sum_{i=1}^N \*X_i (\+\beta^0 - \widehat{\+\beta}_0) (\+\beta^0 - \widehat{\+\beta}_0)' \*X_i'\right\| T^{-\delta/2}\|\widehat{\*F}_1\| \notag\\
& \le O (1)\frac{1}{NT^{(\nu_1+\delta)/2}}\sum_{i=1}^N  \| \*X_i \*D_T\*D_T^{-1}(\+\beta^0 - \widehat{\+\beta}_0)\|^2 \notag\\
& = O_P( T^{1-(\nu_1+\delta)/2} \| \*D_T^{-1}(\+\beta^0 - \widehat{\+\beta}_0 )\|^2 ).
\end{align}

Moreover, since
\begin{eqnarray}
\frac{1}{NT^{\nu_1}} \sum_{i=1}^N \| \*F_{1}^0 \+\gamma_{1,i}^0\|^2 \le \frac{1}{N} \sum_{i=1}^N \|\+\gamma_{1,i}^0\|^2 T^{-\nu_1} \| \*F_{1}^0\|^2 = O_P(1)
\end{eqnarray}
by Assumption \ref{ass:mom}, we can show that
\begin{align}
T^{-\delta/2}\| \*J_{2} \| &\le  O(1)\frac{1}{NT^{(\nu_1+\delta)/2}} \sum_{i=1}^N \| \*X_i (\+\beta^0 - \widehat{\+\beta}_0)\+\gamma_{1,i}^{0\prime} \*F_{1}^{0\prime} \| \nonumber \\
&\le O(1) \left( \frac{1}{NT^{\delta}} \sum_{i=1}^N \| \*X_i \*D_T\*D_T^{-1}(\+\beta^0 - \widehat{\+\beta}_0)\|^2 \right)^{1/2} \left(  \frac{1}{NT^{\nu_1}} \sum_{i=1}^N \| \*F_{1}^0 \+\gamma_{1,i}^0\|^2 \right)^{1/2} \nonumber \\
&= O_P( T^{(1-\delta)/2} \| \*D_T^{-1}(\+\beta^0 - \widehat{\+\beta}_0 )\| ),
\end{align}
and by exactly the same arguments,
\begin{eqnarray}
T^{-\delta/2}\| \*J_{3}\| = O_P( T^{(1-\delta)/2} \|\*D_T^{-1}(\+\beta^0 - \widehat{\+\beta}_0 )\| ).
\end{eqnarray}

For $\*J_{4}$, we use
\begin{align}
\lefteqn{ \frac{1}{NT^{(\nu_1+\delta)/2}} \left\|\sum_{i=1}^N \*X_i\*D_T\*D_T^{-1} (\+\beta^0 - \widehat{\+\beta}_0) ( \*F_{+1}^0 \+\gamma_{+1,i}^0 )' \right\| }\notag\\
&\le \left( \frac{1}{NT^{\delta}} \sum_{i=1}^N \| \*X_i \*D_T\*D_T^{-1}(\+\beta^0 - \widehat{\+\beta}_0)\|^2 \right)^{1/2} \left( \frac{1}{NT^{\nu_1}} \sum_{i=1}^N \|  \*F_{+1}^0 \+\gamma_{+1,i}^0\|^2  \right)^{1/2} \notag\\
&\le  O_P(T^{(1-\delta)/2}\| \*D_T^{-1}(\+\beta^0 - \widehat{\+\beta}_0 )\|)O_P( T^{-\nu_1/2}\| \*C_{+1,T}^{-1} \| ) \notag\\
& = O_P( T^{(1-\delta - \nu_1 +\nu_2)/2} \| \*D_T^{-1}(\+\beta^0 - \widehat{\+\beta}_0 )\|),
\end{align}
where the last equality makes use of the fact that $\|\*C_{+1,T}^{-1}\|= O(T^{\nu_2/2})$, as $\nu_2 > \cdots > \nu_G$ by Assumption \ref{ass:mom}. We can further show that
\begin{align}\label{eqbp1}
\lefteqn{ \frac{1}{NT^{(\nu_1+\delta)/2}} \left\|\sum_{i=1}^N \*X_i \*D_T\*D_T^{-1}(\+\beta^0 - \widehat{\+\beta}_0) \+\varepsilon_i'  \right\| }\notag\\
& =\left\|\frac{1}{NT^{(\nu_1+\delta)/2}} \sum_{i=1}^N \mathrm{vec}\,[ \*X_i \*D_T\*D_T^{-1}(\+\beta^0 - \widehat{\+\beta}_0) \+\varepsilon_i' ] \right\| \nonumber\\
& \le \left\|\frac{1}{NT^{(\nu_1+\delta)/2}} \sum_{i=1}^N (\+\varepsilon_i \otimes \*X_i \*D_T) \right\| \|\*D_T^{-1}(\+\beta^0 - \widehat{\+\beta}_0) \| \nonumber \\
&= O_P( N^{-1/2}T^{(2-\nu_1-\delta)/2} \|\*D_T^{-1}(\+\beta^0 - \widehat{\+\beta}_0)\| ),
\end{align}
where the last equality holds, because by Assumption \ref{ass:eps} and $(\mathrm{tr}\,\*A'\*B)^2 \le (\mathrm{tr}\,\*A'\*A )( \mathrm{tr}\,\*B'\*B)$, we have
\begin{align}
\lefteqn{ E\left\|\frac{1}{NT^{(\nu_1+\delta)/2}} \sum_{i=1}^N (\+\varepsilon_i \otimes \*X_i \*D_T)  \right\| ^2 }\notag\\
&= \frac{1}{N^2T^{\nu_1+\delta}} \sum_{t=1}^T\sum_{s=1}^T\sum_{i=1}^N \sum_{j=1}^N E(  \*x_{j,t}' \*D_T^2 \*x_{i,t} \varepsilon_{i,s} \varepsilon_{j,s} ) \le  \frac{1}{N^2T^{\nu_1+\delta}} \sum_{i=1}^N \sum_{j=1}^N E|\mathrm{tr}( \*D_T \*X_{j}' \*X_{i}\*D_T  )|\cdot |\sigma_{\varepsilon,ij}| \nonumber\\
&\le  \frac{1}{N^2T^{\nu_1+\delta-2}} \sum_{i=1}^N \sum_{j=1}^N \sqrt{T^{-1}E\|\*D_T \*X_{j}\|^2} \sqrt{T^{-1}E\| \*X_{i}\*D_T \|^2} |\sigma_{\varepsilon,ij}| \nonumber\\
&= O(1) \frac{1}{N^2T^{\nu_1+\delta-2}} \sum_{i=1}^N \sum_{j=1}^N |\sigma_{\varepsilon,ij}| =O ( N^{-1}T^{2-\nu_1-\delta}).
\end{align}
Hence, by adding the results,
\begin{align}
T^{-\delta/2}\| \*J_{4} \| & \le   O(1)\frac{1}{NT^{(\nu_1+\delta)/2}} \left\|\sum_{i=1}^N \*X_i (\+\beta^0 - \widehat{\+\beta}_0) ( \*F_{+1}^0 \+\gamma_{+1,i}^0 + \+\varepsilon_i)'  \right\|  \notag\\
&=  O_P(  T^{(1-\delta - \nu_1 +\nu_2)/2} \| \*D_T^{-1}(\+\beta^0 - \widehat{\+\beta}_0 )\|+ N^{-1/2}T^{(2-\nu_1-\delta)/2} \|\*D_T^{-1}(\+\beta^0 - \widehat{\+\beta}_0)\| )\notag\\
& =  o_P ( T^{-(1-\delta)/2}\| \*D_T^{-1}(\+\beta^0 - \widehat{\+\beta}_0 )\|),
\end{align}
where we have used Assumptions \ref{ass:mom} and \ref{ass:weak} ($T/N^2=O(1)$ under $\nu_G < 1$) to show that $T^{(1-\delta - \nu_1 +\nu_2)/2}$ and $N^{-1/2}T^{(2-\nu_1-\delta)/2}$ are $o(1)$. The same steps can be used to show that
\begin{eqnarray}
T^{-\delta/2}\| \*J_{5}\|  \le  o_P( T^{-(1-\delta)/2}\| \*D_T^{-1}(\+\beta^0 - \widehat{\+\beta}_0 )\| ).
\end{eqnarray}

For $\*J_{6}$, we use
\begin{align}
T^{-\delta/2}\| \*J_{6}\| &\le O(1) \left\|\frac{1}{NT^{(\nu_1+\delta)/2}} \sum_{i=1}^N(  \*F_{+1}^0 \+\gamma_{+1,i}^0 + \+\varepsilon_i)(  \*F_{+1}^0 \+\gamma_{+1,i}^0 + \+\varepsilon_i)'\right\|  \notag\\
&\le  O(1)\cdot ( N^{-1}T^{-(\nu_1+\delta)/2}  \| \*F_{+1}^0  \+\Gamma_{+1}^{0\prime} \+\Gamma_{+1}^0 \*F_{+1}^{0\prime} \| + N^{-1}T^{-(\nu_1+\delta)/2}\| \+\varepsilon'\+\varepsilon \|  \notag\\
& + 2N^{-1}T^{-(\nu_1+\delta)/2} \| \*F_{+1}^0 \+\Gamma_{+1}^{0\prime}\+\varepsilon \| ).
\end{align}
By Assumption \ref{ass:mom},
\begin{align}
N^{-1}T^{-(\nu_1+\delta)/2} \| \*F_{+1}^0 \+\Gamma_{+1}^{0\prime} \+\Gamma_{+1}^0 \*F_{+1}^{0\prime} \| =  O_P(T^{\nu_2 - (\nu_1+\delta)/2}) .
\end{align}
Another application of Assumption \ref{ass:mom} and Lemma \ref{applem:mom} gives
\begin{align}
N^{-1}T^{-(\nu_1+\delta)/2}\| \+\varepsilon'\+\varepsilon \| & = O_P( T^{1-(\nu_1+\delta)/2} ( N^{-1/2} \vee T^{-1/2})),\\
N^{-1}T^{-(\nu_1+\delta)/2}\| \*F_{+1}^0 \+\Gamma_{+1}^{0\prime}\+\varepsilon \| & \le N^{-1/2}T^{-(\nu_1+\delta-\nu_2-1)/2} T^{-\nu_2/2}\| \*F_{+1}^0 \| (NT)^{-1/2}\|\+\Gamma_{+1}^{0\prime}\+\varepsilon \| \notag\\
& = O_P(N^{-1/2}T^{-(\nu_1+\delta-\nu_2-1)/2} ).
\end{align}
These results can be inserted into the expression for $T^{-\delta/2}\| \*J_{6}\|$, giving
\begin{eqnarray}
T^{-\delta/2}\| \*J_{6}\| =O_P(  T^{\nu_2 - (\nu_1+\delta)/2} + N^{-1/2} T^{1-(\nu_1+ \delta)/2}).
\end{eqnarray}

Next up is $\*J_{7}$. By using Assumptions \ref{ass:mom} and \ref{ass:weak}, and Lemma \ref{applem:mom}, and the arguments use in evaluating $\*J_6$,
\begin{align}
T^{-\delta/2}\| \*J_{7}\| & \le   O(1) \left\|\frac{1}{NT^{(\nu_1+\delta)/2}}\sum_{i=1}^N\*F_{1}^0 \+\gamma_{1,i}^0( \*F_{+1}^0 \+\gamma_{+1,i}^0+ \+\varepsilon_i)'\right\|  \notag\\
&\le O(1)  N^{-1}T^{-(\nu_1+\delta)/2}\| \*F_{1}^0 \+\Gamma_{1}^{0\prime} \+\Gamma_{+1}^0 \*F_{+1}^{0\prime} \|  +   O(1)N^{-1}T^{-(\nu_1+\delta)/2} \| \*F_{1}^0\+\Gamma_{1}^{0\prime}\+\varepsilon \| \notag\\
&= O_P( T^{(\nu_2-\delta)/2}+ N^{-1/2} T^{(1-\delta)/2})= O_P( T^{(\nu_2-\delta)/2}),
\end{align}
and we can similarly show that
\begin{eqnarray}
T^{-\delta/2}\| \*J_{8}  \| = O_P( T^{(\nu_2-\delta)/2}).
\end{eqnarray}

By putting everything together, \eqref{eq:fexp} becomes
\begin{align}
T^{-\delta/2}\|T^{-(\nu_1+\delta)/2}\widehat{\*F}_1 \*V_{1} - \*J_{9}\|& = O_P( T^{-(\delta-1)/2} \| \*D_T^{-1}(\+\beta^0 - \widehat{\+\beta}_0 )\| ) \notag\\
&+ O_P( N^{-1/2}T^{1-(\nu_1+\delta)/2}) + O_P( T^{-(\delta-\nu_2)/2}) .
\end{align}
We now left multiply \eqref{eq:fexp} by $T^{-(\nu_1+\delta)/2}\widehat{\*F}_{1}^{\prime}$ to obtain that
\begin{align}
&T^{-\nu_1}\*V_1-(T^{-(\nu_1+\delta)/2}  \widehat{\*F}_1'\*F_{1}^{0})(N^{-1}\+\Gamma_{1}^{0\prime}\+\Gamma_{1}^0) (T^{-(\nu_1+\delta)/2}\*F_{1}^{0\prime} \widehat{\*F}_1) \nonumber \\
&=T^{-(\nu_1-\delta)/2}    O_P( T^{-(\delta-1)/2} \| \*D_T^{-1}(\+\beta^0 - \widehat{\+\beta}_0 )\| +N^{-1/2}T^{1-(\nu_1+\delta)/2}+T^{-(\delta-\nu_2)/2})\nonumber \\
&= O_P (  T^{-(\nu_1-1)/2}\| \*D_T^{-1}(\+\beta^0 - \widehat{\+\beta}_0 )\|) + O_P ( T^{-(\nu_1-\nu_2)/2}) \notag\\
& =O_P( T^{-(\nu_1-\nu_2)/2}),
\end{align}
where the third equality follows from Assumption \ref{ass:weak} and Lemma \ref{lem:beta0hat}. This implies that $\*V_1$ is at most of rank $d_1$. Similarly, we can left-multiply \eqref{eq:fexp} by $T^{-\nu_1}\*F_{1}^{0\prime}$ to obtain
\begin{align}
 \|T^{-(\nu_1+\delta)/2}\*F_{1}^{0\prime}\widehat{\*F}_1 (T^{-\nu_1}\*V_{1}) - T^{-\nu_1}\*F_{1}^{0\prime}\*J_{9}\|  = O_P( T^{-(\nu_1-\nu_2)/2}),
\end{align}
which in turn implies that
\begin{align}\label{eqbp2}
\+\Sigma_{F_{1}^0}\+\Sigma_{\Gamma_{1}^0}(T^{-(\nu_1+\delta)/2}\*F_{1}^{0\prime}\widehat{\*F}) = (T^{-(\nu_1+\delta)/2}\*F_{1}^{0\prime}\widehat{\*F})(T^{-\nu_1}\*V_{1}) + o_P(1).
\end{align}
Note that $T^{-(\nu_1+\delta)/2}\*F_{1}^{0\prime}\widehat{\*F}$ is of rank $d_1$, which then further indicates that $\*V_1$ has at least $d_1$ non-zero elements on the main diagonal which converge to the eigenvalues of $\+\Sigma_{F_{1}^0}\+\Sigma_{\Gamma_{1}^0}$. We now can conclude that $\*V_1$ is of rank $d_1$ in limit.

We are now ready to investigate $\widehat{\lambda}_{1,d}$ for $d\le d_1$. Because here $d\le d_1$, we then focus on $\widehat{\*F}_1^0$ and $\*V_{1}^0$, which are defined in Appendix \ref{appsect:not}. Let us write \eqref{eq:fexp} as follows:
\begin{align}
T^{-(\nu_1+\delta)/2}\widehat{\*F}_1^0 \*V_{1}^0 - \*J_{9}^0 = (T^{(\nu_1-\delta)/2}\widehat{\*F}_1^0  - \*F_{1}^0\*H_1^0)(T^{-\nu_1}\*V_{1}^0),
\end{align}
where $\*H_1^0$ is defined in Appendix \ref{appsect:not}. The above expression implies that \eqref{eq:fexp} can be written as
\begin{align}\label{eqbp3}
T^{(\nu_1-\delta)/2}\widehat{\*F}_1^0  - \*F_{1}^0\*H_1^0 = \sum_{j=1}^8 \*J_{j}^0(T^{-\nu_1}\*V_{1}^0)^{-1},
\end{align}
where $\*J_{1}^0,\ldots,\*J_{9}^0$ are $\*J_{1},\ldots,\*J_{9}$ as defined in \eqref{B16}, except that now $d_1$ is taken as known. Note that by the above development
\begin{align}
\sum_{j=1}^8 T^{-\nu_1/2}\|\*J_{j}^0\| = T^{-(\nu_1-\delta)/2}\sum_{j=1}^8 T^{-\delta/2}\|\*J_{j}^0\| = O_P(T^{-(\nu_1-\nu_2)/2}).
\end{align}
Moreover, since $T^{-\nu_1}\*V_{1}^0$ converges to a full rank matrix by the argument under \eqref{eqbp2}, we have $\|(T^{-\nu_1}\*V_{1}^0)^{-1}\| = O_P(1)$, which in turn implies
\begin{align}\label{eq:fcons}
T^{-\nu_1/2}\|T^{(\nu_1-\delta)/2}\widehat{\*F}_1^0  - \*F_{1}^0\*H_1^0\| \le \sum_{j=1}^8 T^{-\nu_1/2} \|\*J_{j}^0\|\|(T^{-\nu_1}\*V_{1}^0)^{-1}\| = O_P( T^{-(\nu_1-\nu_2)/2}).
\end{align}
This is an important result and in what follows we will use it frequently.

Let us now consider $T^{-\nu_1}(\widehat{\lambda}_{1,d} - \lambda_{1,d})$. By the definitions of $\lambda_{1,d}$ and $\widehat{\lambda}_{1,d}$ given in Section \ref{appsect:not},
\begin{align}
T^{-\nu_1}( \widehat{\lambda}_{1,d} -\lambda_{1,d} ) &= ( T^{-\delta/2}\widehat{\*F}_{1,d}^0  -  T^{-\nu_1/2}\*F_1^0\*h_{1,d}^0 + T^{-\nu_1/2}\*F_1^0\*h_{1,d}^0 )' T^{-\nu_1}(\widehat{\+\Sigma}_1 - \+\Sigma_{1}^0+ \+\Sigma_{1}^0) \notag\\
& \times ( T^{-\delta/2}\widehat{\*F}_{1,d}^0  -  T^{-\nu_1/2}\*F_1^0\*h_{1,d}^0 + T^{-\nu_1/2}\*F_1^0\*h_{1,d}^0 ) - T^{-2\nu_1} \*h_{1,d}^{0\prime} \*F_1^{0\prime} \+\Sigma_{1}^0 \*F_1^0\*h_{1,d}^0 \nonumber \\
&= ( T^{-\delta/2}\widehat{\*F}_{1,d}^0  -  T^{-\nu_1/2}\*F_1^0\*h_{1,d}^0)'T^{-\nu_1}(\widehat{\+\Sigma}_1 - \+\Sigma_{1}^0) ( T^{-\delta/2}\widehat{\*F}_{1,d}^0  -  T^{-\nu_1/2}\*F_1^0\*h_{1,d}^0 ) \notag\\
& + 2( T^{-\delta/2}\widehat{\*F}_{1,d}^0 -  T^{-\nu_1/2}\*F_1^0\*h_{1,d}^0 )'T^{-\nu_1}(\widehat{\+\Sigma}_1 - \+\Sigma_{1}^0) T^{-\nu_1/2}\*F_1^0\*h_{1,d}^0 \nonumber \\
&+ ( T^{-\delta/2}\widehat{\*F}_{1,d}^0  -  T^{-\nu_1/2}\*F_1^0\*h_{1,d}^0 )' T^{-\nu_1}\+\Sigma_{1}^0( T^{-\delta/2}\widehat{\*F}_{1,d}^0  -  T^{-\nu_1/2}\*F_1^0\*h_{1,d}^0 ) \notag\\
& +2 ( T^{-\delta/2}\widehat{\*F}_{1,d}^0  -  T^{-\nu_1/2}\*F_1^0\*h_{1,d}^0 )' T^{-3\nu_1/3}\+\Sigma_{1}^0 \*F_1^0\*h_{1,d}^0 \nonumber  \\
&+ T^{-2\nu_1}\*h_{1,d}^{0\prime} \*F_1^{0\prime}( \widehat{\+\Sigma}_1 - \+\Sigma_{1}^0) \*F_1^0\*h_{1,d}^0 \nonumber  \\
&= J_1 +2J_2+J_3+2J_4 +J_5,
\end{align}
with obvious definitions of $J_1,\ldots,J_5$. From \eqref{eq:fcons},
\begin{align}
T^{-\nu_1/2}\|T^{(\nu_1-\delta)/2}\widehat{\*F}_{1,d}^0 - \*F_{1}^0\*h_{1,d}^0 \| = \|T^{-\delta/2}\widehat{\*F}_{1,d}^0 - T^{-\nu_1/2}\*F_{1}^0\*h_{1,d}^0 \| = O_P(  T^{-(\nu_1-\nu_2)/2}),
\end{align}
which is $o_P(1)$ under Assumption \ref{ass:mom}. This implies $|J_1| = o_P(|J_5|)$, $|J_2| = o_P(|J_5|)$ and $|J_3| = o_P(|J_4|)$. It remains to consider $J_4$ and $J_5$. The order of the first of these terms is given by
\begin{align}
|J_4| &\le  \| T^{-\delta/2}\widehat{\*F}_{1,d}^0  -  T^{-\nu_1/2}\*F_1^0\*h_{1,d}^0 \| T^{-3\nu_1/2}\|\+\Sigma_{1}^0 \*F_1^0\*h_{1,d}^0\| \notag\\
&\le  \| T^{-\delta/2}\widehat{\*F}_{1,d}^0  -  T^{-\nu_1/2}\*F_1^0\*h_{1,d}^0 \| T^{-\nu_1/2}\| \*F_{1}^{0}\|\| (N^{-1}\+\Gamma_{1}^{0\prime} \+\Gamma_{1}^0)\|T^{-\nu_1}\|\*F_{1}^{0\prime} \*F_1^0\|\|\*h_{1,d}^0\| \notag\\
& =O_P(  T^{-(\nu_1-\nu_2)/2} ),
\end{align}
where we have made use of the fact that $\|\*H_{g}^0\| = O_P(1)$ (see Appendix \ref{appsect:not}), which implies that $\|\*h_{1,d}^0\|$ is of the same order.

The order of $J_5$ is the same as that of $J_4$. In order to appreciate this, we begin by noting
\begin{align}
T^{-\nu_1} (\widehat{\*\Sigma}_1 - \+\Sigma_{1}^0) &= \frac{1}{NT^{\nu_1}}  \sum_{i=1}^N \*X_i (\+\beta^0 - \widehat{\+\beta}_0) (\+\beta^0 - \widehat{\+\beta}_0)' \*X_i'   \notag\\
& + \frac{1}{NT^{\nu_1}}\sum_{i=1}^N \*X_i (\+\beta^0 - \widehat{\+\beta}_0)\+\gamma_{1,i}^{0\prime} \*F_{1}^{0\prime}  +  \frac{1}{NT^{\nu_1}}\sum_{i=1}^N \*F_{1}^0\+\gamma_{1,i}^0 (\+\beta^0 - \widehat{\+\beta}_0)' \*X_i'  \notag\\
& +\frac{1}{NT^{\nu_1}}\sum_{i=1}^N \*X_i (\+\beta^0 - \widehat{\+\beta}_0) ( \*F_{+1}^0 \+\gamma_{+1,i}^0 + \+\varepsilon_i)'   \notag\\
& + \frac{1}{NT^{\nu_1}}\sum_{i=1}^N ( \*F_{+1}^0 \+\gamma_{+1,i}^0+ \+\varepsilon_i) (\+\beta^0 - \widehat{\+\beta}_0)' \*X_i'  \notag\\
& +\frac{1}{NT^{\nu_1}}\sum_{i=1}^N ( \*F_{+1}^0 \+\gamma_{+1,i}^0 + \+\varepsilon_i)( \*F_{+1}^0\+\gamma_{+1,i}^0 + \+\varepsilon_i)'  \notag\\
& + \frac{1}{NT^{\nu_1}}\sum_{i=1}^N \*F_{1}^0 \+\gamma_{1,i}^0 ( \*F_{+1}^0 \+\gamma_{+1,i}^0 + \+\varepsilon_i)'   + \frac{1}{NT^{\nu_1}}\sum_{i=1}^N (\*F_{+1}^0\+\gamma_{+1,i}^0 + \+\varepsilon_i)  \+\gamma_{1,i}^{0\prime} \*F_{1}^{0\prime}  .
\end{align}
By the proof for each term of \eqref{B16}, it is easy to know that
\begin{eqnarray}
T^{-\nu_1} \|\widehat{\*\Sigma}_1 - \+\Sigma_{1}^0\| =  O_P( T^{-(\nu_1-\nu_2)/2}),
\end{eqnarray}
and so
\begin{eqnarray}
|J_5| \le \|\*h_{1,d}^{0}\|^2 T^{-\nu_1}\|\*F_1^{0} \|^2 T^{-\nu_1}\| \widehat{\+\Sigma}_1 - \+\Sigma_{1}^0\| =O_P( T^{-(\nu_1-\nu_2)/2} ).
\end{eqnarray}

Hence, by putting everything together,
\begin{align}
T^{-\nu_1}| \widehat{\lambda}_{1,d} -\lambda_{1,d} | \le |J_1| +2|J_2|+|J_3|+2|J_4| +|J_5| =O_P(  T^{-(\nu_1-\nu_2)/2} ),
\end{align}
which establishes (a).

\medskip

Consider (b). This proof is based on Lemma \ref{applem:ortho}. We therefore start by introducing some notation in order to make the problem here fit the one in Lemma \ref{applem:ortho}. Let us therefore denote by $\*F_1^{\perp}$ a $T\times (d_{max}- d_1)$ matrix such that $T^{-\nu_1} (\*F_1^{\perp},  \*F_1^0 \*R)'(\*F_1^{\perp},  \*F_1^0\*R) = \mathrm{diag}( \*I_{d_{max}-d_1} , \*I_{d_1} )$, where $\*R$ is a $d_1\times d_1$ rotation matrix. The matrices $T^{-\nu_1/2}\*F_1^{\perp}$, $T^{-\nu_1/2}\*F_1^0 \*R$, $\+\Sigma_{1}^0$ and $\widehat{\+\Sigma}_1  - \+\Sigma_{1}^0$ correspond to $\*Q_1$, $\*Q_2$, $\*A$ and $\*E$ of Lemma \ref{applem:ortho}. Our counterpart of the matrix $\*Q^0_1$ appearing in this other lemma is thus given by
\begin{eqnarray}
\widehat{\*F}^\perp = T^{-\nu_1/2}( \*F_1^{\perp} + \*F_1^0\*R\*P)(\*I_{d_{max}-d_1}+\*P'\*P)^{-1/2},
\end{eqnarray}
where
\begin{align}
\|\*P\|_{2} &\le  \frac{4}{\mathrm{sep}(0, T^{-2\nu_1}\*F_1^{0\prime} \+\Sigma_{1}^0 \*F_1^0 )} T^{-\nu_1}\|\widehat{\+\Sigma}_1  - \+\Sigma_{1}^0\| \le O_P(1) T^{-\nu_1}\|\widehat{\+\Sigma}_1  - \+\Sigma_{1}^0\| \notag\\
& =O_P( T^{-(\nu_1-\nu_2)/2}).
\end{align}
Since $\widehat{\*F}^\perp$ is an orthonormal basis for a subspace that is invariant for $\widehat{\+\Sigma}_1$, we have $\widehat{\lambda}_{1,d_1+d} = \widehat{\*F}_d^{\perp\prime} \widehat{\+\Sigma}_1 \widehat{\*F}_{d}^\perp$, where $d = 1,\ldots,d_{max}-d_1$ and $\widehat{\*F}_{d}^\perp$ is the $d$-th column of $\widehat{\*F}^\perp$. Consider $\|\widehat{\*F}^\perp  -  T^{-\nu_1/2}\*F_1^{\perp} \|_{2}$. By the definition of $\widehat{\*F}^\perp$,
\begin{align}\label{fperpcons}
\lefteqn{ \| \widehat{\*F}^\perp - T^{-\nu_1/2}\*F_1^{\perp} \|_{2} } \nonumber \\
&= T^{-\nu_1/2} \| [ \*F_1^{\perp} + \*F_1^0\*R\*P - \*F_1^{\perp}(\*I_{d_{max}-d_1}+\*P'\*P)^{1/2} ](\*I_{d_{max}-d_1}+\*P'\*P)^{-1/2} \|_{2}\nonumber \\
&\le  T^{-\nu_1/2} \|  \*F_1^{\perp}(\*I_{d_{max}-d_1} - (\*I_{d_{max}-d_1}+\*P'\*P)^{1/2})  (\*I_{d_{max}-d_1}+\*P'\*P)^{-1/2}\|_{2} \nonumber \\
&+  T^{-\nu_1/2} \| \*F_1^0\*R\*P (\*I_{d_{max}-d_1}+\*P'\*P)^{-1/2} \|_{2} \nonumber \\
&\le \|  (\*I_{d_{max}-d_1} - (\*I_{d_{max}-d_1}+\*P'\*P)^{1/2})  (\*I_{d_{max}-d_1}+\*P'\*P)^{-1/2} \|_{2} + \| \*P (\*I_{d_{max}-d_1}+\*P'\*P)^{-1/2}\|_{2} \nonumber \\
&\le \| \*I_{d_{max}-d_1} - (\*I_{d_{max}-d_1}+\*P'\*P)^{1/2}\|_{2} + \| \*P\|_{2}\le 2\|\*P \|_{2}\notag\\
&=O_P( T^{-(\nu_1-\nu_2)/2} ),
\end{align}
where the second and third inequalities follow from \cite[Exercise 1 on page 231]{Magnus}. This last result can be used to show that
\begin{align}
& T^{-\nu_1}|\widehat{\lambda}_{1,d_1+d} | = |\widehat{\*F}_{d}^{\perp\prime} (T^{-\nu_1}\widehat{\+\Sigma}_1 ) \widehat{\*F}^\perp_{d} |\notag\\
& = |(\widehat{\*F}^\perp_{d} - T^{-\nu_1/2}\*F_{1,d}^{\perp} +   T^{-\nu_1/2}\*F_{1,d}^{\perp})' T^{-\nu_1}( \widehat{\+\Sigma}_1 - \+\Sigma_{1}^0+ \+\Sigma_{1}^0 ) (\widehat{\*F}^\perp_{d} -  T^{-\nu_1/2}\*F_{1,d}^{\perp}+  T^{-\nu_1/2}\*F_{1,d}^{\perp} ) | \nonumber \\
&\le \|\widehat{\*F}^\perp_{d} -   T^{-\nu_1/2}\*F_{1,d}^{\perp}\|^2T^{-\nu_1}\|\widehat{\+\Sigma}_1 - \+\Sigma_{1}^0\| \notag\\
& + 2 \|\widehat{\*F}^\perp_{d} -  T^{-\nu_1/2}\*F_{1,d}^{\perp}\|T^{-\nu_1}\|\widehat{\+\Sigma}_1 - \+\Sigma_{1}^0\| T^{-\nu_1/2}\|\*F_{1,d}^{\perp}\| + \|\widehat{\*F}^\perp_{d} -   T^{-\nu_1/2}\*F_{1,d}^{\perp}\|^2 T^{-\nu_1}\|\+\Sigma_{1}^0\| \notag\\
&= O_P( T^{ -(\nu_1-\nu_2)} ),
\end{align}
where $\*F_{1,d}^{\perp}$ is the $d$-th column of $\*F_1^{\perp}$, and the last equality follows from \eqref{fperpcons} and the proof of part (a). This completes the proof of the lemma.\hspace*{\fill}{$\blacksquare$}

\bigskip

\noindent \textbf{Proof of Lemma \ref{applem:f}.}

\bigskip

\noindent For (a), we take the same starting point as in the proof of part (a) in Lemma \ref{applem:lam1}, which is \eqref{eq:fexp} with $\widehat{\*F}_1$ and $\*V_{1}$ based on treating $d_1$ as known. The rationale for doing so is, as already explained in Appendix \ref{sec:outline}, that $\widehat{d}_1$ is consistent. Pre-multiplying this equation through by $T^{-(\nu_1+\delta)/2}\*F_2^{0\prime}$ gives
\begin{align}
\lefteqn{ T^{-(\nu_1+\delta)}\*F_2^{0\prime}\widehat{\*F}_1 \*V_{1} - T^{-(\nu_1+\delta)/2}\*F_2^{0\prime}\*J_{9} }\notag\\
&= T^{-(\nu_1+\delta)}\*F_2^{0\prime}\widehat{\*F}_1 \*V_{1} - T^{-(\nu_1+\delta)/2}\*F_2^{0\prime}\*F_{1}^0(N^{-1}\+\Gamma_{1}^{0\prime}\+\Gamma_{1}^0) (T^{-(\nu_1+\delta)/2}\*F_{1}^{0\prime} \widehat{\*F}_1) \notag\\
& = \sum_{j=1}^8 T^{-(\nu_1+\delta)/2}\*F_2^{0\prime}\*J_{j},
\end{align}
or
\begin{align}
\lefteqn{ T^{-\delta}\*F_2^{0\prime}\widehat{\*F}_1  - T^{-(\nu_1+\delta)/2}\*F_2^{0\prime}\*J_{9}(T^{-\nu_1}\*V_{1})^{-1} }\notag\\
&= T^{-\delta}\*F_2^{0\prime}\widehat{\*F}_1 - T^{-(\nu_1+\delta)/2}\*F_2^{0\prime}\*F_{1}^0(N^{-1}\+\Gamma_{1}^{0\prime}\+\Gamma_{1}^0) (T^{-(\nu_1+\delta)/2}\*F_{1}^{0\prime} \widehat{\*F}_1) (T^{-\nu_1}\*V_{1})^{-1}\notag\\
& = T^{-(\nu_1-\nu_2)/2}\sum_{j=1}^8 T^{-(\delta+\nu_2)/2}\*F_2^{0\prime}\*J_{j}(T^{-\nu_1}\*V_{1})^{-1},
\end{align}
Under Assumption \ref{ass:ortho}, the orders of $\*J_{1},\ldots,\*J_{6}$ are the same as in those in the proof of the first result of Lemma \ref{applem:lam1}. The stated orders of $\*J_{7}$ and $\*J_{8}$ are, however, not sharp and can be improved upon. The order of $T^{-(\delta+\nu_2)/2}\|\*F_2^{0\prime}\*J_{7}\|$ is given by
\begin{align}
T^{-(\delta+\nu_2)/2}\| \*F_2^{0\prime}\*J_{7}\| &\le  O_P(1) T^{-\nu_2/2} \left\|\frac{1}{NT^{(\nu_1+\delta)/2}}\sum_{i=1}^N \*F_2^{0\prime}\*F_1^0 \+\gamma_{1,i}^0( \*F_{+1}^{0} \+\gamma_{+1,i}^0+ \+\varepsilon_i)'\right\|  \notag\\
&\le  O_P(1) \frac{1}{NT^{(\nu_2+\nu_1+\delta)/2}}\|\*F_2^{0\prime} \*F_1^0 \+\Gamma_{1}^{0\prime} \+\Gamma_{+1}^0 \*F_{+1}^{0\prime} \| \notag\\
&+ O_P(1) \frac{1}{NT^{(\nu_2+\nu_1+\delta)/2}} \| \*F_2^{0\prime}\*F_1^0\+\Gamma_{1}^{0\prime}\+\varepsilon \| \notag\\
&= O_P( N^{-(1-p)}T^{q- (\nu_1+\delta)/2} ) + O_P( N^{-1/2}T^{q-(\nu_2+\nu_1+\delta-1)/2} ),
\end{align}
where the last equality follows from Assumption \ref{ass:ortho} and Lemma \ref{applem:mom}. We can similarly show that
\begin{align}
T^{-(\delta+\nu_2)/2}\| \*F_2^{0\prime}\*J_{8}\| &\le O_P(1)T^{-\nu_2/2} \left\|\frac{1}{NT^{(\nu_1+\delta)/2}}\sum_{i=1}^N \*F_2^{0\prime}(\*F_2^0  \+\gamma_{2,i}^0 + \*F_{+2}^{0}\+\gamma_{+2,i}^0 + \+\varepsilon_i)\+\gamma_{1,i}^{0\prime}\*F_1^{0\prime}\right\|  \notag\\
&\le  O_P(1) \frac{1}{NT^{(\nu_2 + \nu_1+\delta)/2}}\| \*F_2^{0\prime}\*F_2^0 \+\Gamma_{2}^{0\prime} \+\Gamma_{1}^0 \*F_1^{0\prime} \|  \notag\\
&+O_P(1)N^{-1}T^{-(\nu_2+\nu_1+\delta)/2}\|\*F_2^{0\prime} \*F_{+2}^{0} \+\Gamma_{+2}^{0\prime} \+\Gamma_{1}^0\*F_1^{0\prime}  \| \notag\\
&+O_P(1)N^{-1}T^{-(\nu_2+\nu_1+\delta)/2}\|\*F_2^{0\prime}\+\varepsilon'\+\Gamma_{1}^0\*F_1^{0\prime} \| \notag\\
&= O_P( N^{-(1-p)}T^{-(\delta -\nu_2)/2})+ O_P(N^{-1/2}T^{-(\delta-1)/2 } ),
\end{align}
where $\*F_{+2}^{0}$ and $\+\Gamma_{+2}^{0}$ are defined analogously to $\*F_{+1}^{0}$ and $\+\Gamma_{+1}^{0}$ in the proof of Lemma \ref{applem:lam1}. By using these last two results together with the orders of $\*J_{1},\ldots,\*J_{6}$ given in the proof of the first result of Lemma \ref{applem:lam1},
\begin{align}
\lefteqn{ \left\| \sum_{j=1}^8T^{-(\delta+\nu_2)/2} \*F_2^{0\prime}\*J_{j} \right\| } \notag\\
&\le O_P( T^{-(\delta-1)/2} \| \*D_T^{-1}(\+\beta^0 - \widehat{\+\beta}_0 )\|) +O_P( T^{\nu_2-(\nu_1+\delta)/2} )  + O_P(N^{-1/2}T^{1-(\nu_1+\delta)/2}) \notag\\
&+ O_P(N^{-(1-p)}T^{-(\delta -\nu_2)/2}) + O_P(N^{-1/2}T^{-(\delta-1)/2 } ) \notag\\
&= O_P(T^{-\delta/2}+N^{-1/2}T^{-(\delta-1)/2} +  N^{-(1-p)}T^{-(\delta -\nu_2)/2}),
\end{align}
where we have used $\|\*D_T^{-1}(\+\beta^0 - \widehat{\+\beta}_0 )\| =O_P(N^{-1/2} \vee T^{-1/2})$ of Lemma \ref{lem:beta0hat}, and Assumptions \ref{ass:eps} and \ref{ass:ortho}. Hence,
\begin{align}
\lefteqn{ \|T^{-\delta}\*F_2^{0\prime}\widehat{\*F}_1  - T^{-(\nu_1+\delta)/2}\*F_2^{0\prime}\*J_{9}(T^{-\nu_1}\*V_{1})^{-1} \| }\notag\\
& \le T^{-(\nu_1-\nu_2)/2}\left\|\sum_{j=1}^8 T^{-(\delta+\nu_2)/2}\*F_2^{0\prime}\*J_{j}\right\|\|(T^{-\nu_1}\*V_{1})^{-1}\|\notag\\
& = T^{-(\nu_1-\nu_2)/2}[O_P(T^{-\delta/2} +N^{-1/2}T^{-(\delta-1)/2} + N^{-(1-p)}T^{-(\delta -\nu_2)/2})] \notag\\
& = O_P( T^{-(\delta+\nu_1-\nu_2)/2}+N^{-1/2}T^{-(\delta+\nu_1 -\nu_2 -1)/2} +N^{-(1-p)}T^{-(\delta +\nu_1-2\nu_2)/2}) ,
\end{align}
which together with Assumption \ref{ass:ortho} yields
\begin{align}
\|T^{-\delta}\*F_2^{0\prime}\widehat{\*F}_1\| & \le \|T^{-(\nu_1+\delta)/2}\*F_2^{0\prime}\*J_{9}\|\|(T^{-\nu_1}\*V_{1})^{-1} \| \notag\\
&+ T^{-(\nu_1-\nu_2)/2}\left\|\sum_{j=1}^8 T^{-(\delta+\nu_2)/2}\*F_2^{0\prime}\*J_{j}\right\|\|(T^{-\nu_1}\*V_{1})^{-1}\|\notag\\
& = O_P(T^{q-(\nu_1+\delta)/2}+ T^{-(\delta+\nu_1-\nu_2)/2} + N^{-1/2}T^{-(\delta+\nu_1 -\nu_2 -1)/2} \notag\\
&+ N^{-(1-p)}T^{-(\delta +\nu_1-2\nu_2)/2}) ,
\end{align}
as was to be shown for (a).

\medskip

Let us now consider (b). Analogously to the proof of (a), by invoking Assumption \ref{ass:ortho} we can improve the orders of $\*J_{7}$ and $\*J_{8}$. For $\*J_7$,
\begin{align}
T^{-\delta/2}\| \*J_{7}\| &\le O_P(1) N^{-1}T^{-(\nu_1+\delta)/2}\| \*F_1^0 \+\Gamma_{1}^{0\prime} \+\Gamma_{+1}^0 \*F_{+1}^{0\prime} \|  + O_P(1) N^{-1}T^{-(\nu_1+\delta)/2} \|\*F_1^0\+\Gamma_{1}^{0\prime}\+\varepsilon \| \notag\\
&= O_P(N^{-(1-p)}T^{-(\delta-\nu_2)/2})+ O_P(N^{-1/2}T^{-(\delta-1)/2}),
\end{align}
where the equality follows from Assumption \ref{ass:ortho} and Lemma \ref{applem:mom}. For $\*J_8$,
\begin{eqnarray}
T^{-\delta/2}\| \*J_{8} \| \le O_P(N^{-(1-p)}T^{-(\delta-\nu_2)/2}) + O_P(N^{-1/2} T^{-(\delta-1)/2}).
\end{eqnarray}
This implies that the result in \eqref{eq:fcons} changes to (after replacing $T^{-\nu_1/2}$ by $T^{-\delta/2}$)
\begin{align}
&T^{-\delta/2}\|T^{(\nu_1-\delta)/2}\widehat{\*F}_1  - \*F_{1}^0\*H_{1}^0\|  \le \sum_{j=1}^8 T^{-\delta/2} \|\*J_{j}\|\|(T^{-\nu_1}\*V_{1})^{-1}\| \notag\\
& = O_P( T^{-\delta/2}+N^{-1/2}T^{-(\delta-1)/2} + N^{-(1-p)}T^{-(\delta -\nu_2)/2}).
\end{align}
Hence, since $\*H_{1}^0$ is invertible with $\|\*H_{1}^0\| = O_P(1)$ and $\*H_1 = T^{-(\nu_1-\delta)/2}\*H_{1}^0$,
\begin{align}\label{eq:fcons2}
T^{-\delta/2}\|\widehat{\*F}_1\*H_{1}^{-1}  - \*F_{1}^0\| = O_P( T^{-\delta/2} + N^{-1/2}T^{-(\delta-1)/2} + N^{-(1-p)}T^{-(\delta -\nu_2)/2}).
\end{align}

We are now ready to consider $\sum_{i=1}^N \| \*F_{1}^{0\prime} \+\gamma_{1,i}^0 - \widehat{\*F}_1 \widehat{\+\gamma}_{1,i}\|^2$.
\begin{align}
& \sum_{i=1}^N \| \*F_{1}^{0\prime} \+\gamma_{1,i}^0 - \widehat{\*F}_1 \widehat{\+\gamma}_{1,i}\|^2  =\sum_{i=1}^N \| \*F_1^0 \+\gamma_{1,i}^0 - T^{-\delta}\widehat{\*F}_1 \widehat{\*F}_1' (\*y_i-\*X_i\widehat{\+\beta}_0)\|^2 \notag\\
&\le O(1)\sum_{i=1}^N[ \| \*P_{\widehat{F}_1}\*X_i (\+\beta^0-\widehat{\+\beta}_0)\|^2 + \| \*M_{\widehat{F}_1} \*F_1^0\+\gamma_{1,i}^0 \|^2 + \|\*P_{\widehat{F}_1}\*F_{+1}^{0}\+\gamma_{+1,i}^0\|^2 + \|\*P_{\widehat{F}_1}\+\varepsilon_i\|^2].
\end{align}
We now evaluate each of the terms on the right-hand side one by one. Making use of Assumption \ref{ass:mom} and Lemma \ref{lem:beta0hat}, we get
\begin{align}
\sum_{i=1}^N \| \*P_{\widehat{F}_1}\*X_i (\+\beta^0-\widehat{\+\beta}_0)\|^2 & \le \sum_{i=1}^N\| \*X_i\*D_T\|^2  \|\*D_T^{-1}(\+\beta^0-\widehat{\+\beta}_0)\|^2 \notag\\
& = O_P(NT)O_P(N^{-1}\vee T^{-1}) = O_P(N\vee T),
\end{align}
and by another application of Lemma \ref{applem:mom},
\begin{align}
\sum_{i=1}^N\|\*P_{\widehat{F}_1}\+\varepsilon_i\|^2 \le \sum_{i=1}^N\|\+\varepsilon_i\|^2 = O_P(N\vee T).
\end{align}
For $\sum_{i=1}^N\| \*M_{\widehat{F}_1} \*F_1^0\+\gamma_{1,i}^0 \|^2$, we use \eqref{eq:fcons2} from which it follows that
\begin{align}\label{eq:mfcons}
\| \*M_{\widehat{F}_1} \*F_1^0 \|^2 &= \| \*M_{\widehat{F}_1} (\*F_1^0 - \widehat{\*F}_{1}\*H_1^{-1})\|^2 \le T^\delta (T^{-\delta}\| \*F_1^0 - \widehat{\*F}_{1}\*H_1^{-1}\|^2)\nonumber \\
&= O_P(1) + O_P(N^{-1}T) + O_P(N^{-2(1-p)}T^{\nu_2}),
\end{align}
which in turn implies
\begin{eqnarray}
\sum_{i=1}^N\| \*M_{\widehat{F}_1} \*F_1^0\+\gamma_{1,i}^0 \|^2 = O_P(N) + O_P(T) + O_P(N^{-(1-2p)}T^{\nu_2}).
\end{eqnarray}

For $\sum_{i=1}^N\|\*P_{\widehat{F}_1}\*F_{+1}^{0} \+\gamma_{+1,i}^0\|^2$, we use the result given in part (a), giving
\begin{align}
\sum_{i=1}^N\|\*P_{\widehat{F}_1}\*F_2^0 \+\gamma_{2,i}^0\|^2 & = \sum_{i=1}^N\|T^{-\delta}\widehat{\*F}_1\widehat{\*F}_1'\*F_2^0 \+\gamma_{2,i}^0\|^2  = O_P(NT^\delta ) \|T^{-\delta}\widehat{\*F}_1' \*F_2^0\|^2 \notag\\
&= O_P(NT^\delta) [O_P(T^{q-(\nu_1+\delta)/2}) + O_P( T^{-(\delta+\nu_1-\nu_2)/2})  \notag\\
&+ O_P(N^{-1/2}T^{-(\delta+\nu_1 -\nu_2 -1)/2})+ O_P(N^{-(1-p)}T^{-(\delta +\nu_1-2\nu_2)/2})]^2 \notag\\
&= O_P(N)[ O_P(T^{-(\nu_1-\nu_2)/2}) + O_P(N^{-1/2}T^{-(\nu_1 -\nu_2 -1)/2}) \notag\\
& + O_P(N^{-(1-p)}T^{-(\nu_1-2\nu_2)/2})+ O_P(T^{q-\nu_1/2}) ]^2.
\end{align}
The order of $\sum_{i=1}^N\|\*P_{\widehat{F}_1}\*F_{+2}^{0} \+\gamma_{+2,i}^0\|^2$ is the same. Hence, by adding the above results, (b) follows after simple algebra. The proof is now complete.\hspace*{\fill}{$\blacksquare$}

\bigskip

\noindent \textbf{Proof of Lemma \ref{applem:lam2}.}

\bigskip

\noindent Let $\*U_i = \*F_{+2}^0 \+\gamma_{+2,i}^0 +\*F_{1}^{0\prime} \+\gamma_{1,i}^0 - \widehat{\*F}_1 \widehat{\+\gamma}_{1,i}$. In this notation,
\begin{align}
\lefteqn{ T^{-(\nu_2+\delta)/2}\widehat{\*F}_2\*V_{2} }\notag\\
&= \frac{1}{NT^{(\nu_2+\delta)/2}} \sum_{i=1}^N[ ( \*X_i (\+\beta^0 - \widehat{\+\beta}_0) + \*F_2^0 \+\gamma_{2,i}^0 + \*U_i + \+\varepsilon_i ) \notag\\
& \times ( \*X_i (\+\beta^0 - \widehat{\+\beta}_0) + \*F_2^0 \+\gamma_{2,i}^0 + \*U_i + \+\varepsilon_i )'] \widehat{\*F}_2 \nonumber \\
&= \frac{1}{NT^{(\nu_2+\delta)/2}} \sum_{i=1}^N \*X_i (\+\beta^0 - \widehat{\+\beta}_0) (\+\beta^0 - \widehat{\+\beta}_0)' \*X_i' \widehat{\*F}_2 \nonumber\\
& + \frac{1}{NT^{(\nu_2+\delta)/2}} \sum_{i=1}^N \*X_i (\+\beta^0 - \widehat{\+\beta}_0) \+\gamma_{2,i}^{0\prime}\*F_2^{0\prime}\widehat{\*F}_2 + \frac{1}{NT^{(\nu_2+\delta)/2}} \sum_{i=1}^N \*F_2^0 \+\gamma_{2,i}^0  (\+\beta^0 - \widehat{\+\beta}_0)' \*X_i' \widehat{\*F}_2 \notag\\
& + \frac{1}{NT^{(\nu_2+\delta)/2}} \sum_{i=1}^N \*X_i (\+\beta^0 - \widehat{\+\beta}_0) \*U_i'\widehat{\*F}_2 + \frac{1}{NT^{(\nu_2+\delta)/2}}  \sum_{i=1}^N \*U_i (\+\beta^0 - \widehat{\+\beta}_0)' \*X_i' \widehat{\*F}_2 \notag\\
& + \frac{1}{NT^{(\nu_2+\delta)/2}} \sum_{i=1}^N \*X_i (\+\beta^0 - \widehat{\+\beta}_0) \+\varepsilon_i' \widehat{\*F}_2 + \frac{1}{NT^{(\nu_2+\delta)/2}}\sum_{i=1}^N \+\varepsilon_i (\+\beta^0 - \widehat{\+\beta}_0)' \*X_i' \widehat{\*F}_2 \nonumber \\
& + \frac{1}{NT^{(\nu_2+\delta)/2}} \sum_{i=1}^N \+\varepsilon_i \+\varepsilon_i' \widehat{\*F}_2 + \frac{1}{NT^{(\nu_2+\delta)/2}} \sum_{i=1}^N  \*F_2^0 \+\gamma_{2,i}^0 \+\varepsilon_i' \widehat{\*F}_2 + \frac{1}{NT^{(\nu_2+\delta)/2}} \sum_{i=1}^N \+\varepsilon_i \+\gamma_{2,i}^{0\prime}\*F_2^{0\prime} \widehat{\*F}_2 \nonumber \\
& + \frac{1}{NT^{(\nu_2+\delta)/2}} \sum_{i=1}^N \+\varepsilon_i \*U_i'\widehat{\*F}_2 +\frac{1}{NT^{(\nu_2+\delta)/2}}  \sum_{i=1}^N \*U_i \+\varepsilon_i' \widehat{\*F}_2 + \frac{1}{NT^{(\nu_2+\delta)/2}} \sum_{i=1}^N \*U_i \+\gamma_{2,i}^{0\prime}\*F_2^{0\prime}\widehat{\*F}_2 \notag\\
&+\frac{1}{NT^{(\nu_2+\delta)/2}} \sum_{i=1}^N  \*F_2^0 \+\gamma_{2,i}^0\*U_i'\widehat{\*F}_2  + \frac{1}{NT^{(\nu_2+\delta)/2}} \sum_{i=1}^N  \*U_i\*U_i'\widehat{\*F}_2  +\frac{1}{NT^{(\nu_2+\delta)/2}}  \sum_{i=1}^N \*F_2^0 \+\gamma_{2,i}^0 \+\gamma_{2,i}^{0\prime} \*F_2^{0\prime} \widehat{\*F}_2 \nonumber \\
&=\sum_{j=1}^{16} \*K_{j},
\end{align}
where $\*K_{1},\ldots,\*K_{16}$ are implicitly defined. Analogously to the proof of the first result of Lemma \ref{applem:lam1} we move $\*K_{16} =  \*F_2^0 (N^{-1}\+\Gamma_{2}^{0\prime} \+\Gamma_{2}^0)(T^{-(\nu_2+\delta)/2} \*F_2^{0\prime} \widehat{\*F}_2)$ over to the left, giving
\begin{eqnarray}\label{eq:f2exp}
T^{-(\nu_2+\delta)/2}\widehat{\*F}_2 \*V_{2} - \*F_2^0(N^{-1}\+\Gamma_{2}^{0\prime} \+\Gamma_{2}^0 )( T^{-(\nu_2+\delta)/2} \*F_2^{0\prime} \widehat{\*F}_2) = \sum_{j=1}^{15} \*K_j.
\end{eqnarray}
By using the same steps employed in the proof of the first result of Lemma \ref{applem:lam1}, we can show that
\begin{eqnarray}
T^{-\delta/2} \| \*K_{1} + \*K_{2} + \*K_{3} \| = O_P( T^{-(\delta-1)/2} \| \*D_T^{-1}(\+\beta^0 - \widehat{\+\beta}_0 )\| ).
\end{eqnarray}

For $\*K_{4}$,
\begin{align}
&T^{-\delta/2}\|\*K_{4}\|   = T^{-\delta/2}\left\|\frac{1}{NT^{(\nu_2+\delta)/2}}  \sum_{i=1}^N \*X_i (\+\beta^0 - \widehat{\+\beta}_0) \*U_i'\widehat{\*F}_2 \right\| \notag\\
&\le O(1) \left( \frac{1}{NT^\delta} \sum_{i=1}^N \| \*X_i (\+\beta^0 - \widehat{\+\beta}_0)\|^2 \right)^{1/2} \notag\\
&\times \left(\frac{1}{NT^{\nu_2}} \sum_{i=1}^N\| \*F_{+2}^0 \+\gamma_{2,i}^0\|^2+ \frac{1}{NT^{\nu_2}} \sum_{i=1}^N \| \*F_{1}^{0\prime} \+\gamma_{1,i}^0 - \widehat{\*F}_1 \widehat{\+\gamma}_{1,i} \|^2 \right)^{1/2} \notag\\
&= O_P( 1) T^{-(\delta-1)/2}\| \*D_T^{-1}(\+\beta^0 - \widehat{\+\beta}_0)\| [ O_P(T^{-(\nu_2-\nu_3)})+ O_P(T^{-\nu_2}) +
O_P(N^{-1}T^{-(\nu_2-1)}) \notag\\
&  + O_P(T^{-(\nu_1-\nu_2)})  + O_P(T^{-(\nu_2+\nu_1 -2q)}) + O_P(N^{-1}T^{-(\nu_1+\nu_2-2)}) + O_P(N^{-2(1-p)})] \notag\\
&= o_P ( T^{-(\delta-1)/2}\| \*D_T^{-1}(\+\beta^0 - \widehat{\+\beta}_0)\|),
\end{align}
where the second equality follows Lemma \ref{applem:f}, and the third follows from Assumptions \ref{ass:mom}, \ref{ass:eps}, and \ref{ass:ortho}. The same arguments can be used to show that
\begin{eqnarray}
T^{-\delta/2}\|\*K_{5} \| = o_P ( T^{-(\delta-1)/2}\| \*D_T^{-1}(\+\beta^0 - \widehat{\+\beta}_0)\|).
\end{eqnarray}
For $\*K_{6}$,
\begin{align}
T^{-\delta/2}\| \*K_{6} \| & \le T^{-\delta/2}\| \widehat{\*F}_2\|\left\|\frac{1}{NT^{(\nu_2+\delta)/2}}  \sum_{i=1}^N \*X_i (\+\beta^0 - \widehat{\+\beta}_0) \+\varepsilon_i'\right\| \notag\\
&=   O_P( N^{-1/2}T^{-(\delta+\nu_2-2)/2} \| \*D_T^{-1}(\+\beta^0 - \widehat{\+\beta}_0)\| ) \notag\\
& = o_P (  T^{-(\delta-1)/2}\| \*D_T^{-1}(\+\beta^0 - \widehat{\+\beta}_0)\| ),
\end{align}
where the development is similar to \eqref{eqbp1}. The order of $T^{-\delta/2}\|\*K_{7}\|$ is the same.

For $\*K_{8}$,
\begin{align}
T^{-\delta/2}\| \*K_{8} \|&= T^{-\delta/2}\left\| \frac{1}{NT^{(\nu_2+\delta)/2}} \sum_{i=1}^N \+\varepsilon_i \+\varepsilon_i' \widehat{\*F}_2  \right\|\le T^{-\delta/2}\| \widehat{\*F}_2\| N^{-1}T^{-(\nu_2+\delta)/2} \|  \+\varepsilon' \+\varepsilon  \| \notag\\
&= O_P( T^{1-(\nu_2+\delta)/2} ( N^{-1/2}\vee T^{-1/2} ) ),
\end{align}
where the last equality holds by Lemma \ref{applem:mom}.

Further use of Lemma \ref{applem:mom} gives
\begin{align}
T^{-\delta/2}\| \*K_{9} \| &= T^{-\delta/2}\left\| \frac{1}{NT^{(\nu_2+\delta)/2}} \sum_{i=1}^N \*F_2^0 \+\gamma_{2,i}^0 \+\varepsilon_i' \widehat{\*F}_2 \right\|\le  N^{-1}T^{-(\nu_2+\delta)/2}\| \*F_2^0 \+\Gamma_{2}^{0\prime}\+\varepsilon \| T^{-\delta/2}\| \widehat{\*F}_2\|\notag\\
&= O_P( N^{-1/2}T^{-(\delta-1)/2}),
\end{align}
and we can show that $T^{-\delta/2}\| \*K_{10} \|$ is of the same order.

$\*K_{11}$ requires more work. We begin by expanding it in the following way:
\begin{align}\label{eq:k11exp}
& T^{-\delta/2}\| \*K_{11} \| = T^{-\delta/2}\left\|  \frac{1}{NT^{(\nu_2+\delta)/2}} \sum_{i=1}^N \+\varepsilon_i (\*F_1^0 \+\gamma_{1,i}^0 - \widehat{\*F}_1 \widehat{\+\gamma}_{1,i} +\*F_{+2}^0 \+\gamma_{2,i}^0 )'\widehat{\*F}_2 \right\| \notag\\
&\le  \left\|  \frac{1}{NT^{(\nu_2+\delta)/2}} \sum_{i=1}^N \+\varepsilon_i (\*F_1^0 \+\gamma_{1,i}^0 - \widehat{\*F}_1 T^{-\delta} \widehat{\*F}_1' (\*y_i-\*X_i\widehat{\+\beta}_0) + \*F_{+2}^0 \+\gamma_{2,i}^0)'\right\| T^{-\delta/2}\| \widehat{\*F}_2\|\notag\\
&\le O(1) \left\| \frac{1}{NT^{(\nu_2+\delta)/2}}\sum_{i=1}^N \+\varepsilon_i (\+\beta^0-\widehat{\+\beta}_0)' \*X_i ' \*P_{\widehat{F}_1}\right\| \notag\\
& + O(1) \left\| \frac{1}{NT^{(\nu_2+\delta)/2}}\sum_{i=1}^N\+\varepsilon_i \+\gamma_{1,i}^{0\prime} \*F_1^{0\prime}\*M_{\widehat{F}_1} \right\| +O(1) \left\| \frac{1}{NT^{(\nu_2+\delta)/2}}\sum_{i=1}^N\+\varepsilon_i  \+\gamma_{+1,i}^{0\prime} \*F_{+1}^{0\prime} \*P_{\widehat{F}_1}\right\|\notag\\
&+ O(1) \left\| \frac{1}{NT^{(\nu_2+\delta)/2}}\sum_{i=1}^N\+\varepsilon_i \+\varepsilon_i'\*P_{\widehat{F}_1} \right\| +O(1)\left\| \frac{1}{NT^{(\nu_2+\delta)/2}}\sum_{i=1}^N\+\varepsilon_i  \+\gamma_{+2,i}^{0\prime} \*F_{+2}^{0\prime} \right\| \notag\\
&= O(1)(K_{111}+K_{112}+K_{113}+K_{114}+K_{115}),
\end{align}
where, similarly to the analysis of $\*K_{6}$ and using $\|\*P_{\widehat F_1}\|_2 = 1$,
\begin{align}
K_{111} &\le  \left\|\frac{1}{NT^{(\nu_2+\delta)/2}}\sum_{i=1}^N \+\varepsilon_i (\+\beta^0-\widehat{\+\beta}_0)' \*X_i ' \right\| =O_P( N^{-1/2}T^{-(\delta+\nu_2-2)/2}\| \*D_T^{-1}(\+\beta^0 - \widehat{\+\beta}_0)\| )  \notag\\
& =  o_P( T^{-(\delta-1)/2} \| \*D_T^{-1}(\+\beta^0 - \widehat{\+\beta}_0)\| ).
\end{align}
Also, making use of \eqref{eq:mfcons}, we can show that
\begin{align}
K_{112} &\le  N^{-1}T^{-(\nu_2+\delta)/2} \|\+\Gamma_{1}^{0\prime}\+\varepsilon \| \|\*M_{\widehat{F}_1}\*F_1^0\|  \notag\\
&= O_P( N^{-1/2}T^{-(\nu_2+\delta-1)/2}) + O_P(N^{-1}T^{1-(\nu_2+\delta)/2}) + O_P(N^{-(3/2-p)}T^{-(\delta-1)/2}).
\end{align}
Similarly, for $K_{113}$, we can show that
\begin{align}
K_{113} = o_P(K_{112})+O_P( N^{-1/2}T^{q-(\nu_1+\nu_2+\delta-1)/2}) ,
\end{align}
where we have used Lemmas \ref{applem:mom} and \ref{applem:f}, and Assumption \ref{ass:mom}. Further use of Lemma \ref{applem:mom} shows that $K_{114}$ is of the following order:
\begin{eqnarray}
K_{114} \le  N^{-1}T^{-(\nu_2+\delta)/2} \| \+\varepsilon' \+\varepsilon \|   =O_P( T^{-(\nu_2+\delta-2)/2} ( N^{-1/2}\vee T^{-1/2} ) ),
\end{eqnarray}
while the order of $K_{115}$ is
\begin{eqnarray}
K_{115} = N^{-1}T^{-(\nu_2+\delta)/2} \|\+\varepsilon'\+\Gamma_{+2}^0 \*F_{+2}^{0\prime}\| = O_P( N^{-1/2} T^{-(\nu_2+\delta-1 -\nu_3)/2} ).
\end{eqnarray}
By inserting the above results into \eqref{eq:k11exp}, we obtain
\begin{align}
T^{-\delta/2}\|  \*K_{11} \| & = O_P(  N^{-(3/2-p)}T^{-(\delta-1)/2}) + O_P( N^{-1/2}T^{q-(\nu_1+\nu_2+\delta-1)/2})  \notag\\
& + O_P( T^{-(\nu_2+\delta-2)/2} (N^{-1/2}\vee T^{-1/2})) + O_P( N^{-1/2} T^{-(\nu_2+\delta-1 -\nu_3)/2} )\notag\\
& + o_P( T^{-(\delta-1)/2}\| \*D_T^{-1}(\+\beta^0 - \widehat{\+\beta}_0)\| ) .
\end{align}
The order of $T^{-\delta/2}\| \*K_{12} \|$ is the same, which can be shown using the above steps.

We move on to $\*K_{13}$, whose order is given by
\begin{align}
& T^{-\delta/2}\|  \*K_{13} \| \le \left\|\frac{1}{NT^{(\nu_2+\delta)/2}} \sum_{i=1}^N(\*F_{+2}^0 \+\gamma_{2,i}^0 +  \*F_{1}^{0} \+\gamma_{1,i}^0 - \widehat{\*F}_1 \widehat{\+\gamma}_{1,i}) \+\gamma_{2,i}^{0\prime}\*F_2^{0\prime} \right\|  T^{-\delta/2}\|\widehat{\*F}_2\|\notag\\
&\le O(1)N^{-1}T^{-(\nu_2+\delta)/2} \| \*F_{+2}^0\+\Gamma_{+2}^{0\prime}\+\Gamma_{2}^0 \*F_2^{0\prime} \| \notag\\
& +O(1)\left\| \frac{1}{NT^{(\nu_2+\delta)/2}} \sum_{i=1}^N (\*F_{1}^{0\prime} \+\gamma_{1,i}^0 - \widehat{\*F}_1 \widehat{\+\gamma}_{1,i})  \+\gamma_{2,i}^{0\prime}\*F_2^{0\prime}  \right\| \notag\\
&=  O_P( N^{-(1-p)}T^{-(\delta-\nu_3)/2})+\left(\frac{1}{NT^\delta} \sum_{i=1}^N \|\*F_{1}^{0} \+\gamma_{1,i}^0 - \widehat{\*F}_1 \widehat{\+\gamma}_{1,i}\|^2 \right)^{1/2}\left(\frac{1}{NT^{\nu_2}} \sum_{i=1}^N \|\*F_2^0 \+\gamma_{2,i}^0\|^2 \right)^{1/2} \notag\\
&= O_P( N^{-1/2}T^{-(\delta -1)/2}) + O_P( T^{-\delta/2}) + O_P( T^{q-(\delta+\nu_1)/2})   + O_P( N^{-(1-p)} T^{-(\delta-\nu_2)/2} ),
\end{align}
where the second equality follows from Lemma \ref{applem:f} and Assumption \ref{ass:mom}. The order of $T^{-\delta/2}\| \*K_{14} \|$ is the same.

For $\*K_{15}$,
\begin{align}
& T^{-\delta/2}\| \*K_{15} \| = T^{-\delta/2}\left\|\frac{1}{NT^{(\nu_2+\delta)/2}} \sum_{i=1}^N \*U_i \*U_i'\widehat{\*F}_2 \right\|\notag\\
&\le O_P(1) \frac{1}{NT^{(\nu_2+\delta)/2}}\sum_{i=1}^N \| \*F_{1}^{0} \+\gamma_{1,i}^0 - \widehat{\*F}_1 \widehat{\+\gamma}_{1,i}\|^2 +O_P(1) \frac{1}{NT^{(\nu_2+\delta)/2}}\sum_{i=1}^N \| \*F_{+2}^0 \+\gamma_{+2,i}^0\|^2 \notag\\
&= O_P (  T^{-(\nu_2+\delta)/2}) +O_P( N^{-1}T^{1-(\nu_2+\delta)/2}) + O_P( T^{2q-\nu_1-(\nu_2+\delta)/2}) \notag\\
&+ O_P( N^{-2(1-p)}T^{-(\delta-\nu_2)/2})  + O_P( T^{\nu_3- (\nu_2+\delta)/2}),
\end{align}
where the second equality follows from Lemma \ref{applem:f}.

We now insert the above results for $\*K_1,\ldots,\*K_{15}$ into \eqref{eq:f2exp}. But first we left multiply by $T^{-\nu_2}\*F_2^{0\prime}$ and $T^{-(\nu_2+\delta)/2}\widehat{\*F}_2$  respectively. It then gives
\begin{align}
&\| T^{-(\nu_2+\delta)/2}\*F_2^{0\prime}\widehat{\*F}_2 (T^{-\nu_2}\*V_{2}) - (T^{-\nu_2}\*F_2^{0\prime}\*F_2^0)(N^{-1}\+\Gamma_{2}^{0\prime} \+\Gamma_{2}^0 )( T^{-(\nu_2+\delta)/2} \*F_2^{0\prime} \widehat{\*F}_2) \|\notag\\
&= O_P( N^{-1/2}T^{-(\nu_2-1)/2})  + O_P( T^{-\nu_2/2})  +  O_P( T^{-(\nu_2-\nu_3)})\notag\\
&  + O_P( T^{q-(\nu_1+\nu_2)/2})+ O_P( N^{-(1-p)})
\end{align}
and
\begin{align}
&\|T^{-\nu_2}\*V_{2} - ( T^{-(\nu_2+\delta)/2}\widehat{\*F}_2' \*F_2^{0} ) (N^{-1}\+\Gamma_{2}^{0\prime} \+\Gamma_{2}^0 )( T^{-(\nu_2+\delta)/2} \*F_2^{0\prime} \widehat{\*F}_2) \|\notag\\
&= O_P( N^{-1/2}T^{-(\nu_2-1)/2})  + O_P( T^{-\nu_2/2})  +  O_P( T^{-(\nu_2-\nu_3)})\notag\\
&  + O_P( T^{q-(\nu_1+\nu_2)/2})+ O_P( N^{-(1-p)}) .
\end{align}

The rest of the proofs of (a) and (b) follows from the same arguments used in Lemma \ref{applem:lam1}. It is therefore omitted.\hspace*{\fill}{$\blacksquare$}

\bigskip

Before continuing onto the proof of Lemma \ref{applem:beta}, we note that the rates given in Lemma \ref{applem:lam2} can actually be improved upon. Suppose that $d_2$ in known. Consider the next term which is a part of $\*K_{15}$:
\begin{align}\label{IPrate}
\lefteqn{ N^{-1}T^{-(\nu_2/2 +\delta})  \| \*F_{+2}^0 \+\Gamma_{+2}^{0\prime} \+\Gamma_{+2}^{0} \*F_{+2}^{0\prime}\widehat{\*F}_2\| } \notag\\
&=N^{-1}T^{-(\nu_2/2 +\delta}) \| \*F_{+2}^0 \+\Gamma_{+2}^{0\prime} \+\Gamma_{+2}^{0} \*F_{+2}^{0\prime}(\widehat{\*F}_2\*V_2^0-\*F_{2}^0 \*H_2\*V_2^0+  \*F_{2}^0 \*H_2\*V_2^0)(\*V_2^0)^{-1}\| \notag\\
&\le N^{-1}T^{-(\nu_2/2 +\delta}) \| \*F_{+2}^0 \+\Gamma_{+2}^{0\prime} \+\Gamma_{+2}^{0} \*F_{+2}^{0\prime}(\widehat{\*F}_2\*V_2^0- \*F_{2}^0 \*H_2 \*V_2^0 )(\*V_2^0)^{-1}\| \notag\\
&+N^{-1}T^{-(\nu_2/2 +\delta}) \| \*F_{+2}^0 \+\Gamma_{+2}^{0\prime} \+\Gamma_{+2}^{0} \*F_{+2}^{0\prime} \*F_{2}^0 \*H_2 \| \notag\\
&\le O_P(1) T^{ \nu_3- \nu_2 } \cdot T^{-(\nu_2/2+\delta)} \|  \widehat{\*F}_2\*V_2^0- \*F_{2}^0 \*H_2 \*V_2^0  \|
+O_P(1) T^{\nu_3/2 -(\nu_2 +\delta/2)}   \| \*F_{+2}^{0\prime} \*F_{2}^0  \|,
\end{align}
where $\*V_2^0$ and $\*H_2$ are defined in Appendix \ref{appsect:not}. Simple algebra shows that the term $T^{-(\nu_2-\nu_3)}$ in $\tau_{NT}$ of Lemma \ref{applem:lam2} can be dropped.

\bigskip

\noindent \textbf{Proof of Lemma \ref{applem:beta}.}

\bigskip

\noindent Consider (a). Let us assume without loss of generality that $\+\beta^0 = \*0_{d_x\times 1}$, as in \cite{Bai}. It then follows that
\begin{align}
\lefteqn{  (NT)^{-1}[\mathrm{SSR}(\widehat{\+\beta}_1, \widehat{\*F})  - \mathrm{SSR}(\widehat{\+\beta}_0, \widehat{\*F})] }\nonumber\\
&= \frac{1}{NT} \sum_{i=1}^N (- \*X_i \widehat{\+\beta}_1 + \*F^{0}\+\gamma_{i}^0 )'\*M_{\widehat{F}}( -\*X_i  \widehat{\+\beta}_1 + \*F^{0}\+\gamma_{i}^0 ) \nonumber\\
& - \widehat{\+\beta}_1'\frac{2}{NT}\sum_{i=1}^N \*X_i'\*M_{\widehat{F}} \+\varepsilon_i   +\frac{2}{NT} \sum_{i=1}^N\+\gamma_{i}^{0\prime} \*F^{0\prime}\*M_{\widehat{F}} \+\varepsilon_i  +\frac{1}{NT}\sum_{i=1}^N\+\varepsilon_i' \*M_{\widehat{F}}\+\varepsilon_i \nonumber\\
&  -\frac{1}{NT} \sum_{i=1}^N (- \*X_i \widehat{\+\beta}_0 + \*F^0\+\gamma_{i}^0 )'\*M_{\widehat{F}}(- \*X_i  \widehat{\+\beta}_0 + \*F^0\+\gamma_{i}^0 ) \nonumber\\
&+ \widehat{\+\beta}_0' \frac{2}{NT}\sum_{i=1}^N \*X_i'\*M_{\widehat{F}} \+\varepsilon_i   - \frac{2}{NT} \sum_{i=1}^N\+\gamma_{i}^{0\prime} \*F^{0\prime}\*M_{\widehat{F}} \+\varepsilon_i  - \frac{1}{NT}\sum_{i=1}^N\+\varepsilon_i' \*M_{\widehat{F}}\+\varepsilon_i \nonumber\\
&\ge  \frac{1}{NT} \sum_{i=1}^N( \widehat{\+\beta}_1-\widehat{\+\beta}_0 ) '\*X_i'\*M_{\widehat{F}}\*X_i ( \widehat{\+\beta}_1-\widehat{\+\beta}_0 )\notag \\
&-2\left(\frac{1}{NT} \sum_{i=1}^N (\widehat{\+\beta}_1-\widehat{\+\beta}_0) '\*X_i'\*M_{\widehat{F}}\*X_i   (\widehat{\+\beta}_1-\widehat{\+\beta}_0)\right)^{1/2}\left(\frac{1}{NT} \sum_{i=1}^N  (\widehat{\+\beta}_0 - \+\beta^0)  '\*X_i'\*M_{\widehat{F}}\*X_i  (\widehat{\+\beta}_0 - \+\beta^0)  \right)^{1/2}\nonumber\\
& - \frac{2}{NT} \sum_{i=1}^N (\widehat{\+\beta}_1 - \widehat{\+\beta}_0) '\*X_i'\*M_{\widehat{F}}\*F^0 \+\gamma_{i}^0-(\widehat{\+\beta}_1- \widehat{\+\beta}_0)' \frac{2}{NT}\sum_{i=1}^N \*X_i'\*M_{\widehat{F}} \+\varepsilon_i.
\end{align}
Note that by expanding the term $\*M_{\widehat{F}}\*F^0$ as in the proof for the second result of this lemma, we can obtain that
\begin{align}
\lefteqn{ \left\|\frac{1}{NT} \sum_{i=1}^N (\widehat{\+\beta}_1 - \widehat{\+\beta}_0) '\*X_i'\*M_{\widehat{F}}\*F^0 \+\gamma_{i}^0\right\| }\notag\\
& \le \|\*D_T^{-1}(\widehat{\+\beta}_1 - \widehat{\+\beta}_0)\|\left\|\frac{1}{NT} \sum_{i=1}^N \*D_T\*X_i'\*M_{\widehat{F}}\*F^0 \+\gamma_{i}^0\right\| \notag\\
& = O_P(\|\*D_T^{-1}(\widehat{\+\beta}_1 - \widehat{\+\beta}_0)\|)[O_P((NT)^{-1/2}) + o_P( \|\*D_T^{-1} ( \widehat{\+\beta}_0-\+\beta^0)\| )].
\end{align}
It follows that
\begin{align}
0 & \ge (NT)^{-1}[\mathrm{SSR}(\widehat{\+\beta}_1, \widehat{\*F})  - \mathrm{SSR}(\widehat{\+\beta}_0, \widehat{\*F})] \nonumber\\
&= \frac{1}{NT} \sum_{i=1}^N( \widehat{\+\beta}_1-\widehat{\+\beta}_0 ) '\*X_i'\*M_{\widehat{F}}\*X_i ( \widehat{\+\beta}_1-\widehat{\+\beta}_0 )\notag\\
&+O_P(\|\*D_T^{-1}(\widehat{\+\beta}_1 - \widehat{\+\beta}_0)\|)[O_P((NT)^{-1/2})  + o_P(\|\*D_T^{-1} ( \widehat{\+\beta}_0-\+\beta^0)\| )] \notag\\
&\ge \lambda_{min}\left(\frac{1}{NT} \sum_{i=1}^N\*D_T \*X_i'\*M_{\widehat{F}}\*X_i \*D_T \right) \| \*D_T^{-1}( \widehat{\+\beta}_1-\widehat{\+\beta}_0 )\|^2  \notag\\
& + O_P(\|\*D_T^{-1}(\widehat{\+\beta}_1 - \widehat{\+\beta}_0)\|)[O_P((NT)^{-1/2}) + o_P( \|\*D_T^{-1} ( \widehat{\+\beta}_0-\+\beta^0)\| )] ,
\end{align}
where the first term on the right is quadratic in $\|\*D_T^{-1}( \widehat{\+\beta}_1-\widehat{\+\beta}_0 )\|$. Hence, to ensure the right-hand side is non-positive, $\|\*D_T^{-1}( \widehat{\+\beta}_1-\widehat{\+\beta}_0 )\|$ cannot converge to zero at a rate faster than $(NT)^{-1/2} \vee \|\*D_T^{-1} ( \widehat{\+\beta}_0-\+\beta^0)\|$. It follows that
\begin{align}
\|\*D_T^{-1}( \widehat{\+\beta}_1-\widehat{\+\beta}_0 )\| = O_P((NT)^{-1/2}\vee \|\*D_T^{-1} ( \widehat{\+\beta}_0-\+\beta^0)\| ),
\end{align}
as was to be shown.

\medskip

We now turn to (b).  Note that
\begin{align}
\widehat{\+\beta}_1-\+\beta^0 &= \*D_T\left( \frac{1}{NT} \sum_{i=1}^N  \*D_T \*X_i'\*M_{\widehat{F}}\*X_i \*D_T \right)^{-1} \nonumber\\
& \times \left(  \frac{1}{NT} \sum_{i=1}^N \*D_T\*X_i'\*M_{\widehat{F}}\*F^0\+\gamma_{i}^0  + \frac{1}{NT}\sum_{i=1}^N\*D_T \*X_i'\*M_{\widehat{F}}\+\varepsilon_i \right) \notag\\
&= \*D_T \*B^{-1} \left(  \*L + \frac{1}{NT}\sum_{i=1}^N\*D_T \*X_i'\*M_{\widehat{F}}\+\varepsilon_i \right),
\end{align}
with obvious definitions of $\*L$ and $\*B$.

Consider $\*L$. Let $\*Q_{g} = (N^{-1}\+\Gamma_g^{0\prime}\+\Gamma_g^0)(T^{-(\nu_g+\delta)/2} \*F_{g}^{0\prime}\widehat{\*F}_g )$, such that $\*H_{g}^{-1} = T^{-(\nu_g+\delta)/2} \*V_{g}^0\*Q_{g}^{-1}$. We also introduce $\*e_{g,i}$, which is defined to be zero for $g=1$ and $\*e_{g,i}= \sum_{j=1}^{g-1} (\*F_{j}^0\gamma_{j,i}^0 - \widehat{\*F}_j \widehat{\+\gamma}_{j,i})$ for $g=2,\ldots,G$. In this notation,
\begin{align}
\*L & = \frac{1}{NT} \sum_{i=1}^N \*D_T\*X_i'\*M_{\widehat{F}}\*F^0\+\gamma_{i}^0 = -\frac{1}{NT}\sum_{g=1}^G  \sum_{i=1}^N \*D_T\*X_i'\*M_{\widehat{F}}(\widehat{\*F}_g\*H_g^{-1} -\*F_{g}^0)\+\gamma_{g,i}^0 \notag\\
&=-(\*L_{1} + \cdots + \*L_{15}),
\end{align}
where
\begin{align*}
\*L_{1} &= \frac{1}{NT}\sum_{g=1}^G  \sum_{i=1}^N \*D_T\*X_i'\*M_{\widehat{F}} \frac{1}{NT^{(\nu_g+\delta)/2}} \sum_{j=1}^N \*X_j (\+\beta^0 - \widehat{\+\beta}_0) (\+\beta^0 - \widehat{\+\beta}_0)' \*X_j' \widehat{\*F}_g \*Q_{g}^{-1}\+\gamma_{g,i}^0,\\
\*L_{2} &= \frac{1}{NT}\sum_{g=1}^G  \sum_{i=1}^N \*D_T\*X_i'\*M_{\widehat{F}}\frac{1}{NT^{(\nu_g+\delta)/2}} \sum_{j=1}^N \*X_j (\+\beta^0 - \widehat{\+\beta}_0) \+\gamma_{g,j}^{0\prime}\*F_{g}^{0\prime}\widehat{\*F}_g \*Q_{g}^{-1}\+\gamma_{g,i}^0,\\
\*L_{3} &= \frac{1}{NT}\sum_{g=1}^G  \sum_{i=1}^N \*D_T\*X_i'\*M_{\widehat{F}}\frac{1}{NT^{(\nu_g+\delta)/2}} \sum_{j=1}^N \*F_{g}^0 \+\gamma_{g,j}^0  (\+\beta^0 - \widehat{\+\beta}_0)' \*X_j' \widehat{\*F}_g \*Q_{g}^{-1}\+\gamma_{g,i}^0,\\
\*L_{4} &= \frac{1}{NT}\sum_{g=1}^G  \sum_{i=1}^N \*D_T\*X_i'\*M_{\widehat{F}}\frac{1}{NT^{(\nu_g+\delta)/2}} \sum_{j=1}^N \*X_j (\+\beta^0 - \widehat{\+\beta}_0) \*e_{g,j}'\widehat{\*F}_g \*Q_{g}^{-1}\+\gamma_{g,i}^0,\\
\*L_{5} &= \frac{1}{NT}\sum_{g=1}^G  \sum_{i=1}^N \*D_T\*X_i'\*M_{\widehat{F}}\frac{1}{NT^{(\nu_g+\delta)/2}}  \sum_{j=1}^N \*e_{g,j} (\+\beta^0 - \widehat{\+\beta}_0)' \*X_j' \widehat{\*F}_g \*Q_{g}^{-1}\+\gamma_{g,i}^0,\\
\*L_{6} &= \frac{1}{NT}\sum_{g=1}^G  \sum_{i=1}^N \*D_T\*X_i'\*M_{\widehat{F}}\frac{1}{NT^{(\nu_g+\delta)/2}} \sum_{j=1}^N \*X_j (\+\beta^0 - \widehat{\+\beta}_0) \+\varepsilon_j' \widehat{\*F}_g \*Q_{g}^{-1}\+\gamma_{g,i}^0,\\
\*L_{7} &= \frac{1}{NT}\sum_{g=1}^G  \sum_{i=1}^N \*D_T\*X_i'\*M_{\widehat{F}} \frac{1}{NT^{(\nu_g+\delta)/2}}\sum_{j=1}^N \+\varepsilon_j (\+\beta^0 - \widehat{\+\beta}_0)' \*X_j' \widehat{\*F}_g \*Q_{g}^{-1}\+\gamma_{g,i}^0,\\
\*L_{8} &= \frac{1}{NT}\sum_{g=1}^G  \sum_{i=1}^N \*D_T\*X_i'\*M_{\widehat{F}}\frac{1}{NT^{(\nu_g+\delta)/2}} \sum_{j=1}^N \+\varepsilon_j \+\varepsilon_j' \widehat{\*F}_g \*Q_{g}^{-1}\+\gamma_{g,i}^0,\\
\*L_{9} &= \frac{1}{NT}\sum_{g=1}^G  \sum_{i=1}^N \*D_T\*X_i'\*M_{\widehat{F}}\frac{1}{NT^{(\nu_g+\delta)/2}} \sum_{j=1}^N  \*F_{g}^0 \+\gamma_{g,j}^0 \+\varepsilon_j' \widehat{\*F}_g \*Q_{g}^{-1}\+\gamma_{g,i}^0, \\
\*L_{10} &= \frac{1}{NT}\sum_{g=1}^G  \sum_{i=1}^N \*D_T\*X_i'\*M_{\widehat{F}}\frac{1}{NT^{(\nu_g+\delta)/2}} \sum_{j=1}^N \+\varepsilon_j \+\gamma_{g,j}^{0\prime}\*F_{g}^{0\prime} \widehat{\*F}_g \*Q_{g}^{-1}\+\gamma_{g,i}^0,\\
\*L_{11} &= \frac{1}{NT}\sum_{g=1}^G  \sum_{i=1}^N \*D_T\*X_i'\*M_{\widehat{F}}\frac{1}{NT^{(\nu_g+\delta)/2}} \sum_{j=1}^N \+\varepsilon_j \*e_{g,j}'\widehat{\*F}_g \*Q_{g}^{-1}\+\gamma_{g,i}^0, \\
\*L_{12} &= \frac{1}{NT}\sum_{g=1}^G  \sum_{i=1}^N \*D_T\*X_i'\*M_{\widehat{F}}\frac{1}{NT^{(\nu_g+\delta)/2}}  \sum_{j=1}^N \*e_{g,j} \+\varepsilon_j' \widehat{\*F}_g \*Q_{g}^{-1}\+\gamma_{g,i}^0,\\
\*L_{13} &= \frac{1}{NT}\sum_{g=1}^G  \sum_{i=1}^N \*D_T\*X_i'\*M_{\widehat{F}}\frac{1}{NT^{(\nu_g+\delta)/2}} \sum_{j=1}^N \*e_{g,j}\+\gamma_{g,j}^{0\prime}\*F_{g}^{0\prime}\widehat{\*F}_g \*Q_{g}^{-1}\+\gamma_{g,i}^0,\\
\*L_{14} &= \frac{1}{NT}\sum_{g=1}^G  \sum_{i=1}^N \*D_T\*X_i'\*M_{\widehat{F}}\frac{1}{NT^{(\nu_g+\delta)/2}} \sum_{j=1}^N  \*F_{g}^0 \+\gamma_{g,j}^0\*e_{g,j}'\widehat{\*F}_g \*Q_{g}^{-1}\+\gamma_{g,i}^0 ,\\
\*L_{15} &= \frac{1}{NT}\sum_{g=1}^G  \sum_{i=1}^N \*D_T\*X_i'\*M_{\widehat{F}} \frac{1}{NT^{(\nu_g+\delta)/2}} \sum_{j=1}^N  \*e_{g,j}\*e_{g,j}'\widehat{\*F}_g \*Q_{g}^{-1}\+\gamma_{g,i}^0.
\end{align*}
We now evaluate each of these terms. From the analysis of $\*L_{2}$ below, it is easy to show that $\|\*L_{1} \| = o_P(\|\*D_T^{-1}( \widehat{\+\beta}_0 - \+\beta^0)\|)$. We therefore start from $\*L_{2}$, which we write as
\begin{align}
\*L_{2}& = \frac{1}{NT}\sum_{g=1}^G \sum_{i=1}^N \*D_T\*X_i'\*M_{\widehat{F}}  \frac{1}{NT^{(\nu_g+\delta)/2}}\sum_{j=1}^N \*X_j\*D_T\*D_T^{-1}(\+\beta^0-\widehat{\+\beta}_0)(\*F_{g}^0\+\gamma_{g,j}^0)'\widehat{\*F}_g\*Q_{g}^{-1} \+\gamma_{g,i}^0 \nonumber  \\
&= \frac{1}{N^2T} \sum_{g=1}^G\sum_{i=1}^N\sum_{j=1}^N \*D_T \*X_i'\*M_{\widehat{F}} \*X_j\*D_T \+\gamma_{g,j}^{0\prime} (
T^{-(\nu_g+\delta)/2}\*F_{g}^{0\prime}\widehat{\*F}_g)(T^{-(\nu_g+\delta)/2}\*F_{g}^{0\prime}\widehat{\*F}_g)^{-1} \notag\\
&\times (N^{-1}\+\Gamma_g^{0\prime}\+\Gamma_g^0)^{-1} \+\gamma_{g,i}^0 \*D_T^{-1} (\+\beta^0-\widehat{\+\beta}_0)\nonumber \\
&=\frac{1}{NT}  \sum_{g=1}^G\sum_{i=1}^N\sum_{j=1}^N \*D_T\*X_i'\*M_{\widehat{F}}  \*X_j \*D_T \+\gamma_{g,j}^{0\prime} (\+\Gamma_g^{0\prime} \+\Gamma_g^0)^{-1} \+\gamma_{g,i}^0  \*D_T^{-1} (\+\beta^0-\widehat{\+\beta}_0)\nonumber \\
&=\frac{1}{NT} \sum_{i=1}^N\sum_{j=1}^N \*D_T\*X_i'\*M_{\widehat{F}}  \*X_j \*D_T a_{ji} \*D_T^{-1} (\+\beta^0-\widehat{\+\beta}_0).
\end{align}
We will use this expression for $\*L_{2}$ later.

Let us now move on to $\*L_{3}$.
\begin{align}
\*L_{3} &=  \frac{1}{NT}\sum_{g=1}^G \sum_{i=1}^N \*D_T\*X_i'\*M_{\widehat{F}} \frac{1}{NT^{(\nu_g+\delta)/2}}\sum_{j=1}^N \*F_{g}^0\+\gamma_{g,j}^{0}(\+\beta^0-\widehat{\+\beta}_0)' \*X_j'\widehat{\*F}_g \*Q_{g}^{-1} \+\gamma_{g,i}^0 \notag\\
&= \frac{1}{NT} \sum_{i=1}^N  \*D_T \*X_i'\*M_{\widehat{F}} \sum_{g=1}^G ( \*F_{g}^0- \widehat{\*F}_g \*H_g^{-1}) \notag\\
& \times \frac{1}{NT^{(\nu_g+\delta)/2}}\sum_{j=1}^N \+\gamma_{g,j}^{0}(\+\beta^0-\widehat{\+\beta}_0)' \*X_j'\widehat{\*F}_g \*Q_{g}^{-1} \+\gamma_{g,i}^0 ,
\end{align}
where, by using arguments that are similar to those used in the proofs of Lemmas \ref{applem:lam1} and \ref{applem:lam2},
\begin{align}
\lefteqn{ \frac{1}{NT}\sum_{i=1}^N \left\| \sum_{g=1}^G ( \*F_{g}^0- \widehat{\*F}_g \*H_g^{-1}) \frac{1}{NT^{(\nu_g+\delta)/2}}\sum_{j=1}^N \+\gamma_{g,j}^{0}(\+\beta^0-\widehat{\+\beta}_0)' \*X_j'\widehat{\*F}_g \*Q_{g}^{-1} \+\gamma_{g,i}^0  \right\|^2 }\notag\\
& \le  \frac{1}{NT}\sum_{i=1}^N\left( \sum_{g=1}^G\|\*F_{g}^0- \widehat{\*F}_g \*H_g^{-1} \| \frac{1}{NT^{(\nu_g+\delta)/2}}\sum_{j=1}^N \| \+\gamma_{g,j}^{0} (\+\beta^0-\widehat{\+\beta}_0)' \*D_T^{-1}\*D_T\*X_j'\widehat{\*F}_g \|   \| \*Q_{g}^{-1} \+\gamma_{g,i}^0\|\right)^2 \notag\\
&\le O_P(1) \sum_{g=1}^GT^{-\nu_g}\|\*F_{g}^0- \widehat{\*F}_g \*H_g^{-1} \| ^2 \|\*D_T^{-1}(\+\beta^0-\widehat{\+\beta}_0) \|^2 \notag\\
&=o_P( \|\*D_T^{-1}(\+\beta^0-\widehat{\+\beta}_0) \|^2)
\end{align}
implying
\begin{align}
\|\*L_{3}\| &\le \left(  \frac{1}{NT} \sum_{i=1}^N  \|\*D_T \*X_i'\*M_{\widehat{F}}\|^2 \right)^{1/2} \notag\\
& \times \left(\frac{1}{NT} \sum_{i=1}^N \left\|\sum_{g=1}^G ( \*F_{g}^0- \widehat{\*F}_g \*H_g^{-1})\frac{1}{NT^{(\nu_g+\delta)/2}}\sum_{j=1}^N \+\gamma_{g,j}^{0}(\+\beta^0-\widehat{\+\beta}_0)' \*X_j'\widehat{\*F}_g \*Q_{g}^{-1} \+\gamma_{g,i}^0\right\|^2\right)^{1/2} \notag\\
&=o_P( \|\*D_T^{-1}(\+\beta^0-\widehat{\+\beta}_0) \|).
\end{align}
The same steps can be used to show that $\*L_{4}$ and $\*L_{5}$ are of the same order.

For $\*L_{6}$, write
\begin{align}
\*L_{6} & =\frac{1}{NT} \sum_{g=1}^G \sum_{i=1}^N \*D_T\*X_i'\*M_{\widehat{F}}  \frac{1}{NT^{(\nu_g+\delta)/2}}\sum_{j=1}^N\*X_j(\+\beta^0-\widehat{\+\beta}_0)\+\varepsilon_j' \widehat{\*F}_g\*Q_{g}^{-1}  \+\gamma_{g,i}^0 \notag\\
&\le  \sum_{g=1}^G \frac{1}{N^2T^{(\nu_g+\delta)/2+1}} \sum_{i=1}^N \sum_{j=1}^N \*D_T\*X_i'\*M_{\widehat{F}} \*X_j(\+\beta^0-\widehat{\+\beta}_0)\+\varepsilon_j' \*F_{g}^0  \*H_g\*Q_{g}^{-1}  \+\gamma_{g,i}^0 \notag\\
& + \sum_{g=1}^G \frac{1}{N^2T^{(\nu_g+\delta)/2+1}}  \sum_{i=1}^N \sum_{j=1}^N \*D_T\*X_i'\*M_{\widehat{F}} \*X_j(\+\beta^0-\widehat{\+\beta}_0) \+\varepsilon_j'(\widehat{\*F}_g-\*F_{g}^0  \*H_g) \*Q_{g}^{-1}  \+\gamma_{g,i}^0,
\end{align}
where the first term on the right is bounded by
\begin{align}
&  \sum_{g=1}^G \frac{1}{N^2T^{(\nu_g+\delta)/2+1}} \sum_{i=1}^N \sum_{j=1}^N \|\*D_T\*X_i'\*M_{\widehat{F}} \*X_j\*D_T \+\varepsilon_j' \*F_{g}^0  \*H_g\*Q_{g}^{-1}  \+\gamma_{g,i}^0 \| \|\*D_T^{-1}( \+\beta^0-\widehat{\+\beta}_0) \| \notag\\
&\le O_P(1)\| \*D_T^{-1}(\+\beta^0-\widehat{\+\beta}_0) \| \sum_{g=1}^G \frac{1}{N^2T^{(\nu_g+\delta)/2+1}} \left( \sum_{i=1}^N \| \*D_T\*X_i'\*M_{\widehat{F}} \|^2\right)^{1/2} \left( \sum_{i=1}^N \| \+\gamma_{g,i}^0 \|^2\right)^{1/2} \notag\\
&\times \left(\sum_{j=1}^N\|\*M_{\widehat{F}}\*X_j\*D_T\|^2\right)^{1/2} \left(\sum_{j=1}^N \|\+\varepsilon_j' \*F_{g}^0\|^2\right)^{1/2}\|  \*H_g\| \notag\\
&\le O_P(1)\| \*D_T^{-1}(\+\beta^0-\widehat{\+\beta}_0) \|\sum_{g=1}^G T^{-(\nu_g+\delta)/2-1} O_P(\sqrt{T}) O_P(\sqrt{T}) O_P(T^{\nu_g/2}) O_P( T^{-(\nu_g-\delta)/2}) \notag\\
&= o_P ( \| \*D_T^{-1}(\+\beta^0-\widehat{\+\beta}_0) \|  ).
\end{align}
The second term is of the same order. Thus, $\|\*L_{6} \| =o_P ( \| \*D_T^{-1}(\+\beta^0-\widehat{\+\beta}_0) \|  )$. The same is true for $\|\*L_{7}\|$.

We now examine $\*L_{8}$. Let us define $\+\Sigma_{\varepsilon} = N^{-1}\sum_{i=1}^N \+\Sigma_{\varepsilon,i}$, in which $\+\Sigma_{\varepsilon,i} $ has been defined in Assumption \ref{ass:eps}.  $\*L_{8}$ can then be written as
\begin{align}
\*L_{8} &=  \sum_{g=1}^G \frac{1}{N T^{(\nu_g+\delta)/2 +1}} \sum_{i=1}^N  \*D_T \*X_i'\*M_{\widehat{F}} \+\Sigma_{\varepsilon} \widehat{\*F}_g \*Q_{g}^{-1}  \+\gamma_{g,i}^0  \notag\\
&+ \sum_{g=1}^G  \frac{1}{N^2T^{(\nu_g+\delta)/2+1}} \sum_{i=1}^N  \sum_{j=1}^N \*D_T \*X_i' ( \+\varepsilon_j  \+\varepsilon_j' -\+\Sigma_{\varepsilon,j}) \widehat{\*F}_g \*Q_{g}^{-1}  \+\gamma_{g,i}^0 \notag\\
&- \sum_{g=1}^G  \frac{1}{N^2T^{(\nu_g+\delta)/2+1}} \sum_{i=1}^N  \sum_{j=1}^N \*D_T \*X_i' \*P_{\widehat{F}} ( \+\varepsilon_j  \+\varepsilon_j' -\+\Sigma_{\varepsilon,j}) \widehat{\*F}_g \*Q_{g}^{-1}  \+\gamma_{g,i}^0 .
\end{align}
For the first term on the right,
\begin{align}
\lefteqn{
\sqrt{NT} \left\|\sum_{g=1}^G \frac{1}{N T^{(\nu_g+\delta)/2 +1}} \sum_{i=1}^N  \*D_T \*X_i'\*M_{\widehat{F}} \+\Sigma_{\varepsilon} \widehat{\*F}_g \*Q_{g}^{-1}  \+\gamma_{g,i}^0\right\| }\notag\\
&\le \sum_{g=1}^G  \frac{1}{\sqrt{N} T^{(\nu_g+\delta+1)/2}}  \sum_{i=1}^N \| \*D_T \*X_i'\*M_{\widehat{F}}  \+\Sigma_{\varepsilon,j} \widehat{\*F}_g \*Q_{g}^{-1}  \+\gamma_{g,i}^0  \| \notag \\
&\le O_P(1)\sum_{g=1}^G N^{-1/2} T^{-(\nu_g+\delta+1)/2} O_P(NT) O_P(T^{(\delta-1)/2}) \notag \\
& \le O_P(1)\sum_{g=1}^G N^{1/2}T^{-\nu_g/2}  =O_P ( \sqrt{N}T^{-\nu_G/2} ) =O_P(1),
\end{align}
where the last equality holds by $NT^{-\nu_G}<\infty $. Let
\begin{align}
\overline{\*A}_1 = \frac{1}{NT} \sum_{i=1}^N  \*D_T \*X_i'\*M_{F_0} \+\Sigma_{\varepsilon} T^{-(\nu_G-1)/2}\*F_{G}^{0} (T^{-\nu_G} \*F_{G}^{0\prime}\*F_{G}^{0} )^{-1} (N^{-1} \+\Gamma_G^{0\prime}\+\Gamma_G^0 )^{-1} \+\gamma_{G,i}^0 .
\end{align}
Note how $\*A_{1} = \plim_{N,T\to\infty}E(\overline{\*A}_1 | \mathcal{C})$. In this notation,
\begin{align}
\lefteqn{ \sqrt{NT}\sum_{g=1}^G \frac{1}{N T^{(\nu_g+\delta)/2 +1}} \sum_{i=1}^N  \*D_T \*X_i'\*M_{\widehat{F}} \+\Sigma_{\varepsilon} \widehat{\*F}_g \*Q_{g}^{-1}  \+\gamma_{g,i}^0 } \notag\\
& =\sqrt{NT}\sum_{g=1}^G \frac{1}{N T^{(\nu_g+\delta)/2 +1}} \sum_{i=1}^N  \*D_T \*X_i'\*M_{F_0} \+\Sigma_{\varepsilon} \widehat{\*F}_g ( T^{-(\nu_g+\delta)/2} \*F_{g}^{0\prime}\widehat{\*F}_g )^{-1}  (N^{-1} \+\Gamma_g^{0\prime}\+\Gamma_g^0 )^{-1} \+\gamma_{g,i}^0 (1+o_P(1)) \notag\\
& =\sum_{g=1}^G \sqrt{\frac{N}{T^{\nu_g}}}  \frac{1}{N T } \sum_{i=1}^N  \*D_T \*X_i'\*M_{F_0} \+\Sigma_{\varepsilon} T^{-(\nu_g-1)/2} \*F_{g}^{0} (T^{-\nu_g} \*F_{g}^{0\prime}\*F_{g}^{0} )^{-1} (N^{-1} \+\Gamma_g^{0\prime}\+\Gamma_g^0 )^{-1} \+\gamma_{g,i}^0 (1+o_P(1)) \notag\\
& = \sqrt{\frac{N}{T^{\nu_G}}} \frac{1}{NT} \sum_{i=1}^N  \*D_T \*X_i'\*M_{F_0} \+\Sigma_{\varepsilon} T^{-(\nu_G-1)/2}\*F_{G}^{0} (T^{-\nu_G} \*F_{G}^{0\prime}\*F_{G}^{0} )^{-1} (N^{-1} \+\Gamma_G^{0\prime}\+\Gamma_G^0 )^{-1} \+\gamma_{G,i}^0 (1+o_P(1))\notag\\
& = \sqrt{N}T^{-\nu_G/2} \overline{\*A}_1(1+o_P(1)),
\end{align}
where the second equality follows from arguments similar to those used in \eqref{eqbp3}. Note also that $\lim \sqrt{N}T^{-\nu_G/2} <\infty$. Further use of the same steps used by \citet{JYGH} establishes that the second and third terms of $\*L_{8}$ are $o_P(\|\*D_T^{-1}(\widehat{\+\beta}_0 -\+\beta^0) \|)+ o_P((NT)^{-1/2})$. Hence,
\begin{align}
\*L_{8} = T^{-(\nu_G+1)/2}\overline{\*A}_1 +o_P(\|\*D_T^{-1}(\widehat{\+\beta}_0 -\+\beta^0) \|)+ o_P((NT)^{-1/2}).
\end{align}

$\*L_{9}$ can be written as
\begin{align}
\*L_{9} &= \sum_{g=1}^G  \frac{1}{N^2T^{(\nu_g+\delta)/2 +1}}\sum_{i=1}^N \sum_{j=1}^N  \*D_T\*X_i'\*M_{\widehat{F}}  (\*F_{g}^0 -\widehat{\*F}_g \*H_g^{-1}) \+\gamma_{g,j}^{0} \+\varepsilon_j' \*F_{g}^0\*H_g \*Q_{g}^{-1}  \+\gamma_{g,i}^0 \notag\\
&+  \sum_{g=1}^G \frac{1}{N^2T^{(\nu_g+\delta)/2 +1}}\sum_{i=1}^N  \sum_{j=1}^N \*D_T\*X_i'\*M_{\widehat{F}} (\*F_{g}^0 -\widehat{\*F}_g \*H_g^{-1}) \+\gamma_{g,j}^{0} \+\varepsilon_j' (\widehat{\*F}_g-\*F_{g}^0\*H_g) \*Q_{g}^{-1}  \+\gamma_{g,i}^0 ,
\end{align}
where the first term on the right-hand side is bounded by
\begin{align}
& \sum_{g=1}^G \frac{1}{N^2T^{(\nu_g+\delta)/2+1}} \sum_{i=1}^N  \| \*D_T\*X_i'\*M_{\widehat{F}}\| \| \*F_{g}^0 -\widehat{\*F}_g \*H_g^{-1} \|  \|T^{-(\nu_g-1)/2} \+\Gamma_g^{0\prime}\+\varepsilon \*F_{g}^0 \| \notag\\
& \times T^{(\nu_g-1)/2}\|\*H_g \| \| \*Q_{g}^{-1} \+\gamma_{g,i}^0\| \notag\\
&\le O_P(1) \sum_{g=1}^G N^{-1} T^{-(\nu_g+\delta)/2-1} O_P( \sqrt{T} ) \| \*F_{g}^0 -\widehat{\*F}_g \*H_g^{-1} \|  O_P(\sqrt{NT}) O_P(T^{ (\nu_g-1)/2}) O_P(T^{-(\nu_g-\delta)/2}) \notag \\
&\le O_P(1) \sum_{g=1}^G (NT)^{-1/2} T^{-\nu_g/2}  \| \*F_{g}^0 -\widehat{\*F}_g \*H_g^{-1} \| \notag\\
& = o_P( (NT)^{-1/2} ),
\end{align}
where the first inequality follows from $\|T^{-(\nu_g-1)/2} \+\Gamma_g^{0\prime}\+\varepsilon \*F_{g}^0 \|=O_P(\sqrt{NT})$. The second term on the right-hand side of $\*L_{9}$ is also $o_P( (NT)^{-1/2} )$. We therefore conclude that $\|\*L_{9}\| = o_P( (NT)^{-1/2} )$.

$\*L_{10}$ can be written more compactly as
\begin{align}
\*L_{10} &= \frac{1}{NT}\sum_{g=1}^G  \sum_{i=1}^N \*D_T\*X_i'\*M_{\widehat{F}}\frac{1}{NT^{(\nu_g+\delta)/2}} \sum_{j=1}^N \+\varepsilon_j \+\gamma_{g,j}^{0\prime}\*F_{g}^{0\prime} \widehat{\*F}_g \*Q_{g}^{-1}\+\gamma_{g,i}^0 \notag\\
&= \frac{1}{NT}\sum_{i=1}^N \sum_{j=1}^N\*D_T\*X_i'\*M_{\widehat{F}} \+\varepsilon_j a_{ij} ,
\end{align}
which we will again make use of later.

Let us move on to $\*L_{11}$. We begin by rewriting $\*e_{g,j}$ as
\begin{align}
\*e_{g,j} &= \sum_{d=1}^{g-1} [ \*F_{d}^0\+\gamma_{d,j}^0 - \widehat{\*F}_d T^{-\delta} \widehat{\*F}_d ' (\*y_j -\*X_j \widehat{\+\beta}_0 -\widehat{\*F}_d \widehat{\+\gamma}_{d,j} ) ] \nonumber \\
&= \sum_{d=1}^{g-1}[-\*P_{\widehat{F}_d } \*X_j(\+\beta^0-\widehat{\+\beta}_0) + \*M_{\widehat{F}_d } \*F_{d}^0\+\gamma_{d,j}^0 - \*P_{\widehat{F}_d } \*F_{+d}^0 \+\gamma_{+d,j}^0 - \*P_{\widehat{F}_d } \*e_{d,j} - \*P_{\widehat{F}_d } \+\varepsilon_j],
\end{align}
where $\*F_{+d}^0 = (\*F_{d+1}^0,\ldots, \*F_{G}^0)$ and $\+\gamma_{+d,j}^0 = (\+\gamma_{d+1,j}^{0\prime},\ldots, \+\gamma_{G,j}^{0\prime})'$, as in Appendix \ref{appsect:not}. By inserting this into $\*L_{11}$, we get
\begin{align}
\*L_{11}&= \sum_{g=1}^G \frac{1}{NT} \sum_{i=1}^N \*D_T\*X_i'\*M_{\widehat{F}} \frac{1}{NT^{(\nu_g+\delta)/2}} \sum_{j=1}^N \+\varepsilon_j  \sum_{d=1}^{g-1}[-\*P_{\widehat{F}_d } \*X_j(\+\beta^0-\widehat{\+\beta}_0) \notag\\
& + \*M_{\widehat{F}_d } \*F_{d}^0\+\gamma_{d,j}^0 - \*P_{\widehat{F}_d } \*F_{+d}^0 \+\gamma_{+d,j}^0 - \*P_{\widehat{F}_d } \*e_{d,j} - \*P_{\widehat{F}_d } \+\varepsilon_j]' \widehat{\*F}_g  \*Q_{g}^{-1}  \+\gamma_{g,i}^0.
\end{align}
Hence, $\*L_{11}$ can be written as a sum of five terms. There is no need to study the forth term, the one due to $\*P_{\widehat{F}_d } \*e_{d,j}$, as we can keep expanding $\*e_{d,j}$ until we cannot. The first and fifth terms are $o_P(\|\*D_T^{-1}(\widehat{\+\beta}_0 -\+\beta^0) \|)$ and $o_P((NT)^{-1/2})$, respectively, by the same arguments used for evaluating $\*L_{7}$ and $\*L_{8}$. Moreover, the steps used for evaluating $\*L_{9}$ can be used to show that
\begin{align}
\lefteqn{ \left\|\sum_{g=1}^G \frac{1}{NT} \sum_{i=1}^N \*D_T\*X_i'\*M_{\widehat{F}} \frac{1}{NT^{(\nu_g+\delta)/2}} \sum_{j=1}^N \+\varepsilon_j \sum_{d =1}^{g-1}\+\gamma_{d,j}^{0\prime}\+F_{d}^{0\prime}\*M_{\widehat{F}_d }  \widehat{\*F}_g  \*Q_{g}^{-1}  \+\gamma_{g,i}^0\right\| }\notag\\
&= \left\|\sum_{g=1}^G \frac{1}{NT} \sum_{i=1}^N \*D_T\*X_i'\*M_{\widehat{F}} \frac{1}{NT^{(\nu_g+\delta)/2}} \sum_{d =1}^{g-1} \+\varepsilon' \+\Gamma_{d}^0 \*F_{d}^{0\prime}\*M_{\widehat{F}_d }  \widehat{\*F}_g  \*Q_{g}^{-1}  \+\gamma_{g,i}^0\right\| \notag\\
&= \left\|\sum_{g=1}^G \frac{1}{N^2T^{(\nu_g+\delta)/2+1}}  \sum_{i=1}^N  \sum_{d =1}^{g-1} \*D_T\*X_i' \+\varepsilon' \+\Gamma_{d}^0 (\*F_{d}^0-\widehat{\*F}_d \*H_{d}^{-1})'\*M_{\widehat{F}_d }  \widehat{\*F}_g  \*Q_{g}^{-1}  \+\gamma_{g,i}^0\right\| \notag\\
&+ \left\|\sum_{g=1}^G \frac{1}{N^2T^{(\nu_g+\delta)/2+1}}  \sum_{i=1}^N \sum_{d =1}^{g-1} \*D_T\*X_i'\*P_{\widehat{F}} \+\varepsilon' \+\Gamma_{d}^0(\*F_{d}^0-\widehat{\*F}_d \*H_{d}^{-1})'\*M_{\widehat{F}_d }  \widehat{\*F}_g  \*Q_{g}^{-1}  \+\gamma_{g,i}^0\right\|\notag\\
& = o_P( (NT)^{-1/2} ).
\end{align}
The second term in $\*L_{11}$ is therefore negligible. It remains to consider the third term, which is
\begin{align}
\lefteqn{ \left\|\sum_{g=1}^G \frac{1}{NT} \sum_{i=1}^N \*D_T\*X_i'\*M_{\widehat{F}} \frac{1}{NT^{(\nu_g+\delta)/2}} \sum_{j=1}^N \+\varepsilon_j \sum_{d =1}^{g-1}\+\gamma_{+d,j}^{0\prime} \*F_{+d}^{0\prime}\*P_{\widehat{F}_d }  \widehat{\*F}_g  \*Q_{g}^{-1}  \+\gamma_{g,i}^0\right\| } \notag\\
&= \left\|\sum_{g=1}^G \frac{1}{NT} \sum_{i=1}^N \*D_T\*X_i'\*M_{\widehat{F}} \frac{1}{NT^{(\nu_g+\delta)/2}}  \sum_{d =1}^{g-1}  \+\varepsilon'\+\Gamma_{+d}^{0} \*F_{+d}^{0\prime} \*P_{\widehat{F}_d }  \widehat{\*F}_g  \*Q_{g}^{-1}  \+\gamma_{g,i}^0\right\| \notag\\
&= o_P(\| \*L_{10}\| ) = o_P( (NT)^{-1/2} )
\end{align}
where we have used Assumption \ref{ass:ortho}. Therefore, $\|\*L_{11}\| = o_P( (NT)^{-1/2} )$.

For $\*L_{12}$,
\begin{align}
\*L_{12}&= \sum_{g=1}^G \frac{1}{NT} \sum_{i=1}^N \*D_T\*X_i'\*M_{\widehat{F}} \frac{1}{NT^{(\nu_g+\delta)/2}} \sum_{j=1}^N\*e_{g,j} \+\varepsilon_j '\widehat{\*F}_g  \*Q_{g}^{-1}  \+\gamma_{g,i}^0 \notag\\
&=\sum_{g=1}^G \frac{1}{NT} \sum_{i=1}^N \*D_T\*X_i'\*M_{\widehat{F}} \frac{1}{NT^{(\nu_g+\delta)/2}} \sum_{j=1}^N \*F_{g}^0\+\gamma_{i}^0 \+\varepsilon_j '\widehat{\*F}_g  \*Q_{g}^{-1}  \+\gamma_{g,i}^0= \*L_{9},
\end{align}
where the second equality follows from the construction of $\*e_{g,j}$. Hence, since $\*L_{9}$ is negligible, $\*L_{12}$ is also negligible.

Next up is $\*L_{13}$, whose order can be worked out in the following way:
\begin{align}
\|\*L_{13}\|  &\le  \sum_{g=1}^G \frac{1}{N^2T} \sum_{i=1}^N \left\| \*D_T\*X_i' \*M_{\widehat{F}}   \sum_{d =1}^{g-1} \*F_{d}^0\+\Gamma_{d}^{0\prime}  \+\Gamma_g^0(N^{-1}\+\Gamma_g^{0\prime}\+\Gamma_g^0)^{-1}\+\gamma_{g,i}^0\right\| \notag\\
&\le \sum_{g=1}^G \frac{1}{N^2T} \sum_{i=1}^N \left\| \*D_T\*X_i' \*M_{\widehat{F}}   \sum_{d =1}^{g-1} (\*F_{d}^0 -\widehat{\*F}_d \*H_{d}^{-1} ) \+\Gamma_{d}^{0\prime} \+\Gamma_g^0(N^{-1}\+\Gamma_g^{0\prime}\+\Gamma_g^0)^{-1} \+\gamma_{g,i}^0\right\| \notag\\
&\le  O_P(1) \sum_{g=1}^G  \frac{1}{N^2T }  \sum_{d =1}^{g-1}  \sum_{i=1}^N  \| \*D_T\*X_i' \*M_{\widehat{F}} \| \| \*F_{d}^0 -\widehat{\*F}_d \*H_{d}^{-1} \| \| \+\Gamma_{d}^{0\prime} \+\Gamma_g^0\| \| \+\gamma_{g,i}^0 \| \notag\\
&\le  O_P(1) \sum_{g=1}^G  \frac{1}{N T}  \sum_{d =1}^{g-1} O_P(\sqrt{T})\| \*F_{d}^0 -\widehat{\*F}_d \*H_{d}^{-1} \|  O_P( N^p) \notag\\
&\le  O_P(1) \sum_{g=1}^G  \frac{1}{N^{1-p}\sqrt{T}}  \sum_{d =1}^{g-1} \| \*F_{d}^0 -\widehat{\*F}_d \*H_{d}^{-1} \| = o_P((NT)^{-1/2}),
\end{align}
where the first inequality follows from the construction of $\*e_{g,j}$, and the last equality follows by going through a development similar to the first result of Lemma \ref{applem:lam2} and further using a development similar to \eqref{IPrate}. The same arguments show that $\|\*L_{14}\|$ and $\|\*L_{15}\|$ are negligible, too.

We now put everything together. This yields
\begin{align}
\*L & = -(\*L_{1} + \cdots + \*L_{15}) \notag \\
&= -\*L_{2} - \*L_8 - \*L_{10}  + o_P(\|\*D_T^{-1}(\widehat{\+\beta}_0 -\+\beta^0) \|)+ o_P((NT)^{-1/2}) \notag\\
& = \left( - \frac{1}{NT} \sum_{i=1}^N\sum_{j=1}^N \*D_T\*X_i'\*M_{\widehat{F}}  \*X_j \*D_T a_{ji} + o_P(1) \right)\*D_T^{-1} (\+\beta^0-\widehat{\+\beta}_0) - T^{(\nu_G+1)/2}\overline{\*A}_1 \notag\\
& - \frac{1}{NT} \sum_{i=1}^N \sum_{j=1}^N\*D_T\*X_i'\*M_{\widehat{F}}a_{ji} \+\varepsilon_j  + o_P((NT)^{-1/2}),
\end{align}
which in turn implies
\begin{align}
\*D_T^{-1}(\widehat{\+\beta}_1-\+\beta^0) &= \*B^{-1} \left( \*L + \frac{1}{NT}\sum_{i=1}^N\*D_T \*X_i'\*M_{\widehat{F}}\+\varepsilon_i \right) \notag\\
&= (\*B^{-1}\*N + o_P(1))\*D_T^{-1} (\widehat{\+\beta}_0- \+\beta^0)+ \*B^{-1} \frac{1}{NT}\sum_{i=1}^N\*D_T \*Z_i(\widehat{\*F})'\+\varepsilon_i\notag \\
&- T^{-(\nu_G+1)/2}\*B^{-1}\overline{\*A}_1 + o_P((NT)^{-1/2}),
\end{align}
where $\*Z_i(\*F) $ is defined in Assumption \ref{ass:id}, and
\begin{align*}
\*N = \frac{1}{NT} \sum_{i=1}^N\sum_{j=1}^N \*D_T\*X_i'\*M_{\widehat{F}}  \*X_j \*D_T a_{ij}.
\end{align*}
This expression for $\*D_T^{-1}(\widehat{\+\beta}_1-\+\beta^0)$ can be inserted into $\*D_T^{-1}(\widehat{\+\beta}_0 -\+\beta^0)$, giving
\begin{align}
\*D_T^{-1}(\widehat{\+\beta}_0 -\+\beta^0) &=\*D_T^{-1}(\widehat{\+\beta}_1-\+\beta^0) - \*D_T^{-1}( \widehat{\+\beta}_1-\widehat{\+\beta}_0) \notag\\
&= (\*B^{-1}\*N + o_P(1))\*D_T^{-1} (\widehat{\+\beta}_0 - \+\beta^0) + \*B^{-1} \frac{1}{NT}\sum_{i=1}^N\*D_T \*Z_i(\widehat{\*F})'\+\varepsilon_i\notag\\
& - \*D_T^{-1} ( \widehat{\+\beta}_1-\widehat{\+\beta}_0) -T^{-(\nu_G+1)/2}\*B^{-1}\overline{\*A}_1 + o_P((NT)^{-1/2}) ,
\end{align}
which can be solved for $\*D_T^{-1}(\widehat{\+\beta}_0 -\+\beta^0)$
\begin{align}
 \*D_T^{-1} (\widehat{\+\beta}_0 - \+\beta^0) &= \*B(\widehat{\*F})^{-1}\left( \frac{1}{NT}\sum_{i=1}^N\*D_T \*Z_i(\widehat{\*F})'\+\varepsilon_i - \*B\*D_T^{-1} ( \widehat{\+\beta}_1-\widehat{\+\beta}_0) -T^{-(\nu_G+1)/2} \overline{\*A}_1\right) \notag\\
& + o_P(\|\*D_T^{-1} ( \widehat{\+\beta}_1-\widehat{\+\beta}_0)\|)  + o_P((NT)^{-1/2}).
\end{align}
Also, making use of Assumption \ref{ass:eps}, it is not difficult to show that $(NT)^{-1}\sum_{i=1}^N\*D_T \*Z_i(\widehat{\*F})'\+\varepsilon_i=O_P((NT)^{-1/2})$. Hence,
\begin{align}\label{eq:hatbeta0exp}
\lefteqn{ \*D_T^{-1} (\widehat{\+\beta}_0 - \+\beta^0) + \*B(\widehat{\*F})^{-1}\*B\*D_T^{-1} ( \widehat{\+\beta}_1-\widehat{\+\beta}_0) }\notag\\
& = \*B(\widehat{\*F})^{-1} \frac{1}{NT}\sum_{i=1}^N\*D_T \*Z_i(\widehat{\*F})'\+\varepsilon_i -T^{-(\nu_G+1)/2}\*B(\widehat{\*F})^{-1}\overline{\*A}_1\notag \\
& + o_P(\|\*D_T^{-1} ( \widehat{\+\beta}_1-\widehat{\+\beta}_0)\|) + o_P((NT)^{-1/2}).
\end{align}
By using the fact that
\begin{align}
\*B(\widehat{\*F})^{-1}\*B\*D_T^{-1} = \*D_T^{-1} \left(  \sum_{i=1}^N \*Z_i(\widehat{\*F})'\*Z_i(\widehat{\*F}) \right)^{-1} \sum_{i=1}^N  \*X_i'\*M_{\widehat{F}} \*X_i ,
\end{align}
the left-hand side of this last equation can be written as
\begin{align}
\lefteqn{ \*D_T^{-1} (\widehat{\+\beta}_0 - \+\beta^0) + \*B(\widehat{\*F})^{-1}\*B\*D_T^{-1} ( \widehat{\+\beta}_1-\widehat{\+\beta}_0) }\notag\\
& = \*D_T^{-1} \left[\widehat{\+\beta}_0 +  \left( \sum_{i=1}^N \*Z_i(\widehat{\*F})'\*Z_i(\widehat{\*F}) \right)^{-1} \sum_{i=1}^N  \*X_i'\*M_{\widehat{F}} \*X_i ( \widehat{\+\beta}_1-\widehat{\+\beta}_0) - \+\beta^0 \right] .
\end{align}
It follows that
\begin{align}\label{eq:hatbetaexp1}
\lefteqn{ \sqrt{NT}\*D_T^{-1} \left[\widehat{\+\beta}_0 +  \left( \sum_{i=1}^N \*Z_i(\widehat{\*F})'\*Z_i(\widehat{\*F}) \right)^{-1} \sum_{i=1}^N  \*X_i'\*M_{\widehat{F}} \*X_i ( \widehat{\+\beta}_1-\widehat{\+\beta}_0) - \+\beta^0 \right] }\notag\\
& = \*B(\widehat{\*F})^{-1} \frac{1}{\sqrt{NT}}\sum_{i=1}^N\*D_T \*Z_i(\widehat{\*F})'\+\varepsilon_i - \sqrt{N}T^{-\nu_G/2}  \*B(\widehat{\*F})^{-1}\overline{\*A}_1  \notag\\
& + o_P( \sqrt{NT}\|\*D_T^{-1} ( \widehat{\+\beta}_1-\widehat{\+\beta}_0)\| ) + o_P(1) .
\end{align}
Consider the $o_P( \sqrt{NT}\|\*D_T^{-1} ( \widehat{\+\beta}_1-\widehat{\+\beta}_0)\| )$ reminder term. By the first result of this lemma, $\|\*D_T^{-1}( \widehat{\+\beta}_1-\widehat{\+\beta}_0 )\| = O_P((NT)^{-1/2}\vee \|\*D_T^{-1} ( \widehat{\+\beta}_0-\+\beta^0)\| )$. This can be inserted into \eqref{eq:hatbeta0exp}, giving
\begin{align}
\*D_T^{-1} (\widehat{\+\beta}_0 - \+\beta^0) & = \*B(\widehat{\*F})^{-1} \frac{1}{NT}\sum_{i=1}^N\*D_T \*Z_i(\widehat{\*F})'\+\varepsilon_i + o_P((NT)^{-1/2}\vee \|\*D_T^{-1} ( \widehat{\+\beta}_0-\+\beta^0)\| ) \notag\\
& = O_P((NT)^{-1/2}) + o_P(\|\*D_T^{-1} ( \widehat{\+\beta}_0-\+\beta^0)\| ),
\end{align}
which in turn implies
\begin{align}
\|\*D_T^{-1} ( \widehat{\+\beta}_0-\+\beta^0)\| = O_P((NT)^{-1/2}).
\end{align}
The second result then follows.\hspace*{\fill}{$\blacksquare$}

\section{Proofs of main results}\label{appsect:proof}

\noindent \textbf{Proof of Lemma \ref{lem:beta0hat}.}

\bigskip

\noindent Without loss of generality, we assume that $\+\beta^0 = \*0_{d_x\times 1}$. This implies
\begin{align} \label{eq:ssrexp}
\lefteqn{ (NT)^{-1}[\mathrm{SSR}( \+\beta, \*F) - \mathrm{SSR}(\+\beta^0, \*F^0)] } \nonumber\\
&= \frac{1}{NT} \sum_{i=1}^N ( \*X_i  \+\beta + \*F^0\+\gamma_{i}^0 )'\*M_{F}( \*X_i  \+\beta + \*F^0\+\gamma_{i}^0 ) \nonumber\\
& + \+\beta' \frac{2}{NT}\sum_{i=1}^N \*X_i'\*M_{F} \+\varepsilon_i   + \frac{2}{NT} \sum_{i=1}^N\+\gamma_{i}^{0\prime} \*F^{0\prime}\*M_{F} \+\varepsilon_i  +\frac{1}{NT}\sum_{i=1}^N\+\varepsilon_i' (\*P_{F^0}-\*P_F) \+\varepsilon_i \nonumber\\
&= \frac{1}{NT} \sum_{i=1}^N ( \*X_i \*D_T\*D_T^{-1} \+\beta + \*F^0\+\gamma_{i}^0 )'\*M_{F}( \*X_i \*D_T\*D_T^{-1}\+\beta + \*F^0\+\gamma_{i}^0 ) \nonumber\\
& + \+\beta' \*D_T^{-1}\frac{2}{NT}\sum_{i=1}^N \*D_T \*X_i'\*M_{F} \+\varepsilon_i + \frac{2}{NT} \sum_{i=1}^N\+\gamma_{i}^{0\prime} \*F^{0\prime}\*M_{F} \+\varepsilon_i +O_P(N^{-1}\vee T^{-1})\nonumber \\
&= \+\beta' \*D_T^{-1} \*B(\*F) \*D_T^{-1} \+\beta+ \+\theta'\*B\+\theta + \+\beta'\*D_T^{-1}\frac{2}{NT}\sum_{i=1}^N \*D_T\*X_i'\*M_{F} \+\varepsilon_i  \nonumber \\
& + \frac{2}{NT} \sum_{i=1}^N \+\gamma_{i}^{0\prime}\*F^{0\prime} \*M_{F} \+\varepsilon_i+O_P(N^{-1}\vee T^{-1}),
\end{align}
where the second equality follows from Lemma \ref{applem:mom}, and $\+\theta = \+\eta + \*B^{-1}\*C\+\beta$ and $\*B = N^{-1}\+\Gamma^{0\prime}\+\Gamma^0 \otimes \*I_T$ are defined as on page 1265 of \cite{Bai}. Note that for $\+\beta\in \mathbb{R}^{d_x}$, we may have
\begin{eqnarray}
\sup_{\+\beta\in \mathbb{R}^{d_x}, \*F\in  \mathbb{D}_F}\left| \+\beta' \*D_T^{-1}\frac{2}{NT}\sum_{i=1}^N\*D_T \*X_i'\*M_{F} \+\varepsilon_i \right|\ne o_P(1).
\end{eqnarray}
In our proof of consistency, we consider two cases; (i) $\| \*D_T^{-1}\+\beta\| \le C$ and (ii) $\| \*D_T^{-1}\+\beta\| > C$, where $C$ is a large positive constant. Under (i),
\begin{eqnarray}
\sup_{\| \*D_T^{-1}\+\beta \|\le C, \*F\in  \mathbb{D}_F}\left| \+\beta' \*D_T^{-1}\frac{2}{NT}\sum_{i=1}^N\*D_T \*X_i'\*M_{F} \+\varepsilon_i \right|= O_P( N^{-1/2}\vee T^{-1/2} )
\end{eqnarray}
by Lemma \ref{applem:mom}. The expression given in \eqref{eq:ssrexp} for $(NT)^{-1}[\mathrm{SSR}( \+\beta, F) - \mathrm{SSR}(\+\beta^0, \*F^0)]$ therefore reduces to
\begin{align}\label{eq:rhsi}
(NT)^{-1}[\mathrm{SSR}( \+\beta, \*F) - \mathrm{SSR}(\+\beta^0, \*F^0)] & = \+\beta'\*D_T^{-1}\*B(\*F)\*D_T^{-1}\+\beta + \+\theta' \*B\+\theta +  \frac{2}{NT} \sum_{i=1}^N\+\gamma_{i}^{0\prime} \*F^{0\prime}\*M_{F} \+\varepsilon_i \nonumber\\
&  + O_P( N^{-1/2}\vee T^{-1/2} ),
\end{align}
where $\+\beta'\*D_T^{-1}\*B(\*F)\*D_T^{-1}\+\beta$ does not involve $\*F^0$ and $\sum_{i=1}^N\+\gamma_{i}^{0\prime} \*F^{0\prime} \*M_{F} \+\varepsilon_i$ is independent of $\+\beta$. Hence, provided $d_{max}  \ge d_f$, the consistency of $\*D_T^{-1}\widehat{\+\beta}_0$ in case (i) follows from the same arguments as in \cite{Bai}.

Under (ii), \eqref{eq:ssrexp} can be written as
\begin{align}\label{eq:rhsii}
\lefteqn{ (NT)^{-1}[\mathrm{SSR}( \+\beta, \*F) - \mathrm{SSR}(\+\beta^0, \*F^0)] } \nonumber\\
& = \+\beta'\*D_T^{-1}\*B(\*F)\*D_T^{-1}\+\beta+ \+\theta'\*B\+\theta +\+\beta'\*D_T^{-1}\frac{2}{NT}\sum_{i=1}^N \*D_T\*X_i'\*M_{F} \+\varepsilon_i + \frac{2}{NT} \sum_{i=1}^N \+\gamma_{i}^{0\prime} \*F^{0\prime}\*M_{F} \+\varepsilon_i\nonumber \\
& + O_P( N^{-1/2}\vee T^{-1/2} ) \nonumber \\
&\ge c_0 \| \*D_T^{-1}\+\beta\|^2 + \+\beta'\*D_T^{-1}\frac{2}{NT}\sum_{i=1}^N \*D_T\*X_i'\*M_{F} \+\varepsilon_i  + \+\theta' \*B\+\theta + \frac{2}{NT} \sum_{i=1}^N\+\gamma_{i}^{0\prime}\*F^{0\prime} \*M_{F} \+\varepsilon_i \notag\\
& + O_P( N^{-1/2}\vee T^{-1/2} ) \nonumber \\
&\ge \frac{c_0}{2} C^2  + \+\theta' \*B\+\theta + \frac{2}{NT} \sum_{i=1}^N\+\gamma_{i}^{0\prime}\*F^{0\prime}\*M_{F} \+\varepsilon_i + O_P( N^{-1/2}\vee T^{-1/2} ),
\end{align}
where $c_0$ is defined in Assumption \ref{ass:id}, and the second inequality follows from the fact that the quadratic term dominates the linear one for large values of $C$. Hence, $(NT)^{-1}[\mathrm{SSR}( \+\beta, \*F) - \mathrm{SSR}(\+\beta^0, \*F^0)] > 0$, but from the definition of $\widehat{\+\beta}_0$ we also know that $\mathrm{SSR}( \widehat{\+\beta}_0, \widehat{\*F}) - \mathrm{SSR}(\+\beta^0, \*F^0) \le 0$, which means that $\*D_T^{-1}\widehat{\+\beta}_0$ cannot belong to (ii).

Note that since $\|(NT)^{-1}\sum_{i=1}^N \*D_T\*X_i'\*M_{F} \+\varepsilon_i \|=O_P( N^{-1/2}\vee T^{-1/2} )$ by Lemma \ref{applem:mom}, all we need is $C = C^0 (N^{-1/2}\vee T^{-1/2})$ for some large constant $C^0$ in order to ensure that the last inequality of \eqref{eq:rhsii} holds. This implies $\*D_T^{-1}(\widehat{\+\beta}_0-\+\beta^0) = O_P( N^{-1/2}\vee T^{-1/2} )$, so the proof is complete. \hspace*{\fill}{$\blacksquare$}

\bigskip

\noindent \textbf{Proof of Lemma \ref{lem:d1hat}.}

\bigskip

\noindent Suppose first that $d_1=0$, such that $d_f=0$. In this case,
\begin{align}
\frac{1}{NT } \sum_{i=1}^N \|\*y_i -\*X_i\widehat{\+\beta}_0 \|^2  &= \frac{1}{NT } \sum_{i=1}^N \|\*X_i\*D_T \*D_T^{-1}(\+\beta^0-\widehat{\+\beta}_0) +\+\varepsilon_i\|^2 \nonumber\\
&= \frac{1}{NT } \sum_{i=1}^N\| \+\varepsilon_i \|^2 +o_P(1).
\end{align}
This implies $\tau_N \asymp 1/\ln (T\vee N)$, which is much larger than $\widehat{\lambda}_{1,1}/\widehat{\lambda}_{1,0}$. The result then follows immediately.

Suppose now instead that $d_1>0$. Straightforward algebra reveals that
\begin{align}\label{eq:mockexp}
\frac{1}{NT^{\nu_1}} \sum_{i=1}^N \|\*y_i -\*X_i\widehat{\+\beta}_0 \|^2  &= \frac{1}{NT^{\nu_1}} \sum_{i=1}^N \|\*X_i\*D_T \*D_T^{-1}(\+\beta^0-\widehat{\+\beta}_0) + \*F^0\+\gamma_{i}^0 +\+\varepsilon_i\|^2 \nonumber\\
&= \frac{1}{NT^{\nu_1}} \sum_{i=1}^N\| \*F_1^0 \+\gamma_{1,i}^0 \|^2 +o_P(1) \nonumber\\
&=  (\mathrm{vec}\,\+\Sigma_{\Gamma_1^0})' \mathrm{vec}\,\+\Sigma_{F_1^0} +o_P(1),
\end{align}
which together with Assumption \ref{ass:mom} implies $\tau_N \asymp 1/\ln (T\vee N)$. Note that for $d=1,\ldots, d_1$,
\begin{align}\label{eqbp5}
\frac{T^{\nu_1}}{\lambda_{1,d}} &= \frac{T^{2\nu_1}}{ \*h_{1,d}^{0\prime}\*F_{1}^{0\prime}\+\Sigma_{1}^0\*F_{1}^{0}\*h_{1,d}^0}= \frac{NT^{2\nu_1}}{ \*h_{1,d}^{0\prime}\*F_{1}^{0\prime} \*F_{1}^{0} \+\Gamma_{1}^{0\prime} \+\Gamma_{1}^0\*F_{1}^{0\prime}\*F_{1}^{0}\*h_{1,d}^0}\notag\\
&=\frac{1}{ ( T^{-\nu_1}\*h_{1,d}^{0\prime} \*F_{1}^{0\prime}\*F_{1}^{0})(N^{-1} \+\Gamma_{1}^{0\prime} \+\Gamma_{1}^0)  (T^{-\nu_1} \*F_{1}^{0\prime}\*F_{1}^{0} \*h_{1,d}^0 ) }\notag\\
&=\frac{1}{ (T^{-(\nu_1+\delta)/2} \widehat{\*F}_{1,d}^{0\prime}\*F_{1}^{0} )(N^{-1} \+\Gamma_{1}^{0\prime} \+\Gamma_{1}^0) (T^{-(\nu_1+\delta)/2} \*F_{1}^{0\prime}\widehat{\*F}_{1,d}^{0}) } (1+o_P(1)) \asymp 1,
\end{align}
where the fourth equality follows from \eqref{eq:fcons} and the last step is dues to \eqref{eqbp2} and Assumption \ref{ass:mom}. Using Lemma \ref{applem:lam1}, \eqref{eq:mockexp} and \eqref{eqbp5}, we obtain that
\begin{align}
\frac{\widehat{\lambda}_{1,1}}{\widehat{\lambda}_{1,0}} &= \frac{T^{-\nu_1}\widehat{\lambda}_{1,1}}{(NT^{\nu_1})^{-1} \sum_{i=1}^N \|\*y_i -\*X_i\widehat{\+\beta}_0 \|^2} \asymp 1, \\ \nonumber \\
\frac{ \widehat{\lambda}_{1,d+1}}{ \widehat{\lambda}_{1,d}} & = \frac{T^{-\nu_1}\widehat{\lambda}_{1,d+1}}{T^{-\nu_1}\widehat{\lambda}_{1,d}}  \asymp 1
\end{align}
for $d=1,\ldots, d_1-1$, and $\frac{\widehat{\lambda}_{1,d_1+1} }{\widehat{\lambda}_{1,d_1} } = O_P( T^{-(\nu_1-\nu_2)})$. Moreover, for $d=d_1+1,\ldots, d_{max}$, $\frac{T^{-\nu_1}\widehat{\lambda}_{1,d}}{T^{-\nu_1}\widehat{\lambda}_{1,0}} = O_P(T^{-(\nu_1-\nu_2)})$, which is less than $\tau_N $ by \eqref{eq:mockexp} and Lemma \ref{applem:lam1}. The required result follows from this and the definition of $\widehat{d}_1$. \hspace*{\fill}{$\blacksquare$}

\bigskip

\noindent \textbf{Proof of Lemma \ref{lem:dghat}.}

\bigskip

\noindent Each step in the sequential procedure of Step \ref{step2} introduces additional remainder terms that all converge to zero under Assumptions \ref{ass:eps} and \ref{ass:ortho} by Lemmas \ref{applem:lam1} and \ref{applem:lam2}. This proves (a). Part (b) follows from the (rotational) consistency of $\widehat{\*F}_g$ established as a part of the proofs of Lemmas  \ref{applem:lam1} and \ref{applem:lam2}.\hspace*{\fill}{$\blacksquare$}

\bigskip

\noindent \textbf{Proof of Theorem \ref{thm:betahat}.}

\bigskip

\noindent By Lemma \ref{applem:beta}, the rate of convergence given in Lemma \ref{lem:beta0hat} is therefore not the best one possible. The improved rate implies that
\begin{align}\label{eqnew}
\|\*D_T^{-1}( \widehat{\+\beta}_1-\widehat{\+\beta}_0 )\| = O_P((NT)^{-1/2}\vee \|\*D_T^{-1} ( \widehat{\+\beta}_0-\+\beta^0)\| ) = O_P((NT)^{-1/2}),
\end{align}
which can be inserted back into \eqref{eq:hatbetaexp1}, leading to
\begin{align}\label{eq:hatbetaexp2}
\lefteqn{ \sqrt{NT}\*D_T^{-1} \left[\widehat{\+\beta}_0 +  \left( \sum_{i=1}^N \*Z_i(\widehat{\*F})'\*Z_i(\widehat{\*F}) \right)^{-1} \sum_{i=1}^N  \*X_i'\*M_{\widehat{F}} \*X_i ( \widehat{\+\beta}_1-\widehat{\+\beta}_0) - \+\beta^0 \right] }\notag\\
& = \*B(\widehat{\*F})^{-1} \frac{1}{\sqrt{NT}}\sum_{i=1}^N\*D_T \*Z_i(\widehat{\*F})'\+\varepsilon_i- \sqrt{N}T^{-\nu_G/2}\*B(\widehat{\*F})^{-1}\overline{\*A}_1  + o_P(1) .
\end{align}
Note that $\*Z_i(\widehat{\*F})$ is $\widehat{\*Z}_i$ with $\widehat a_{ij}$ replaced by $a_{ij}$. We now show that the effect of the estimation of $a_{ij}$ is negligible. We begin by noting how
\begin{align}
\lefteqn{\frac{1}{NT} \sum_{i=1}^N [\*D_T\*Z_i(\widehat{\*F})'\*Z_i(\widehat{\*F})\*D_T - \*D_T\widehat{\*Z}_i'\widehat{\*Z}_i\*D_T ] }\notag\\
& = \frac{1}{NT} \sum_{i=1}^N\sum_{j=1}^N \*D_T\*X_i'\*M_{\widehat{F}}  \*X_j \*D_T\widehat a_{ij} - \frac{1}{NT} \sum_{i=1}^N\sum_{j=1}^N \*D_T\*X_i'\*M_{\widehat{F}}  \*X_j \*D_T a_{ij}.
\end{align}
Here,
\begin{align}
\lefteqn{ \frac{1}{NT} \sum_{i=1}^N\sum_{j=1}^N \*D_T\*X_i'\*M_{\widehat{F}}  \*X_j \*D_T\widehat a_{ij} } \notag\\
& =  \frac{1}{NT} \sum_{i=1}^N\sum_{j=1}^N \*D_T\*X_i'\*M_{\widehat{F}}  \*X_j \*D_T\sum_{g=1}^G  (\*y_j -\*X_j\widehat{\+\beta}_0 -\widehat{\*F}_{-g} \widehat{\+\gamma}_{-g,j})' \widehat{\*F}_g\notag\\
& \times ( \widehat{\*F}_g' \widehat{\+\Sigma}_g\widehat{\*F}_g)^{-1} \widehat{\*F}_g'(\*y_i -\*X_i\widehat{\+\beta}_0 -\widehat{\*F}_{-g} \widehat{\+\gamma}_{-g,i})\notag\\
&= \frac{1}{NT} \sum_{i=1}^N\sum_{j=1}^N \*D_T\*X_i'\*M_{\widehat{F}}  \*X_j \*D_T \sum_{g=1}^G  [ \*X_j(\+\beta^0 -\widehat{\+\beta}_0) +\*F_{g}^0\+\gamma_{g,j}^{0}+\*e_{g,j} +\+\varepsilon_j]'\widehat{\*F}_g\notag\\
& \times  T^{-\delta}\*V_{g}^{-1} \widehat{\*F}_g'[ \*X_j(\+\beta^0 -\widehat{\+\beta}_0)  +\*F_{g}^0\+\gamma_{g,i}^0+\*e_{g,i} +\+\varepsilon_i]\notag\\
&= \frac{1}{NT} \sum_{i=1}^N\sum_{j=1}^N \*D_T\*X_i'\*M_{\widehat{F}}  \*X_j \*D_T \sum_{g=1}^G  [ \*X_j(\+\beta^0 -\widehat{\+\beta}_0) +\*F_{g}^0\+\gamma_{g,j}^{0}+\*e_{g,j} +\+\varepsilon_j ]' \widehat{\*F}_g \notag\\
& \times T^{-(\nu_g+\delta)} (T^{-\nu_g}\*V_{g})^{-1} \widehat{\*F}_g'[ \*X_j(\+\beta^0 -\widehat{\+\beta}_0)  +\*F_{g}^0\+\gamma_{g,i}^0+\*e_{g,i} +\+\varepsilon_i](1+o_P(1)) \notag\\
&= \frac{1}{NT} \sum_{i=1}^N\sum_{j=1}^N \*D_T\*X_i'\*M_{\widehat{F}}  \*X_j \*D_T a_{ij} (1+o_P(1)),
\end{align}
where the second equality here follows from the definition of $\*V_{g}^0$, which is again based on taking $d_g$ as known, while the last equality follows from direct calculation. The effect of the estimation of $a_{ij}$ in $\widehat{\*Z}_i$ is therefore negligible, which in turn implies that the right-hand side of \eqref{eq:hatbetaexp2} becomes
\begin{align}
\lefteqn{ \sqrt{NT}\*D_T^{-1} \left[\widehat{\+\beta}_0 +  \left( \sum_{i=1}^N \*Z_i(\widehat{\*F})'\*Z_i(\widehat{\*F}) \right)^{-1} \sum_{i=1}^N  \*X_i'\*M_{\widehat{F}} \*X_i ( \widehat{\+\beta}_1-\widehat{\+\beta}_0) - \+\beta^0 \right] }\notag\\
& = \sqrt{NT}\*D_T^{-1} \left[\widehat{\+\beta}_0 +  \left( \sum_{i=1}^N \widehat{\*Z}_i'\widehat{\*Z}_i \right)^{-1} \sum_{i=1}^N  \*X_i'\*M_{\widehat{F}} \*X_i ( \widehat{\+\beta}_1-\widehat{\+\beta}_0) - \+\beta^0 \right] + o_P(1) \notag\\
& = \sqrt{NT}\*D_T^{-1} (\widehat{\+\beta} - \+\beta^0 ) + o_P(1).
\end{align}
It follows that
\begin{align}\label{eq:hatbetaexp3}
\sqrt{NT}\*D_T^{-1} (\widehat{\+\beta} - \+\beta^0 ) &= \*B(\widehat{\*F})^{-1} \frac{1}{\sqrt{NT}}\sum_{i=1}^N\*D_T \*Z_i(\widehat{\*F})'\+\varepsilon_i -\sqrt{N}T^{-\nu_G/2}\*B(\widehat{\*F})^{-1}\overline{\*A}_1 \notag\\
& + o_P(1).
\end{align}
Consider $(NT)^{-1/2}\sum_{i=1}^N\*D_T \*Z_i(\widehat{\*F})'\+\varepsilon_i$. From the definition of $ \*Z_i(\*F)$,
\begin{align}
\frac{1}{\sqrt{NT}}\sum_{i=1}^N\*D_T \*Z_i(\widehat{\*F})'\+\varepsilon_i & =\frac{1}{\sqrt{NT}}\sum_{i=1}^N\*D_T \*Z_i(\*F^0)'\+\varepsilon_i  +\frac{1}{\sqrt{NT}}\sum_{i=1}^N\*D_T [ \*Z_i(\widehat{\*F}) - \*Z_i(\*F^0) ]'\+\varepsilon_i\notag\\
& =\frac{1}{\sqrt{NT}}\sum_{i=1}^N\*D_T \*Z_i(\*F^0)'\+\varepsilon_i  - \sqrt{NT}( \*R_1 - \*R_2 ), \label{zdiffe}
\end{align}
where
\begin{align}
\*R_1 & = \frac{1}{NT }\sum_{i=1}^N \*D_T\*X_i'(\*P_{\widehat F} - \*P_{F^0})\+\varepsilon_i , \\
\*R_2 & =\frac{1}{NT }\sum_{i=1}^N \sum_{j=1}^N a_{ij}\*D_T\*X_j'(\*P_{\widehat F} - \*P_{F^0})\+\varepsilon_i. \end{align}
Here,
\begin{align}
\*R_1  &=\frac{1}{NT^{1+\delta} }\sum_{i=1}^N \*D_T\*X_i'(\widehat{\*F}-\*F^0\*H) \*H'\*F^{0\prime}\+\varepsilon_i
+\frac{1}{NT^{1+\delta} }\sum_{i=1}^N  \*D_T\*X_i'(\widehat{\*F}-\*F^0\*H) (\widehat{\*F}-\*F^0\*H)'\+\varepsilon_i \notag\\
&+\frac{1}{NT^{1+\delta} }\sum_{i=1}^N  \*D_T\*X_i'\*F^0\*H (\widehat{\*F}-\*F^0\*H)'\+\varepsilon_i +\frac{1}{NT^{1+\delta} }\sum_{i=1}^N \*D_T\*X_i'\*F^0  [\*H\*H' - T^\delta(\*F^{0\prime}\*F^0)^{-1}] \*F^{0\prime}\+\varepsilon_i\notag\\
&= \*R_{11} + \*R_{12} + \*R_{13} + \*R_{14},
\end{align}
with obvious implicit definitions of $\*R_{11},\ldots,\*R_{14}$ and $\*H  = \mathrm{diag}(\*H_{1}, \ldots, \*H_{G} )$. Let $\*R_{1m,j}$ be the $j$-th row of $\*R_{1m}$ for $m\in\{1,\ldots,4\}$. In this notation,
\begin{align}
\| \*R_{11,j}\| &\le  \left\| \frac{1}{NT^{\delta/2+1} }\sum_{i=1}^N  (\+\varepsilon_i'\*F^0\*H)\otimes (T^{-\kappa_j/2}\*X_{j,i})'  \right\|  \|T^{-\delta/2}\mathrm{vec}(\widehat{\*F}-\*F^0\*H) \| \notag\\
&= O_P ( (NT)^{-1/2} ) T^{-\delta/2}\|\widehat{\*F}-\*F^0\*H \| = o_P ( (NT)^{-1/2} ),
\end{align}
where the equality follows from Lemma \ref{applem:mom} and the fact that $T^{(\nu_g-\delta)/2}\*H_g =O_P(1)$. Similarly, for $\*R_{12}$,
\begin{align}
\| \*R_{12,j} \| &\le \left\| \frac{1}{NT}\sum_{i=1}^N  \+\varepsilon_i'\otimes (T^{-\kappa_j/2}\*X_{j,i})'\right\|  \|T^{-\delta} \mathrm{vec}[(\widehat{\*F}-\*F^0\*H)(\widehat{\*F}-\*F^0\*H)'] \| \notag\\
&= O_P( N^{-1/2})T^{-\delta}\|\widehat{\*F}-\*F^0\*H \|^2 =o_P( (NT)^{-1/2} ),
\end{align}
where the second equality is due to \eqref{eq:fcons2}.

Consider $\*R_{14}$. From $T^{-\delta/2}\|\widehat{\*F}-\*F^0\*H \|=o_P(1)$ and  $\|T^{(\nu_g-\delta)/2}\*H_g\| =O_P(1)$, we obtain
\begin{align}
\|\*I_{d_f} - T^\delta(\*H' \*F^{0\prime}\*F^0 \*H)^{-1} \| & =o_P(1),\\
\|T^{\nu_g -\delta}\*H_g \*H_g' - (T^{-\nu_g } \*F_g^{0\prime}\*F_g^{0})^{-1} \| & =o_P(1) \label{eqbp6}.
\end{align}
Together with Lemma \ref{applem:mom} this implies
\begin{align}
\|\*R_{14,j}\| &= \left\|\frac{1}{NT^{1+\delta} }\sum_{i=1}^N T^{-\kappa_j/2}\*X_{j,i}'\*F^0 [\*H\*H' -T^\delta\*H(\*H' \*F^{0\prime}\*F^0 \*H)^{-1}\*H'] \*F^{0\prime}\+\varepsilon_i\right\| \notag\\
&= \left\|\frac{1}{NT^{1+\delta} }\sum_{i=1}^N T^{-\kappa_j/2}\*X_{j,i}'\*F^0\*H[\*I_{d_f} - T^\delta(\*H' \*F^{0\prime}\*F^0 \*H)^{-1}] \*H'\*F^{0\prime}\+\varepsilon_i\right\|\notag \\
&\le \left\| \frac{1}{NT^{1+\delta}}\sum_{i=1}^N  (\+\varepsilon_i'\*F^0\*H)\otimes (T^{-\kappa_j/2}\*X_{j,i}'\*F^0\*H) \right\|  \|\*I_{d_f} - T^\delta(\*H' \*F^{0\prime}\*F^0 \*H)^{-1} \| \notag\\
&= O_P( (NT)^{-1/2} )  \|\*I_{d_f} - T^\delta(\*H' \*F^{0\prime}\*F^0 \*H)^{-1} \|= o_P( (NT)^{-1/2} ) .
\end{align}

Finally, let us consider $\*R_{13}$.
\begin{align}
\*R_{13} & = \frac{1}{NT^{1+\delta} }\sum_{i=1}^N  \*D_T\*X_i'\*F^0\*H (\widehat{\*F}-\*F^0\*H)'\+\varepsilon_i \notag \\
& = \sum_{g=1}^G\frac{1}{NT^{1+\delta} }\sum_{i=1}^N  \*D_T\*X_i'\*F_g^0\*H_g (\widehat{\*F}_g-\*F_g^0\*H_g)'\+\varepsilon_i \notag \\
& = \sum_{g=1}^G\frac{1}{NT^{1+\delta} }\sum_{i=1}^N  \*D_T\*X_i'\*F_g^0 \*H_g \*H_g' (\widehat{\*F}_g \*H_g^{-1}-\*F_g^0 )'\+\varepsilon_i
\end{align}
Expanding $\widehat{\*F}_g \*H_g^{-1}-\*F_g^0 $ as we did for $\*L$ above, we then just need to focus on the leading terms equivalent to $\*L_2$, $\*L_{8}$, and $\*L_{10}$.
\begin{align}
\*R_{13} & =\sum_{g=1}^G\frac{1}{NT^{1+\delta} }\sum_{i=1}^N  \*D_T\*X_i'\*F_g^0 \*H_g \*H_g' \left(\frac{1}{NT^{(\nu_g+\delta)/2}} \sum_{j=1}^N \*X_j (\+\beta^0 - \widehat{\+\beta}_0) \+\gamma_{g,j}^{0\prime}\*F_{g}^{0\prime}\widehat{\*F}_g \*Q_{g}^{-1}\right)'\+\varepsilon_i \notag \\
&+\sum_{g=1}^G\frac{1}{NT^{1+\delta} }\sum_{i=1}^N  \*D_T\*X_i'\*F_g^0 \*H_g \*H_g' \left( \frac{1}{NT^{(\nu_g+\delta)/2}} \sum_{j=1}^N \+\varepsilon_j \+\gamma_{g,j}^{0\prime}\*F_{g}^{0\prime} \widehat{\*F}_g \*Q_{g}^{-1}\right)'\+\varepsilon_i \notag \\
&+\sum_{g=1}^G\frac{1}{NT^{1+\delta} }\sum_{i=1}^N  \*D_T\*X_i'\*F_g^0 \*H_g \*H_g' \left( \frac{1}{\sqrt{T^{\nu_g+1}}} \+\Sigma_{\varepsilon} \frac{\*F_{g}^{0}}{T^{(\nu_g-1)/2}} \left(\frac{\*F_{g}^{0\prime}\*F_{g}^{0}}{T^{\nu_g}} \right)^{-1}  \left(\frac{\+\Gamma_g^{0\prime}\+\Gamma_g^0}{N}\right)^{-1} \right)'\+\varepsilon_i  \notag\\
&+o_P\left(\frac{1}{\sqrt{NT}}\right)\notag \\
&=\*R_{131}+\*R_{132}+\*R_{133}+o_P\left(\frac{1}{\sqrt{NT}}\right)\notag.
\end{align}
It is easy to see that  $\| \*R_{131}\|=o_P(\|\+\beta^0 -\widehat{\+\beta}_0 \|)$, so negligible. We now examine $\*R_{133}$. Then we can write
\begin{align}
\*R_{133} &= \sum_{g=1}^G\frac{1}{NT^{1+\nu_g} }\sum_{i=1}^N  \*D_T\*X_i'\*F_g^0 \left(\frac{\*F_g^{0\prime}\*F_g^{0}}{T^{\nu_g}}\right)^{-1} \left( \frac{\+\Sigma_{\varepsilon} }{\sqrt{T^{\nu_g+1}}} \frac{\*F_{g}^{0}}{T^{(\nu_g-1)/2}} \left(\frac{\*F_{g}^{0\prime}\*F_{g}^{0}}{T^{\nu_g}} \right)^{-1}  \left(\frac{\+\Gamma_g^{0\prime}\+\Gamma_g^0}{N}\right)^{-1} \right)'\+\varepsilon_i \notag\\
&= \sum_{g=1}^G \frac{1 }{T^{\nu_g+1 }}\cdot\frac{1}{NT }\sum_{i=1}^N  \*D_T\*X_i' \frac{\*F_{g}^{0}}{T^{(\nu_g-1)/2}} \left(\frac{\*F_g^{0\prime}\*F_g^{0}}{T^{\nu_g}}\right)^{-1}   \left(\frac{\+\Gamma_g^{0\prime}\+\Gamma_g^0}{N}\right)^{-1}  \left(\frac{\*F_{g}^{0\prime}\*F_{g}^{0}}{T^{\nu_g}} \right)^{-1} \frac{\*F_{g}^{0\prime}}{T^{(\nu_g-1)/2}} \+\Sigma_{\varepsilon}\+\varepsilon_i \notag\\
&=o_P\left(\frac{1}{\sqrt{NT}}\right),
\end{align}
where the first equality follows from \eqref{eqbp6}, and the last step follows from some routine analysis using Assumption \ref{ass:eps}. Below, we investigate $\*R_{132}$, which is one source of the bias term.
\begin{align}
\*R_{132}  &=\sum_{g=1}^G\frac{1}{NT^{1+\delta} }\sum_{i=1}^N  \*D_T\*X_i'\*F_g^0 \*H_g \*H_g' \left( \frac{1}{NT^{(\nu_g+\delta)/2}} \sum_{j=1}^N \+\varepsilon_j \+\gamma_{g,j}^{0\prime}\*F_{g}^{0\prime} \widehat{\*F}_g \*Q_{g}^{-1}\right)'\+\varepsilon_i\notag\\
&=\sum_{g=1}^G\frac{1}{NT^{1+\nu_g} }\sum_{i=1}^N  \*D_T\*X_i'\*F_g^0 (T^{\nu_g -\delta}\*H_g \*H_g' ) \left( \frac{1}{N} \sum_{j=1}^N \+\varepsilon_j \+\gamma_{g,j}^{0\prime} \left(\frac{\+\Gamma_g^{0\prime}\+\Gamma_g^0}{N}\right)^{-1}\right)'\+\varepsilon_i\notag\\
&=\sum_{g=1}^G\frac{1}{NT^{1+\nu_g} }\sum_{i=1}^N  \*D_T\*X_i'\*F_g^0  \left(\frac{\*F_g^{0\prime} \*F_g^0}{T^{\nu_g }} \right)^{-1} \left( \frac{1}{N} \sum_{j=1}^N \+\varepsilon_j \+\gamma_{g,j}^{0\prime} \left(\frac{\+\Gamma_g^{0\prime}\+\Gamma_g^0}{N}\right)^{-1}\right)'\+\varepsilon_i \cdot (1+o_P(1))\notag \\
&=\sum_{g=1}^G \frac{1}{NT^{ (\nu_g-1)/2}}\cdot\frac{1}{N }\sum_{i=1}^N \sum_{j=1}^N  \frac{\*D_T \*X_i'\*F_g^0}{T^{(\nu_g+1)/2}}  \left(\frac{\*F_g^{0\prime} \*F_g^0}{T^{\nu_g }} \right)^{-1} \left( \frac{\+\Gamma_g^{0\prime}\+\Gamma_g^0}{N}\right)^{-1}  \+\gamma_{g,j}^{0}\frac{ \+\varepsilon_j ' \+\varepsilon_i}{T}\cdot (1+o_P(1)),
\end{align}
where the second equality follows from the definition of $\*Q_g$, and the third equality follows from \eqref{eqbp6}.
Then we are able to further write
\begin{align}
\sqrt{NT}\*R_{32}  & =\sum_{g=1}^G \frac{T}{\sqrt{N}T^{ \nu_g/2}}\cdot\frac{1}{N }\sum_{i=1}^N \sum_{j=1}^N  \frac{\*D_T \*X_i'\*F_g^0 }{T^{(\nu_g+1)/2}}  \left(\frac{\*F_g^{0\prime} \*F_g^0}{T^{\nu_g }} \right)^{-1} \left( \frac{\+\Gamma_g^{0\prime}\+\Gamma_g^0}{N}\right)^{-1}  \+\gamma_{g,j}^{0}\frac{ \+\varepsilon_j ' \+\varepsilon_i}{T}\cdot (1+o_P(1))\notag \\
&= \frac{T}{\sqrt{N}T^{ \nu_G/2}} \cdot\frac{1}{N }\sum_{i=1}^N \sum_{j=1}^N  \frac{\*D_T \*X_i'\*F_G^0 }{T^{(\nu_G+1)/2}}  \left(\frac{\*F_G^{0\prime} \*F_G^0}{T^{\nu_G }} \right)^{-1} \left( \frac{\+\Gamma_G^{0\prime}\+\Gamma_G^0}{N}\right)^{-1}  \+\gamma_{G,j}^{0}\frac{ \+\varepsilon_j ' \+\varepsilon_i}{T}\cdot (1+o_P(1)).
\end{align}

Hence, by adding the results,
\begin{align}
\sqrt{NT} \*R_1 &= \sqrt{NT}(\*R_{11} + \*R_{12} + \*R_{13} + \*R_{14}) \notag\\
& = T^{(2-\nu_G)/2}N^{-1/2} \frac{1}{N }\sum_{i=1}^N \sum_{j=1}^N T^{-(\nu_G+1)/2} \*D_T \*X_i'\*F_G^0 (T^{-\nu_G} \*F_G^{0\prime} \*F_G^0 )^{-1} \notag\\
& \times  (N^{-1} \+\Gamma_G^{0\prime}\+\Gamma_G^0 )^{-1} \+\gamma_{G,j}^{0}T^{-1} \+\varepsilon_j' \+\varepsilon_i + o_P(1).
\end{align}
$\sqrt{NT} \*R_2$ can be evaluated in exactly the same way, and the limiting representation is analogous to the one given above for $\sqrt{NT} \*R_1$ with $\*X_i$ replaced by $-\sum_{j=1}^{N}\*X_ja_{ij}$. Moreover, $\|\*B(\widehat{\*F}) - \*B(\*F^0)\| = o_P(1).$ It follows that if we define
\begin{align}
\overline{\*A}_{2} = \frac{1}{N }\sum_{i=1}^N \sum_{j=1}^N T^{-(\nu_G+1)/2} \*D_T \*Z_i(0)'\*F_G^0 (T^{-\nu_G} \*F_G^{0\prime} \*F_G^0 )^{-1}  (N^{-1} \+\Gamma_G^{0\prime}\+\Gamma_G^0 )^{-1} \+\gamma_{G,j}^{0}T^{-1} \+\varepsilon_j' \+\varepsilon_i ,
\end{align}
where $\*Z_i(0) = \*X_i - \sum_{j=1}^{N}\*X_ja_{ij}$, then \eqref{zdiffe} reduces to
\begin{align}
\frac{1}{\sqrt{NT}}\sum_{i=1}^N\*D_T \*Z_i(\widehat{\*F})'\+\varepsilon_i & =\frac{1}{\sqrt{NT}}\sum_{i=1}^N\*D_T \*Z_i(\*F^0)'\+\varepsilon_i  - T^{(2-\nu_G)/2}N^{-1/2}\overline{\*A}_{2} + o_P(1),
\end{align}
which in turn implies that \eqref{eq:hatbetaexp3} becomes
\begin{align}\label{eq:hatbetaexp4}
\sqrt{NT}\*D_T^{-1} (\widehat{\+\beta} - \+\beta^0 ) & = \*B(\*F^0)^{-1} \frac{1}{\sqrt{NT}}\sum_{i=1}^N\*D_T \*Z_i(\*F^0)'\+\varepsilon_i -\sqrt{N}T^{-\nu_G/2} \*B(\*F^0)^{-1}\overline{\*A}_1 \notag\\
& -T^{(2-\nu_G)/2}N^{-1/2}  \*B(\*F^0)^{-1}\overline{\*A}_{2} + o_P(1) .
\end{align}
The required result is now implied by Assumption \ref{ass:norm}. \hspace*{\fill}{$\blacksquare$}

\bigskip

\noindent \textbf{Proof of Corollary \ref{appcor:betahat}.}

\bigskip

\noindent Under $NT^{-\nu_G}\to 0$, \eqref{eq:hatbetaexp3} in the proof of Theorem \ref{thm:betahat} may be written as
\begin{align*}
\sqrt{NT}\*D_T^{-1} (\widehat{\+\beta} - \+\beta^0 ) = \*B(\*F^0)^{-1} \frac{1}{\sqrt{NT}}\sum_{i=1}^N\*D_T \*Z_i(\widehat{\*F})'\+\varepsilon_i + o_P(1) .
\end{align*}
Consider $(NT)^{-1} \sum_{i=1}^N \*D_T\*X_i'\*M_{\widehat{F}} \+\varepsilon_i$, which we can write as
\begin{align}
\frac{1}{NT}\sum_{i=1}^N \*D_T\*X_i'\*M_{\widehat{F}}\+\varepsilon_i &= \frac{1}{NT }\sum_{i=1}^N \*D_T\*X_i'\*M_{F^0}\+\varepsilon_i - \frac{1}{NT }\sum_{i=1}^N \*D_T\*X_i'(\*P_{\widehat F} - \*P_{F^0})\+\varepsilon_i \notag\\
&= \frac{1}{NT}\sum_{i=1}^N \*D_T\*X_i'\*M_{F^0}\+\varepsilon_i - \*R_1.
\end{align}
As in the proof of Theorem \ref{thm:betahat},
\begin{align}
\*R_1 &= \frac{1}{NT}\sum_{i=1}^N  \*D_T\*X_i'(T^{-\delta}\widehat{\*F} \widehat{\*F}'  - \*P_{F^0} )\+\varepsilon_i \notag\\
&=\frac{1}{NT^{1+\delta} }\sum_{i=1}^N \*D_T\*X_i'(\widehat{\*F}-\*F^0\*H) \*H'\*F^{0\prime}\+\varepsilon_i
+\frac{1}{NT^{1+\delta} }\sum_{i=1}^N  \*D_T\*X_i'(\widehat{\*F}-\*F^0\*H) (\widehat{\*F}-\*F^0\*H)'\+\varepsilon_i \notag\\
&+\frac{1}{NT^{1+\delta} }\sum_{i=1}^N  \*D_T\*X_i'\*F^0\*H (\widehat{\*F}-\*F^0\*H)'\+\varepsilon_i +\frac{1}{NT^{1+\delta} }\sum_{i=1}^N \*D_T\*X_i'\*F^0  [\*H\*H' -(\*F^{0\prime}\*F^0)^{-1}] \*F^{0\prime}\+\varepsilon_i\notag\\
&= \*R_{11} + \*R_{12} + \*R_{13} + \*R_{14},
\end{align}
where $\|\*R_{11}\|$, $\|\*R_{12}\|$ and $\|\*R_{14}\|$ are all $o_P((NT)^{-1/2})$, just as before. Let us therefore consider $\*R_{13}$, which is the source of the bias in the asymptotic distribution of $\sqrt{NT}\*D_T^{-1} (\widehat{\+\beta} - \+\beta^0 )$. The purpose of Assumption \ref{ass:bias} is to control this term. Suppose first that condition (b) holds. Let $x_{j,k,t}$ denote the $j$-th row of $\*x_{k,t}$. In this notation,
\begin{align}
\lefteqn{ E\left\| \frac{1}{NT^{\delta/2 +1}}\sum_{i=1}^N  \+\varepsilon_i'\otimes (T^{-\kappa_j/2}\*X_{j,i}' \*F_{g}^0T^{-(\nu_g-\delta)/2}) \right\|^2 }\notag\\
&= \frac{1}{N^2T^{\delta+2}} \sum_{s=1}^T E\left\| \sum_{i=1}^N  \varepsilon_{i,s}\sum_{t=1}^T T^{-\kappa_j/2} x_{j,i,t} T^{-(\nu_g -1)/2}\*f_{g,t}^{0\prime} T^{(\nu_g -1)/2} T^{-(\nu_g-\delta)/2}\right\|^2 \notag\\
&= \frac{1}{(NT)^2} \sum_{i=1}^N\sum_{k=1}^NE\left[ \left(\sum_{t=1}^T T^{-\kappa_j/2} x_{j,i,t} T^{-(\nu_g -1)/2}\*f_{g,t}^{0\prime}\right)\left( \sum_{t=1}^T T^{-(\nu_g -1)/2}\*f_{g,t}^0 T^{-\kappa_j/2}x_{j,k,t} \right)\right] \sigma_{\varepsilon,ik} \notag\\
&=O(N^{-1}).
\end{align}
Moreover, by applying Assumption \ref{ass:norm} (b) to the expression given for $\|T^{-\delta/2}\widehat{\*F}_{2,d} - T^{-\nu_2/2}\*F_2^0\*h_{2,d}^{0}\|$ in the proof of Lemma \ref{applem:lam2}, we can show that
\begin{align}
\|T^{-\delta/2}\mathrm{vec}(\widehat{\*F}-\*F^0\*H)' \| = o_P( T^{-1/2} ).
\end{align}
Making use of these results, we obtain
\begin{align}
\|\*R_{3,j} \| &\le \left\| \frac{1}{NT^{\delta/2 +1}}\sum_{i=1}^N \+\varepsilon_i'\otimes (T^{-\kappa_j/2} \*X_{j,i}'\*F^0\*H)\right\| \|T^{-\delta/2}\mathrm{vec}(\widehat{\*F}-\*F^0\*H)' \| \notag\\
& = o_P( (NT)^{-1/2} ).
\end{align}
Alternatively, we may invoke Assumption \ref{ass:norm} (a) to arrive at the same result. In this case, $\|T^{-\delta/2}\mathrm{vec}(\widehat{\*F}-\*F^0\*H)' \| = o_P( 1 )$, but we also have
\begin{align}
\lefteqn{ E\left\| \frac{1}{NT^{\delta/2 +1}}\sum_{i=1}^N  \+\varepsilon_i'\otimes (T^{-\kappa_j/2}\*X_{j,i}' \*F_{g}^0T^{-(\nu_g-\delta)/2}) \right\|^2 }\notag\\
&= \frac{1}{(NT)^2} \sum_{i=1}^N\sum_{k=1}^NE\left[ \left(\sum_{t=1}^T T^{-\kappa_j/2} x_{j,i,t} T^{-(\nu_g -1)/2}\*f_{g,t}^{0\prime}\right)\left( \sum_{t=1}^T T^{-(\nu_g -1)/2}\*f_{g,t}^0 T^{-\kappa_j/2}x_{j,k,t} \right)\right] \sigma_{\varepsilon,ik} \notag\\
&= \frac{1}{(NT)^2}\sum_{t=1}^T \sum_{i=1}^N\sum_{k=1}^NE[ T^{-\kappa_j} x_{j,i,t}x_{j,k,t} T^{-(\nu_g -1)} E(\|\*f_{g,t}^{0}\|^2| \+\varepsilon_t)] \sigma_{\varepsilon,ik} \notag\\
&+ \frac{2}{(NT)^2}\sum_{t=2}^T\sum_{s<t} \sum_{i=1}^N\sum_{k=1}^NE[ T^{-\kappa_j} x_{j,i,t}x_{j,k,s} T^{-(\nu_g -1)} E(\*f_{g,t}^{0\prime}\*f_{g,s}^{0}| \+\varepsilon_t,\+\varepsilon_s)] \sigma_{\varepsilon,ik} \notag\\
&=O(1) \frac{1}{(NT)^2}\sum_{t=2}^T\sum_{s<t} \sum_{i=1}^N\sum_{k=1}^N |\phi_{ts}| |\sigma_{\varepsilon,ik}| = O((NT)^{-1}),
\end{align}
and so $\|\*R_{3,j} \|$ is of the same order as before. The proof under Assumption \ref{ass:norm} (c) is simpler and is therefore omitted.

Hence, by adding the results,
\begin{align}
\sqrt{NT}\|\*R_1\| \le \sqrt{NT}(\|\*R_{11}\| + \|\*R_{12}\| + \|\*R_{13}\| + \|\*R_{14}\|) = o_P(1).
\end{align}
We have therefore shown that
\begin{align}
\sqrt{NT}\left\|\frac{1}{NT} \sum_{i=1}^N \*D_T\*X_i'(\*P_{\widehat{F}} - \*P_{F^0})\+\varepsilon_i \right\| = o_P(1),
\end{align}
and we can similarly show that
\begin{align}
\sqrt{NT}\left\|\frac{1}{NT} \sum_{i=1}^N\sum_{j=1}^N \*D_T\*X_i'(\*P_{\widehat{F}} - \*P_{F^0}) \*X_j \*D_T a_{ij}\right\| = o_P(1).
\end{align}
These results can be inserted into \eqref{eq:hatbetaexp3}, giving
\begin{align}
\sqrt{NT}\*D_T^{-1} (\widehat{\+\beta} - \+\beta^0 ) = \*B(\*F^0)^{-1} \frac{1}{\sqrt{NT}}\sum_{i=1}^N\*D_T \*Z_i(\*F^0)'\+\varepsilon_i + o_P(1) .
\end{align}
The sought result now follows from Assumption \ref{ass:norm}.\hspace*{\fill}{$\blacksquare$}

\end{document}